\documentclass[a4,center,fleqn]{NAR}

\usepackage{textcomp,booktabs}
\usepackage[usenames,dvipsnames]{color}
\usepackage{colortbl}
\definecolor{mygray}{gray}{.9}

\usepackage[section] {placeins}

\usepackage{float}
\usepackage{mathrsfs}
\usepackage{multirow}
\usepackage{caption}

\begin{document}

\title{Random Fragments Classification of Microbial Marker Clades with Multi-class SVM and N-Best Algorithm}

\author{%
Jingwei Liu
\footnote{To whom correspondence should be addressed.
 Email: liujingwei03@tsinghua.org.cn}
 }

\address{%
School of Mathematics and System Sciences,Beihang University,Beijing,100191,P.R. China
}

\history{%
}

\maketitle

\begin{abstract}

Microbial clades modeling is a challenging problem in biology based on microarray genome sequences, especially in new species gene isolates discovery and category. Marker family genome sequences play important roles in describing specific microbial clades within species, a framework of support vector machine (SVM) based microbial species classification with N-best algorithm is constructed to classify the centroid marker genome fragments randomly generated from marker genome sequences on MetaRef. A time series feature extraction method is proposed by segmenting the centroid gene sequences and mapping into different dimensional spaces. Two ways of data splitting are investigated according to random splitting fragments along genome sequence (DI) , or separating genome sequences into two parts (DII).Two strategies of fragments recognition tasks, dimension-by-dimension and sequence--by--sequence, are investigated. The k-mer size selection, overlap of segmentation and effects of random split percents are also discussed. Experiments on 12390 maker genome sequences belonging to marker families of 17 species from MetaRef show that, both for DI and DII in dimension-by-dimension and sequence-by-sequence recognition, the recognition accuracy rates can achieve above 28\% in top-1 candidate, and above 91\% in top-10 candidate both on training and testing sets overall.

\end{abstract}

\section{Introduction}

Tremendous genomic sequences measured by new generation sequencing machines from second-generation (2G) to third-generation (3G) and fourth-generation (4G) platforms [1] put forward high level requirement of automatical classification and identification for genomic analysis techniques. Designing more accurate and effective models for clades or species classification and disease diagnose are challenging problems facing the sea amount of genomic  fragments. In recent three decades, many machine learning and statistical learning methods are developed in genomic analysis, such as BLAST[2-4], hidden Markov model (HMM) [5-15], support vector machine (SVM)[16-34], combination of linear discriminant analysis (LDA) and artificial neural network (ANN) [35], etc.

HMM is a powerful model in microbial clade classification and genome associate disease analysis [5,14,15]. However, HMM has a limitation in genomic analysis that the number of states in HMM model ($4^k$) grows exponentially with the increasing of k-mer size, and consequently leading to spareness of gene segments for training each state, this phenomenon is distinct especially in rare microbes and diseases. SVM is another popular technique to solve the relative sparse data case in pattern recognition and machine learning [36-39]. Many SVM literatures address the prediction of the microbial genomes classification and prediction [16,23,28,31]. The standard SVM [36-39] is adopted in our fragments modeling.

Feature selection is the first and key step in genomic information analysis. k-mer based sequence binning method and K-means clustering method are two widely used techniques in genomic analysis. k-mer binning method is widely adopted in genomic analysis[40-43], SVM--based genomic analysis[31], and LDA \& ANN based genomic analysis platform[35]. K--means clustering is utilized in both HMM--based [7] and SVM--based genomic analysis[28]. Although these two preprocessing methods can obtain statistical property of genome sequences, they may mask and neglect the subtle or particular local information along genome sequence. In genomic analysis, the genomic information in some species or diseases DNA sequences may exist in some local position under some kinds of measurements. The difference among various cancers DNA sequences from a same human being may appear in some local places of human DNA sequences under an appropriate measurement. For microbial classification, the ideal model of classification and clustering is to make the fragments in same specie more closer than those from other species. Hence, the real fragments along gene sequences are taken into account in this paper. One of our motivation is to examine the accuracy of each fragment belonging to considering species, regardless of the big data problem and time computation consumption. This kind of classification task is called dimension-by-dimension classification. The k-mer size and overlap size are two important issues in this kind of feature extraction method, and investigated in experiments. Furthermore, in another point of view, if all the extracted fragments from a same sequence are taken as a whole of connected subsequence, the classification task to determine the specie of the pseudo subsequence is called sequence-by-sequence classification.
In previous metagenomic sequences analysis literatures, too
\enlargethispage{-65.1pt}
short length sequences are cut off, this limitation is not valid in our model framework.

Precision and recall is a popular criterion in statistical biology information analysis besides accuracy [44]. Based on precision and recall, area under curve (AUC) of receiver operating characteristics  (ROC) is a metric for classification[45]. [46]points out that the practical application of ROC curve to determine parameter optimization is the optimal operating point (OOP) . Limited to over--fitting problem in machine learning, Precision \& recall criterion and ROC curve learning also fall into over--optimism[47]. Based on the multi-class SVM employed in experiments [39], the maximum results over all parameters of SVM on both training and testing data sets are proposed to show the performance of classification on fragmental genome sequence in both dimension-by-dimension or sequence-by-sequence tasks and avoid the over--optimization on training data sets. Ranking is a popular technique in Bioinformatics [48,49] and speech recognition, where it is called N-best algorithm [50-52], the SVM combined with N-best algorithm framework is put forward  to report the experimental results and give an intuitive grasp of the confusion of  microbial species.

Additionally, fragments from genome sequences of maker families from all microbial species on MetaRef \{http://metaref.org\} involved in the experiments show the performance of SVM with N-best model and the effectiveness of k-mer size and overlap in microbial genomic information analysis.

\section{MATERIALS AND METHODS}
\subsection{Sample preparation}

The genomic sequences are manually extracted from MetaRef and centroids\_v.1.0.fna file[53], the marker genomic sequences of all microbes are totally included in 17 species \{Mycoplasma gallisepticum, Alkaliphilus metalliredigens, Streptococcus gallolyticus, Enterococcus gallinarum, Geobacter metallireducens, Treponema pallidum, Phaeobacter gallaeciensis, Bifidobacterium gallicum, Cupriavidus metallidurans, Isosphaera pallida, Burkholderia mallei, Burkholderia pseudomallei, Eubacterium hallii, Leuconostoc fallax, Mycoplasma alligatoris, Prevotella pallens, Vibrio coralliilyticus\}, and 12390 pure DNA \{A, C, T, G\} sequences  of marker families without ambiguous DNA [54] are involved in the experiments. The frequency distribution of 12390 gene sequence lengths is shown in Fig.1. All of the DNA sequences are involved in the experiments from minimum 51 to maximum 14298, and average length is 748.9265, no short length DNA sequence is abandoned. In pattern recognition task, the 17 species are treated as 17 classes, and the total multi--class number M is set to 17. The \{A, C, T, G\} is mapped into \{1,2,3,4\} respectively. According to the definitions of marker family gene and centroid [53], the selected 12390 genes represent all the clades under the 17 species.

\begin{figure}[!htbp]
\begin{center}
\begin{minipage}{6.68cm}
\includegraphics[width=6.68cm]{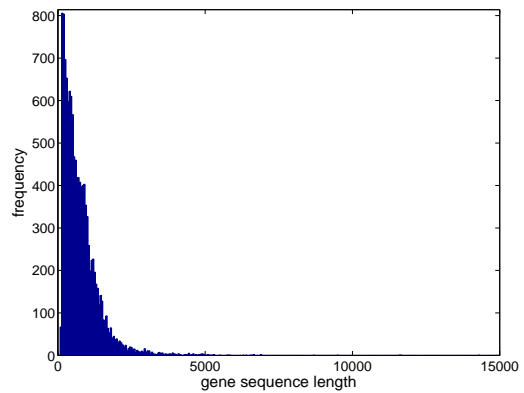}
\caption{The frequency of 12390 gene sequences length (bp).}
\end{minipage}
\end{center}
\end{figure}

\begin{figure}[!htbp]
\begin{center}
\begin{tabular}{cccc}
\includegraphics[width=6.68cm]{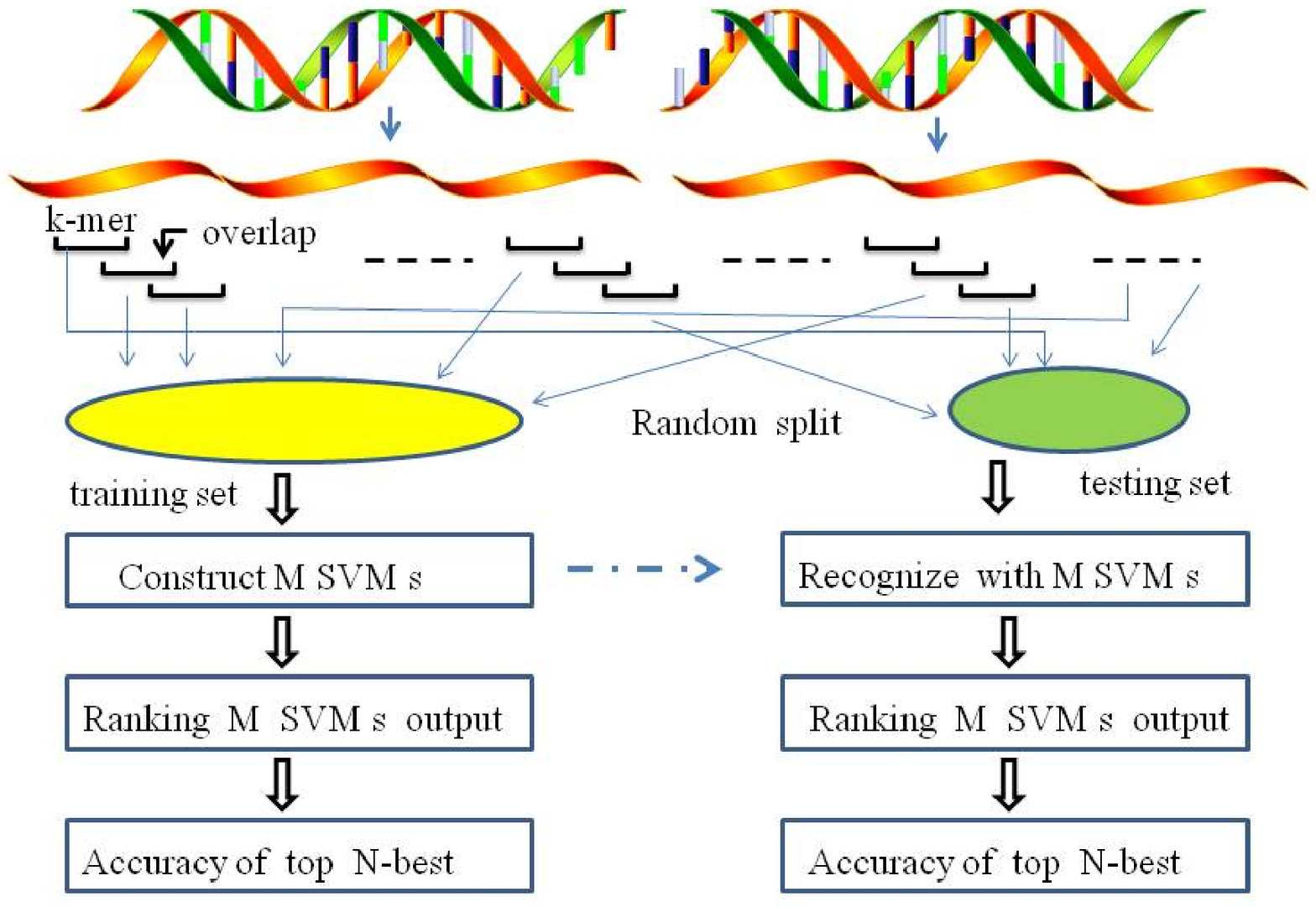}\\ (a)\\
\includegraphics[width=6.68cm]{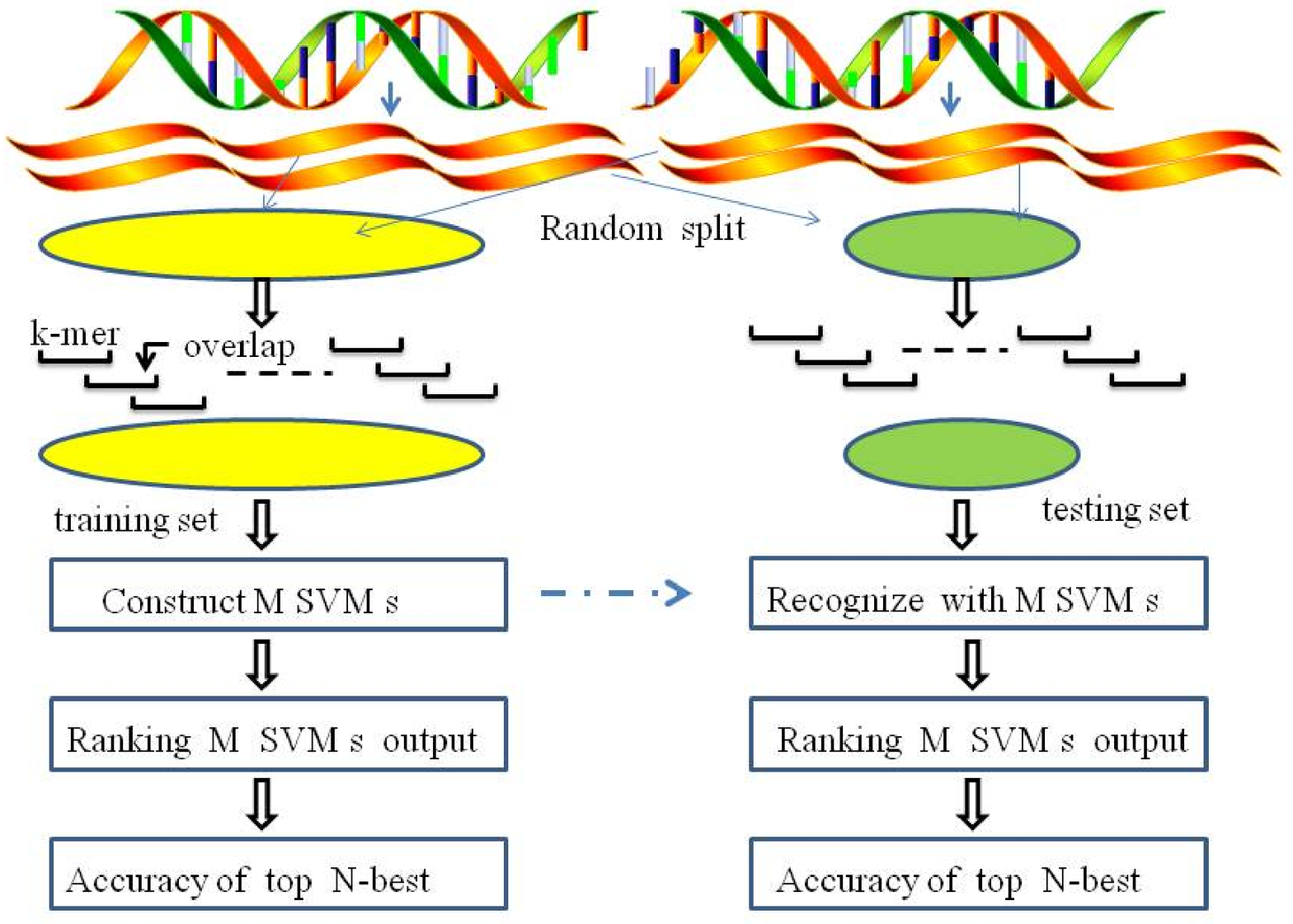}\\(b)\\
\end{tabular}
\caption{Framework of SVM with N-best candidates recognition (a) on DI  (b) on DII.}
\end{center}
\end{figure}

To investigate k-mer size of gene sequences,  the k-mer sizes of \{10, 20, 30, 40, 50\} are discussed, and the overlaps of \{0, 25\%,  50\%, 75\% \} percent of k-mer length are examined. Two types of fragment split strategies are discussed as follows. Taking all genome sequences in each specie into account, for fixed  k-mer size , overlap size  and  split percent , the first strategy (DI) is to segment each genome sequence into k--dimension (bp) sample space, and randomly select the fragments according to given split percent for training data and the rest part (1-split percent)  is denoted as testing data(Fig.2(a)).
The second strategy (DII) is to randomly split all genome sequences into training and testing parts according to given split percent, then all the sequences in training and testing data sets are mapped into k--dimension sample space separatively according to given k-mer size and overlap size(Fig.2(b)) .
DI type data set is obtained along each sequence, although the training and testing data sets are separated according to split percent, the fragments from same sequence still keep the inherent biology information. However, training and testing data sets in DII type never contain this kind of biological information except that these fragments are in same specie property.

After the above segmentation process, all the fragments in the above data sets are treated as dimension-by-dimension segment of genomic sequence. The pattern recognition task on each k--length fragment is called dimension-by-dimension classification.

Taking a further consideration, especially when overlap size is equal to 0, the fragments in training and testing data sets from a same sequence could be treated as two pseudo subsequences of one genomic sequence. As they are randomly selected, the original genomic order information along the same sequence in DI is broken down. The pattern recognition tasks on randomly pseudo genome subsequences in DI and normal order genome subsequences in DII are called sequence-by-sequence classification. While the overlap percent is larger than 0, all the subsequences in DI and DII are pseudo subsequences, the pattern recognition task can still be performed on, it is also named as sequence-by-sequence classification.

In brief, the genomic sequences are mapped into k--dimension sample space, which means the k--bp genomic fragments. To construct the SVM model, the split percents involved in experiments are set as \{60\%, 80\%, 100\% \}, When split proportion is equal to 100\% , the training and testing set are set as same one.
For both types of DI and DII, the training sets and testing sets are denoted as $\mathscr{L}(k-mer,overlap,split\ percent)$ and $\mathscr{T}(k-mer,overlap,1-split\ percent)$ respectively.
Some main size of samples in $\mathscr{L}(k-mer,overlap,split\ percent)$ and $\mathscr{T}(k-mer,overlap,1-split\ percent)$ involved in experiments are listed in Supplementary Table 1.

\subsection{Multi-class SVM}

Support Vector Machine (SVM) is a popular statistical learning and machine learning method based on structural risk minimization and VC dimensions theory, and overcomes the dimension disaster of neural network [36]. It has efficient performance and high accuracy in many classification tasks, hence widely used in various information process fields. The traditional SVM is defined for binary classification, the multi-class SVM is defined on binary SVM with one-versus-one max-wins voting strategy or one-versus-all winner-takes-all strategy [37,38].
Suppose the samples data are $\mathscr{D}=\{ (x_i,y_i)| x_i \in R^{k}, y_i=\pm 1,  1\leq i \leq S \} $. Given a nonlinear mapping function $\phi(\cdot)$, the sample data of $x_i$ is projected to high dimensional space to obtain well separation of different samples. The standard C-SVM [40] solves
\begin{equation}
\begin{array}{ll}
\displaystyle {\min_{w,b,\xi}} & \displaystyle (\frac{1}{2} w^{T}w+C\sum_{i=1}^{S}\xi_{i}) \\
\displaystyle \mbox{subject to:}           & \displaystyle y_{i}(w^{T}\phi(x_i)+b)\geq 1-\xi_{i}, \  \  \ \    \xi_{i}\geq 0,i=1,\cdots,S.
\end{array}
\end{equation}
where $w,b$ are the parameters of optimal linear hyperplane $f(x)=w^{T}\phi(x)+b$, C is the penalty parameter of the error term.  $K(x,x')=(\mathbf{x},\mathbf{x'})=(\phi(x),\phi(x'))$ is the kernel function. The key problem in solving $w$ is involved in the dual problem of above optimization problem and selection of kernel function. The radial basis function(RBF) kernel function is adopted in the experiments, where
\begin{equation*}
  K(x_i,x_j)= \mbox{exp}(-\gamma ||x_{i}-x_{j}||^{2}), \gamma >0.
\end{equation*}

Standard LibSVM 3.18 \{ http://www.csie.ntu.edu.tw/~cjlin/ libsvm/ \}is utilized in modeling the fragment vector space, and the parameter range of C-SVM  with RBF kernel is  $C\times \gamma $ $\in$ \{0.0625, 0.125, 0.25, 0.5, 1, 2, 3, 4 \} $ \times $\{0.0625, 0.125, 0.25, 0.5, 1, 2, 3, 4 \}.
Totally 64 parameter choices are employed in experiments.

\subsection{Classification Accuracy}

For genomic fragment classification, two categories of classification criteria are involved in experiments. The first one is for dimension-by-dimension classification, its aim is to classify the correct accuracy of fragment in k--bp status.

\begin{equation}
\begin{array}{ll}
\displaystyle frg_k= \displaystyle \frac{\#\{correct\ vectors\ for\ right\ class\}}{\#\{total\ vectors\}}
\end{array}
\end{equation}

The other criterion is to recognize the pseudo subsequence composed of randomly selected fragments from a same sequence for right specie,

\begin{equation}
\begin{array}{ll}
\displaystyle frs_k= \displaystyle \frac{\#\{correct\ sequences\ for\ right\ class\}}{\#\{total\ equences\}}
\end{array}
\end{equation}

The criteria of $frg_k$ and $frs_k$ examine the different aspect of gene fragment sequences, both of them address the fragments of partial pieces of genomic sequences.

\subsection{Multi--candidate Accuracy}

Ranking is a popular technique in genomic analysis and information sciences. In order to show the relationship among the species, the top n $(1\leq n\leq 10)$ ranking classes are considered in the recognition stage according to the top-n probability outputs of multi-class SVM voting. The criteria of classification accuracy with N-best algorithm are revised as follows respectively,

\begin{equation}
\begin{array}{ll}
\displaystyle frg_{k|n}= \displaystyle \frac{\#\{correct\ vectors\ in\ top-n\ candidates \}}{\#\{total\ vectors\}}
\end{array}
\end{equation}

\begin{equation}
\begin{array}{ll}
\displaystyle frs_{k|n}= \displaystyle \frac{\#\{correct\ sequences\ in\ top-n\ candidates\}}{\#\{total\ sequences\}}
\end{array}
\end{equation}

\section{RESULTS}

\subsection{Accuracy on DI with different k-mer size and Overlap}

As there are many combinations of k-mer size, overlap and split percent. Firstly, we investigate the effectiveness of these parameters in a special case that all data is involved in training, that is split percent=100\% on DI. And, the case of overlap percent=0 is examined first of all, denoted as $\mathscr{L}(*,0,100\%) $. The dimension--by-dimension recognition experimental results show that on training set $\mathscr{L}(*,0,100\%) $ , the accuracy rates trend highly along the increasing of number of N-candidates and k-mer size value (Fig.3(a), Supplementary Fig.1(a)).

Secondly, we examine the performance on $\mathscr{L}(50,*,100\%)$ with k=50, the accuracy rates increase along number of N-candidates (Fig.3(b)) and decrease along overlap percent (Supplementary Fig.1(b)).

\begin{figure}[!htbp]
\begin{center}
\begin{tabular}{cccc}
\includegraphics[width=4cm]{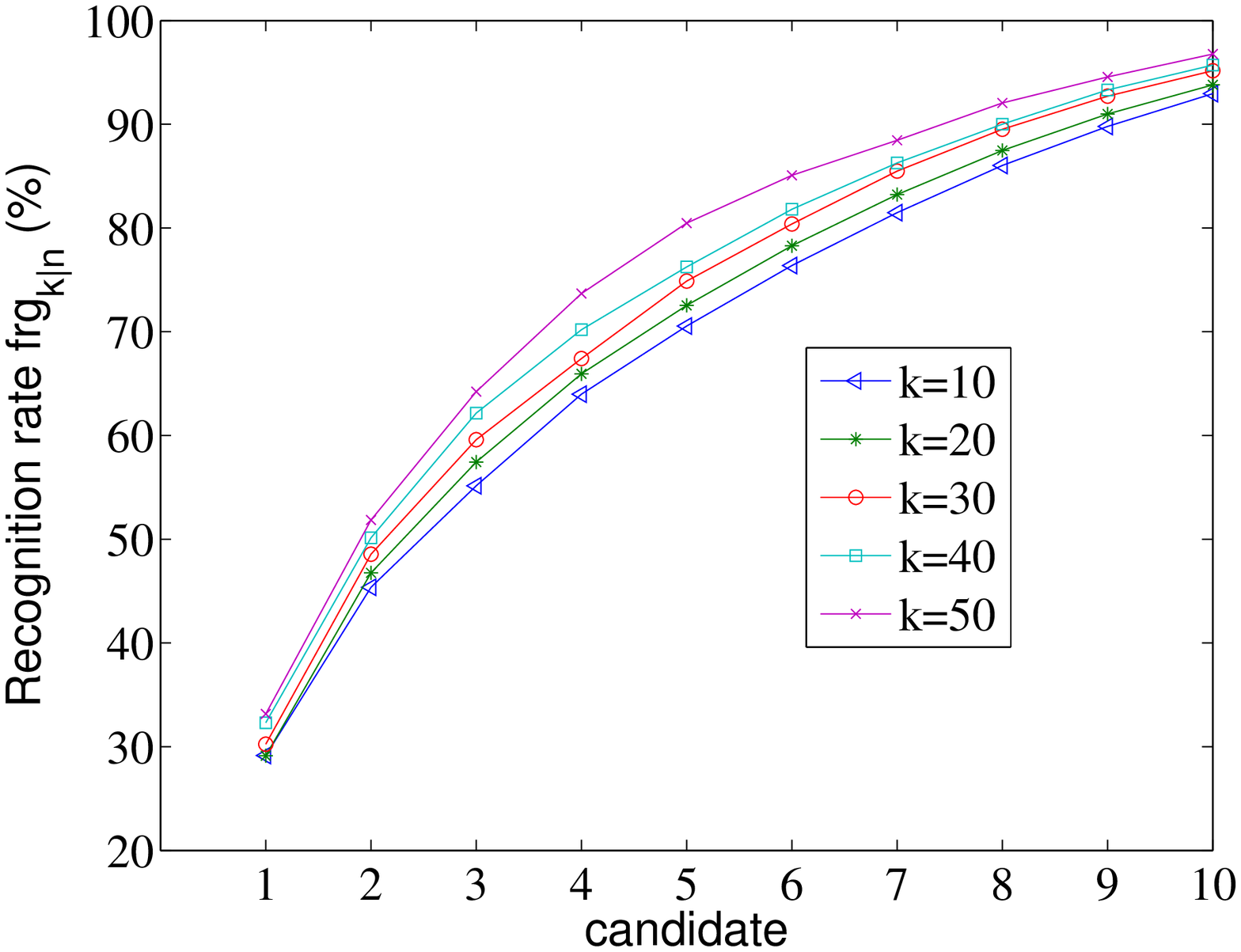} &\ \  \includegraphics[width=4cm]{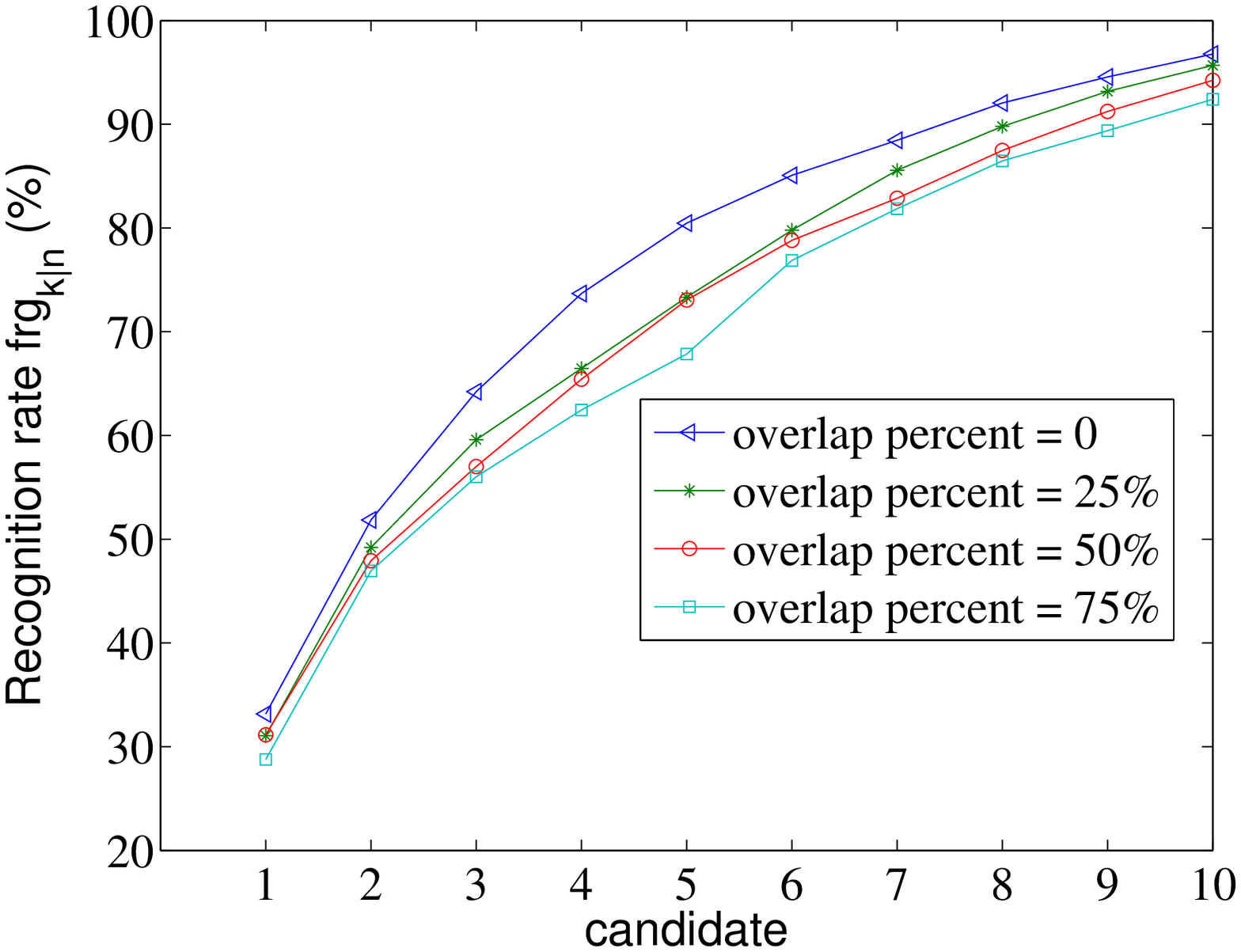} \\(a)&(b)\\
\includegraphics[width=4cm]{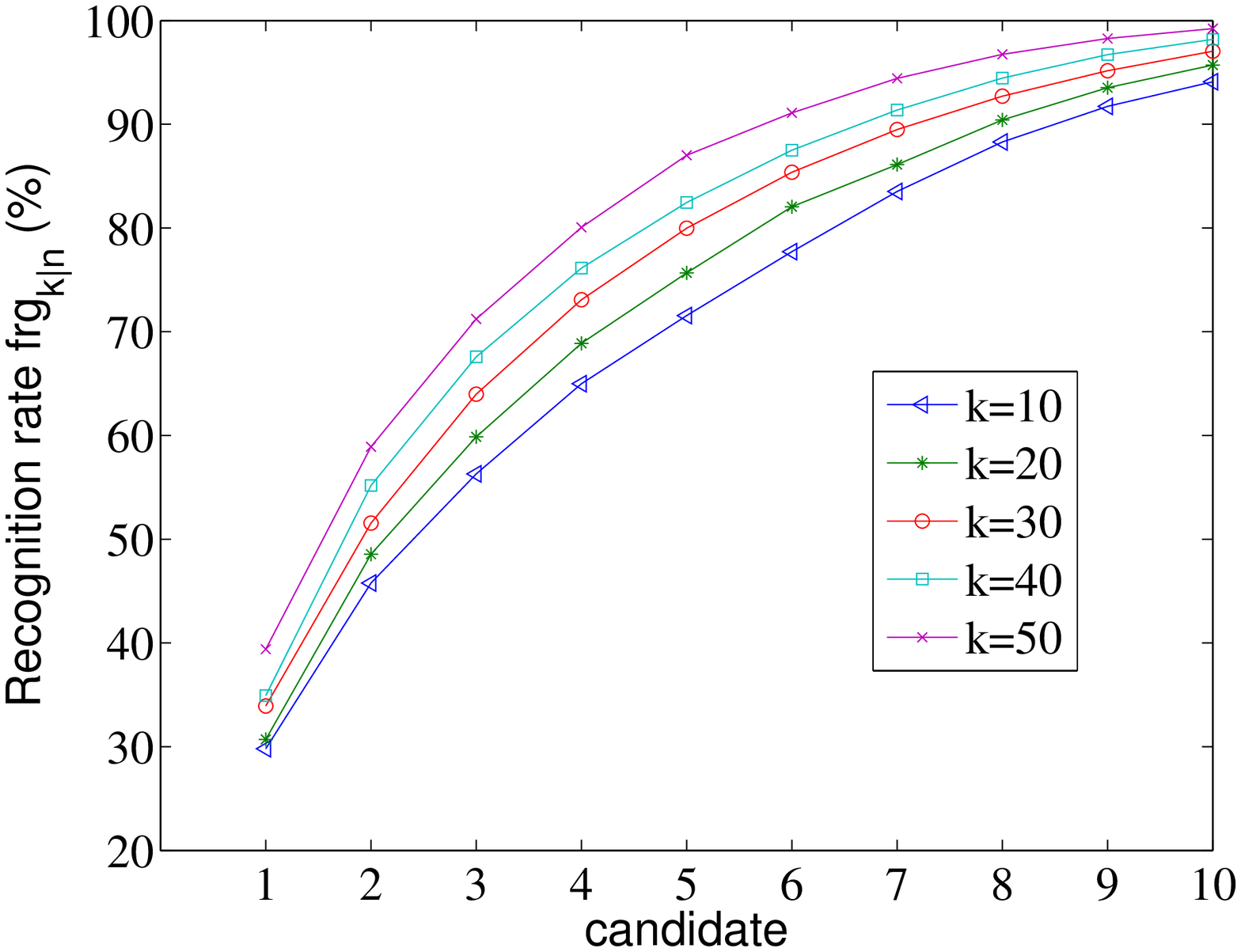}  &\ \ \includegraphics[width=4cm]{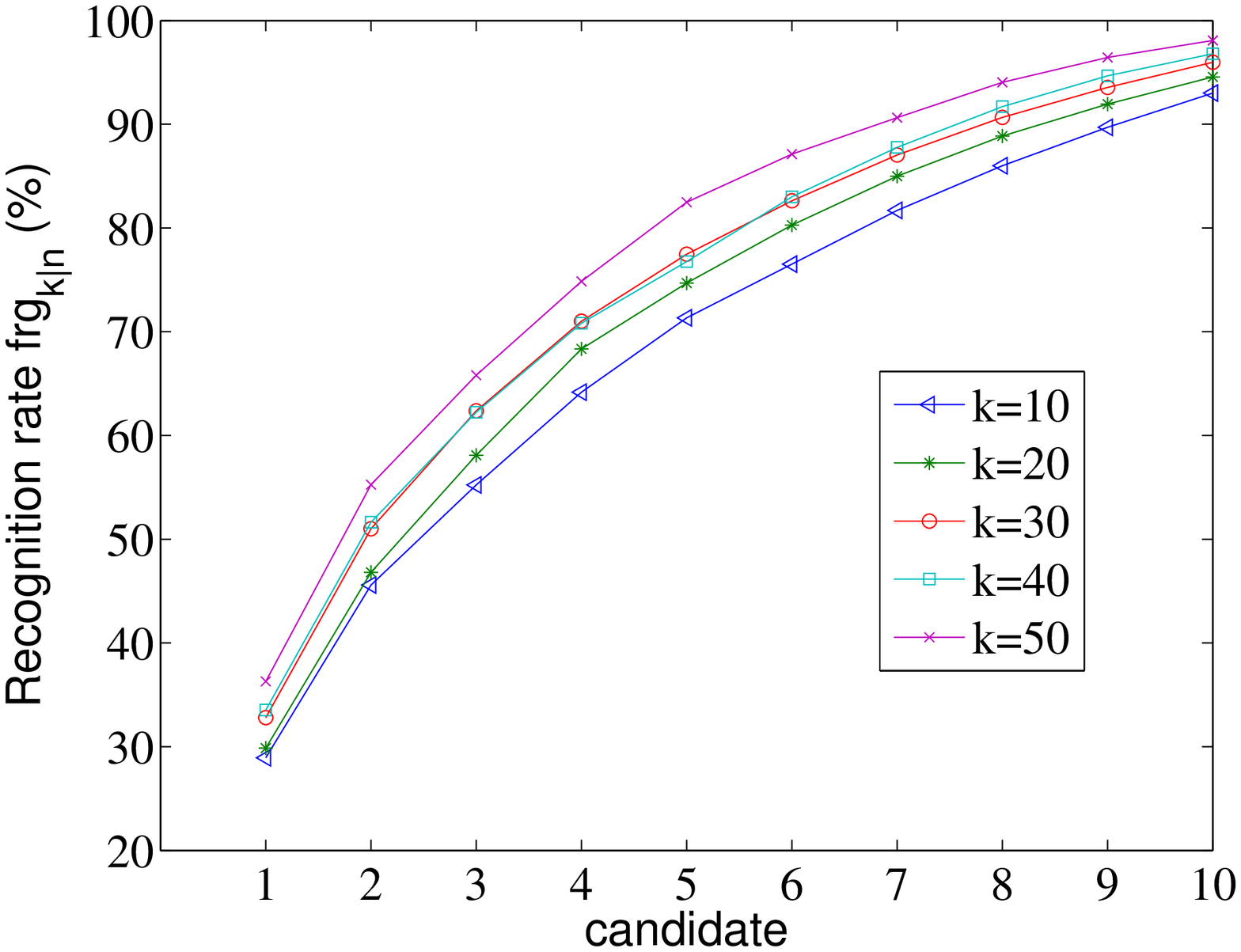} \\(c)&(d)\\
\includegraphics[width=4cm]{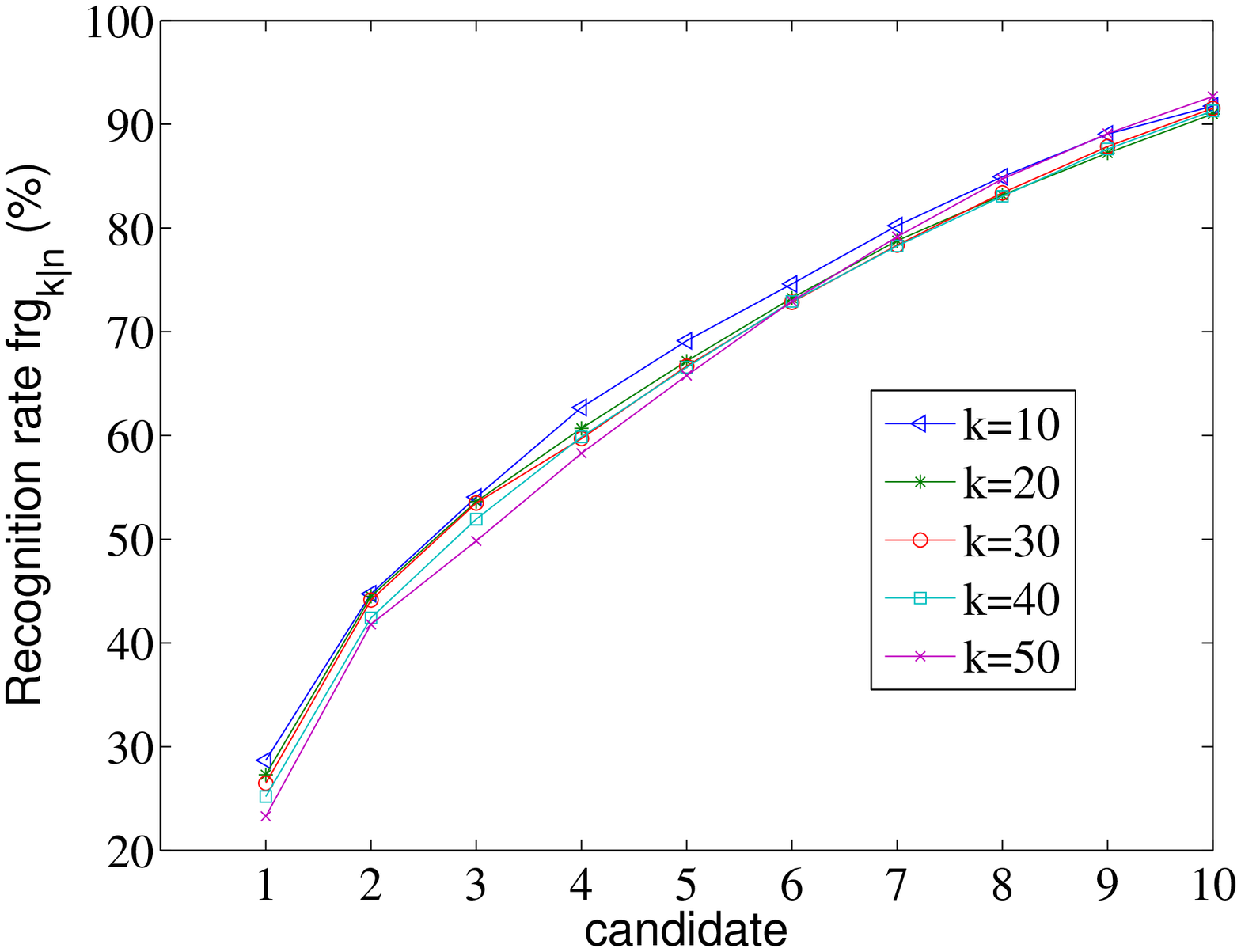} &\ \  \includegraphics[width=4cm]{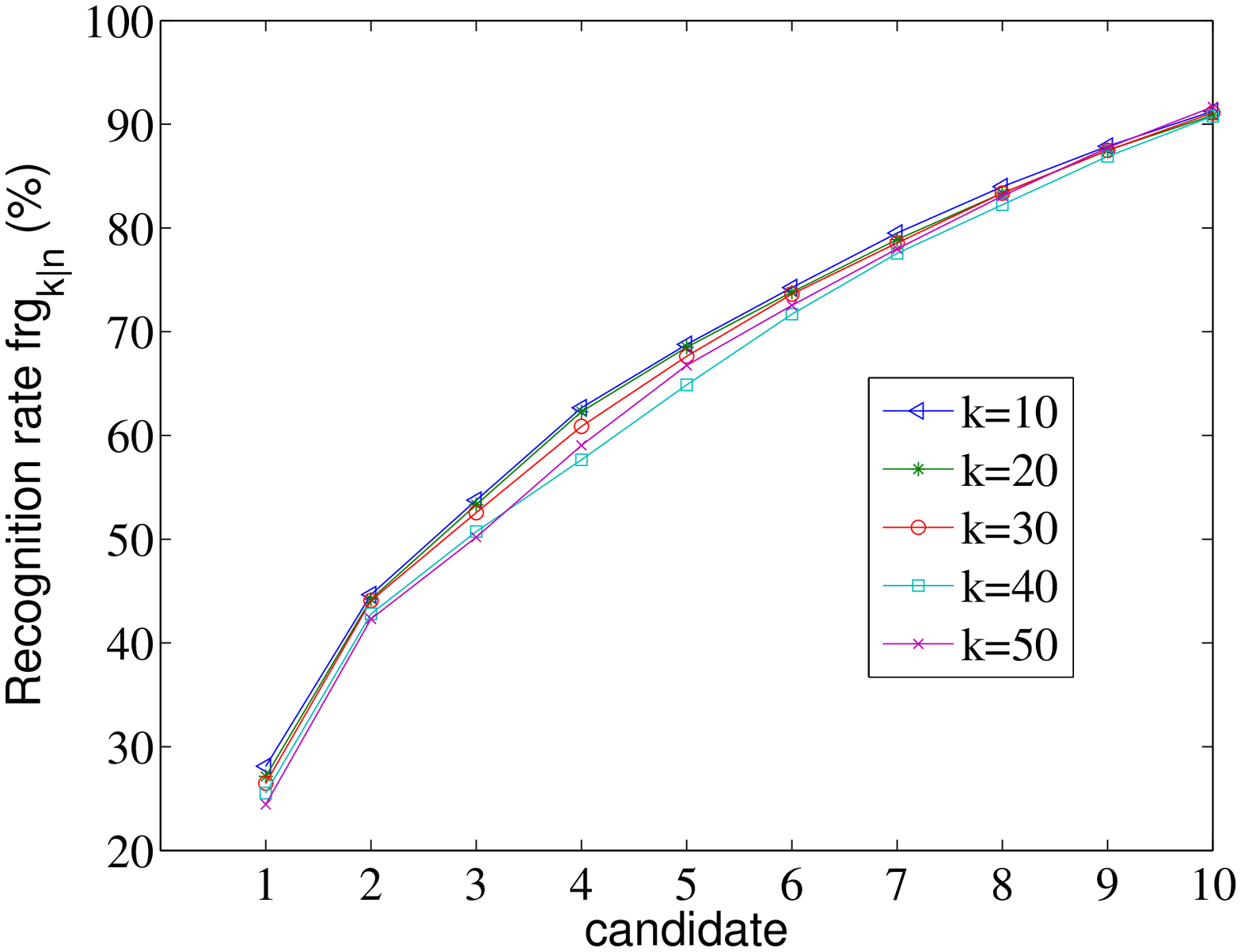} \\(e)&(f)\\
\end{tabular}
\caption{Dimension-by-Dimension recognition rates on DI along multi--candidates. (a) $\mathscr{L}(*,0,100\%)$ (b) $\mathscr{L}(50,*,100\%)$ (c) $\mathscr{L}(*,0,60\%)$ (d) $\mathscr{L}(*,0,80\%)$ (e) $\mathscr{T}(*,0,40\%)$ (f)$\mathscr{T}(*,0,20\%)$. }
\end{center}
\end{figure}

Thirdly, we perform multi-class SVM with N-best algorithm on DI with split percent=60\% and 80\% cases. In the dimension-by-dimension case, in training sets, the accuracy rates increase along number of candidates (Fig.3(c)(d)) and k-mer size (Supplementary Fig.1(c)(d)).
While in testing sets, the accuracy rates increase along number of candidates (Fig.3(e)(f)) and slightly decrease along k-mer size (Supplementary Fig.1(e)(f)) especial in low candidate values.


Fourthly, the multi-class SVM with N-best algorithm are performed in sequence-by-sequence cases of  $\mathscr{L}(*,0,100\%)$ and $\mathscr{L}(50,*,100\%)$, the same conclusions hold as dimension-by-dimension cases (Fig.4(a)(b),Supplementary Fig.2(a)(b)). At same time, the accuracy rate of recognition in pseudo sequence-by-sequence form is more higher than in dimension-by-dimension fragment form.

Fifthly, in the sequence-by-sequence cases on DI with split percent=60\% and 80\% cases,  the accuracy rates increase along number of candidates (Fig.4(c)(d)), but will not do so along k-mer size (Supplementary Fig.2(c)(d)) in training sets. The accuracy rates increase from 10 to 30 of k-mer size from top-1 to top-6 candidate cases, however, decrease in top-10 case within k-mer size from 10 to 30 (Supplementary Fig.2(c)(d)). From 30 to 50 k-mer size, the accuracy rates decrease from top-2 to top-10 in $\mathscr{L}(*,0,60\%)$ (Supplementary Fig.2(c)) and from top-3 to top-10 in $\mathscr{T}(*,0,80\%)$ (Supplementary Fig.2(d)). These experimental results demonstrate the tendency of accuracy rates with k-mer size and top-n candidates.
In each fixed k-mer size, the principle that the larger the number of top-n candidates, the higher the accurate rates still holds.
While, in testing data, The accuracy rates under each fixed k-mer size increases along top-n candidates (Fig.4(e)(f)), however, the  accuracy rates under each fixed top-n candidates decreases along k-mer size (Supplementary Fig.2(e)(f)). The appropriate choice of k-mer size for genomic fragments and sequences predictions would be 10.

\begin{figure}[!htbp]
\begin{center}
\begin{tabular}{cccc}
\includegraphics[width=4cm]{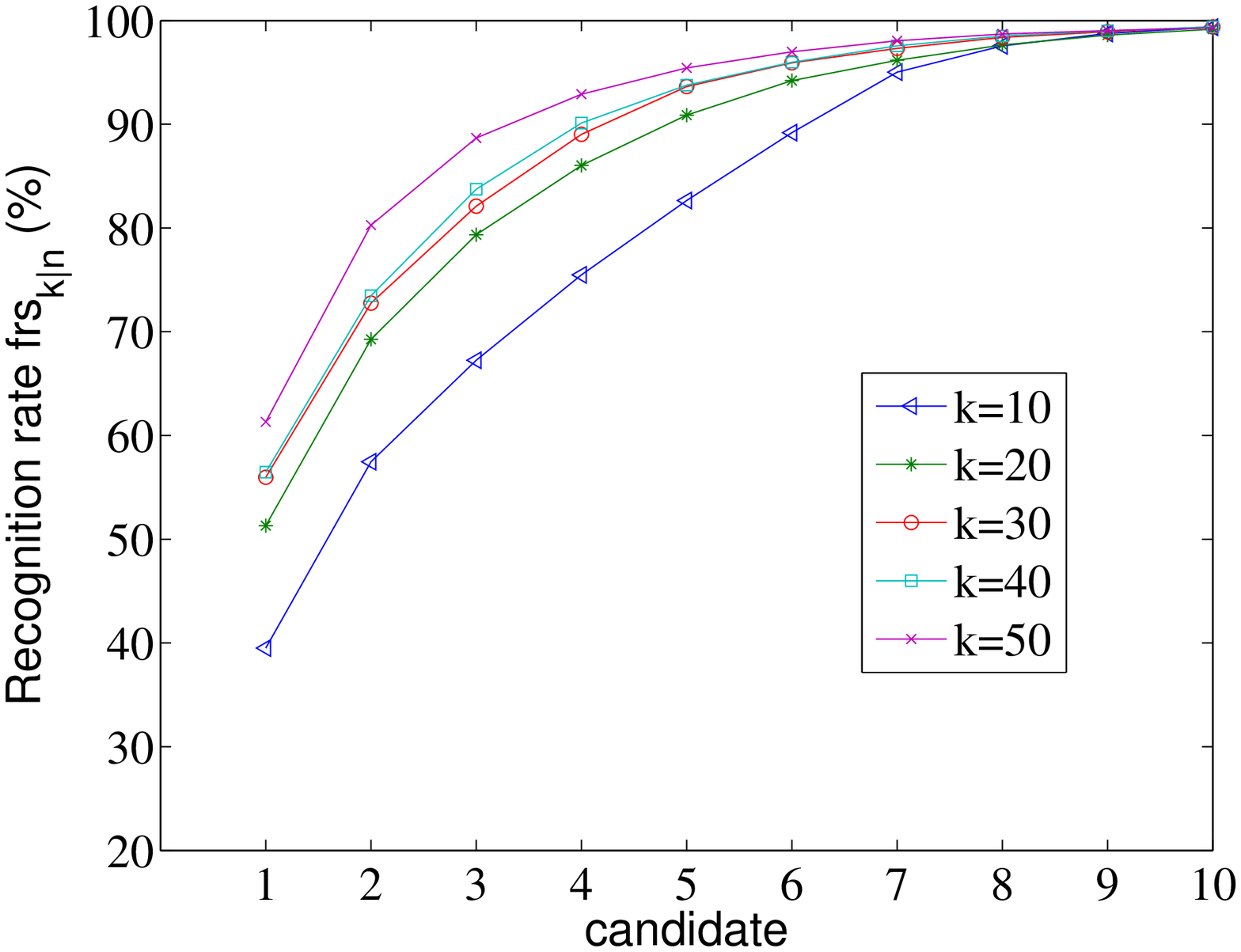}  &\ \ \includegraphics[width=4cm]{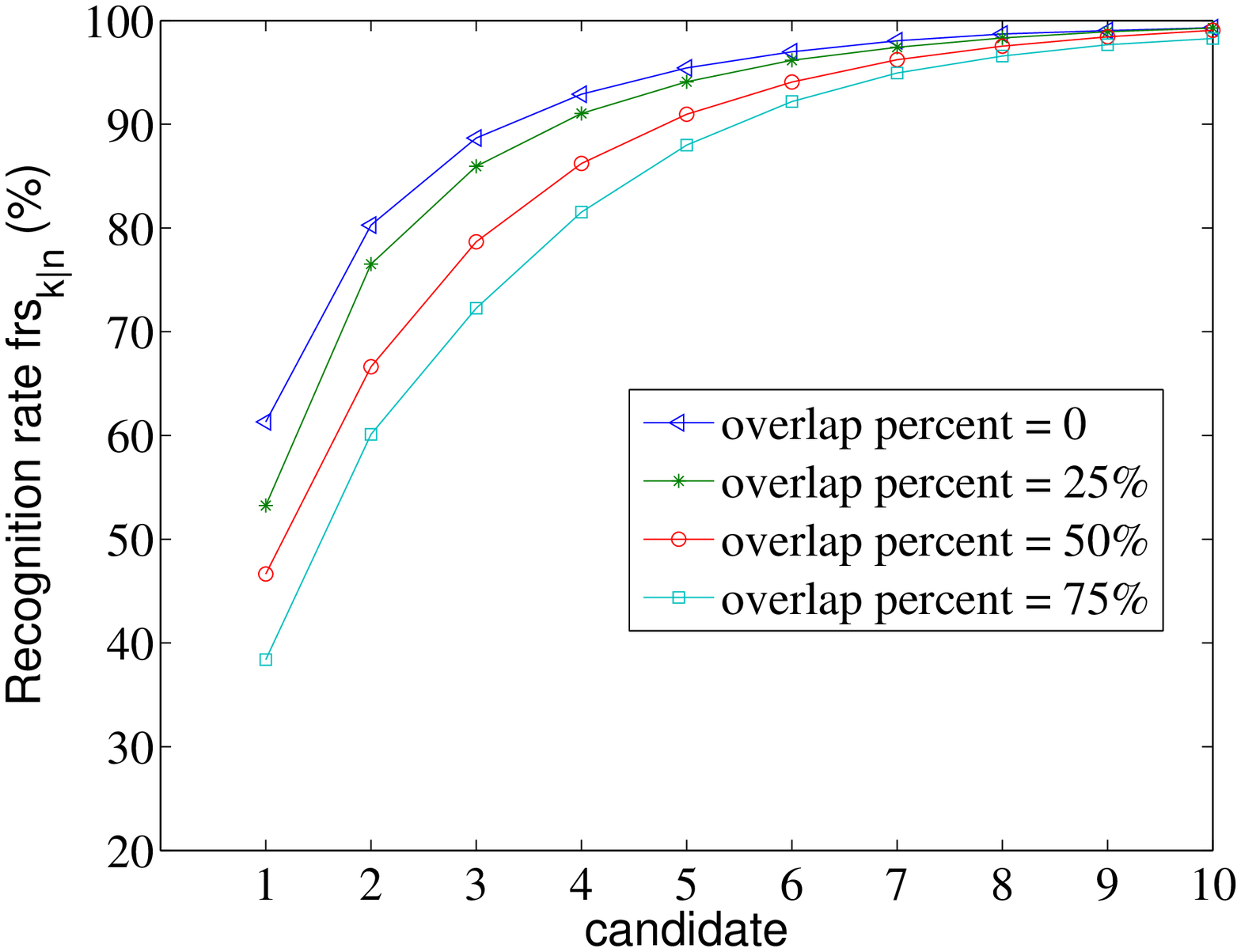} \\(a)&(b)\\
\includegraphics[width=4cm]{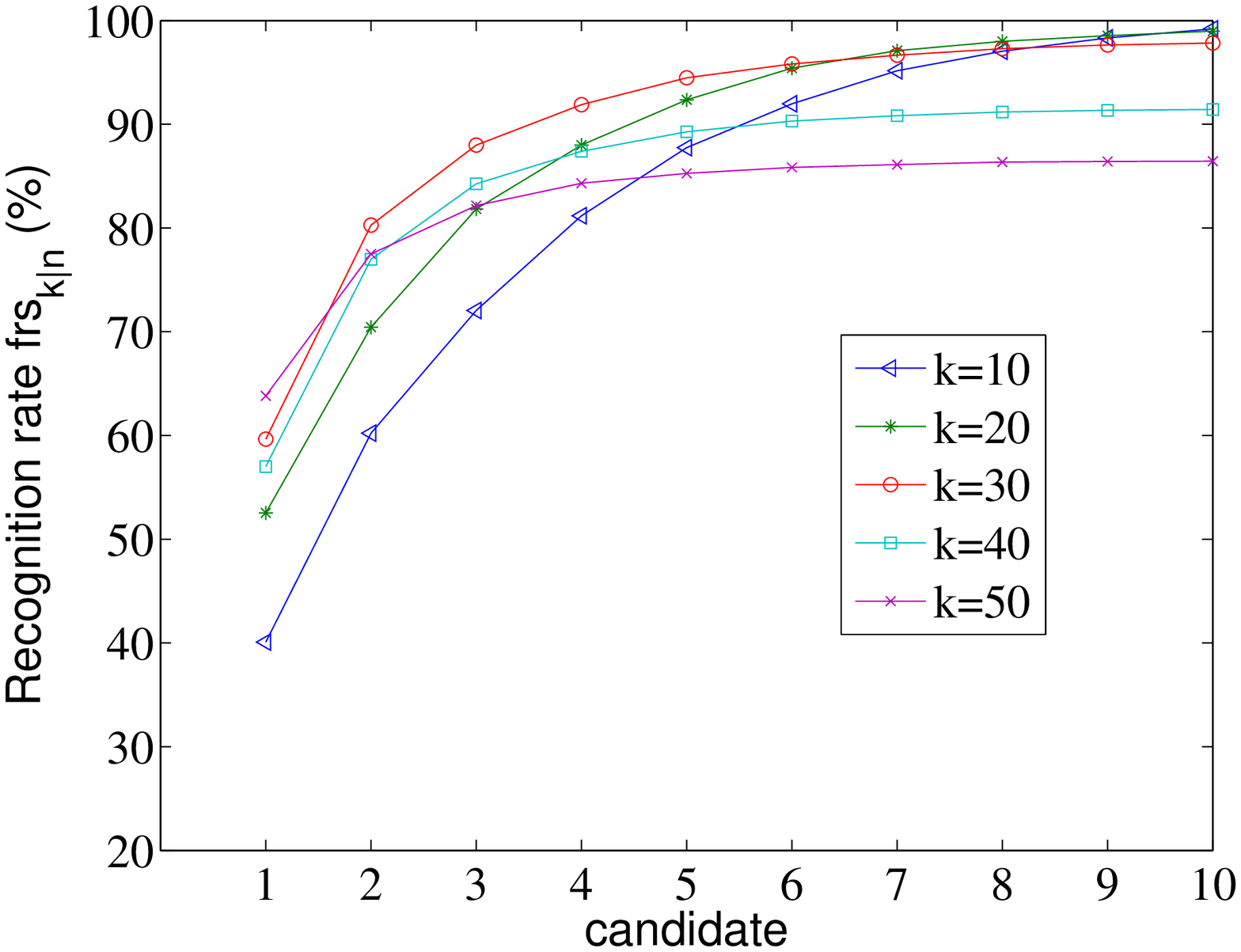}  &\ \ \includegraphics[width=4cm]{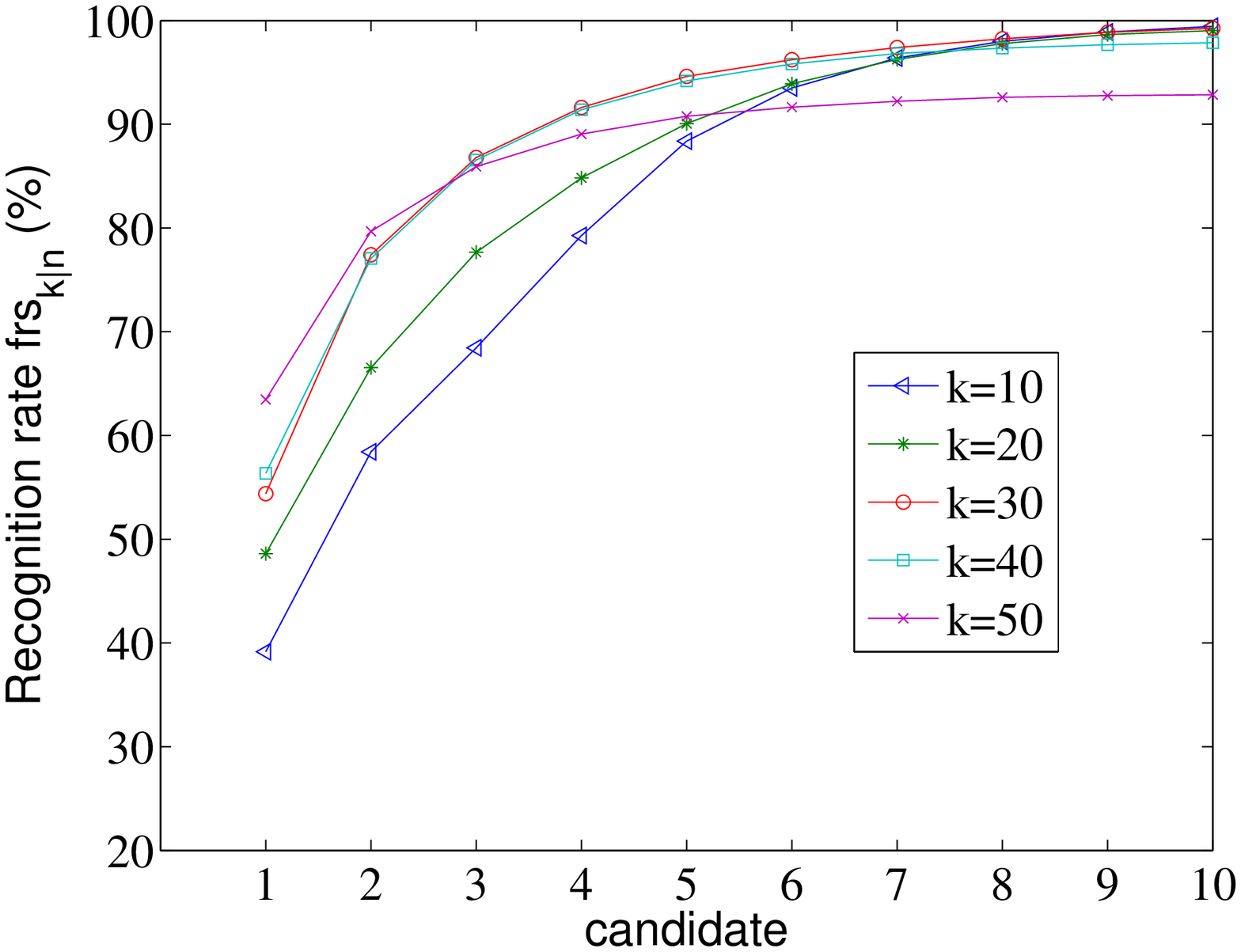} \\(c)&(d)\\
\includegraphics[width=4cm]{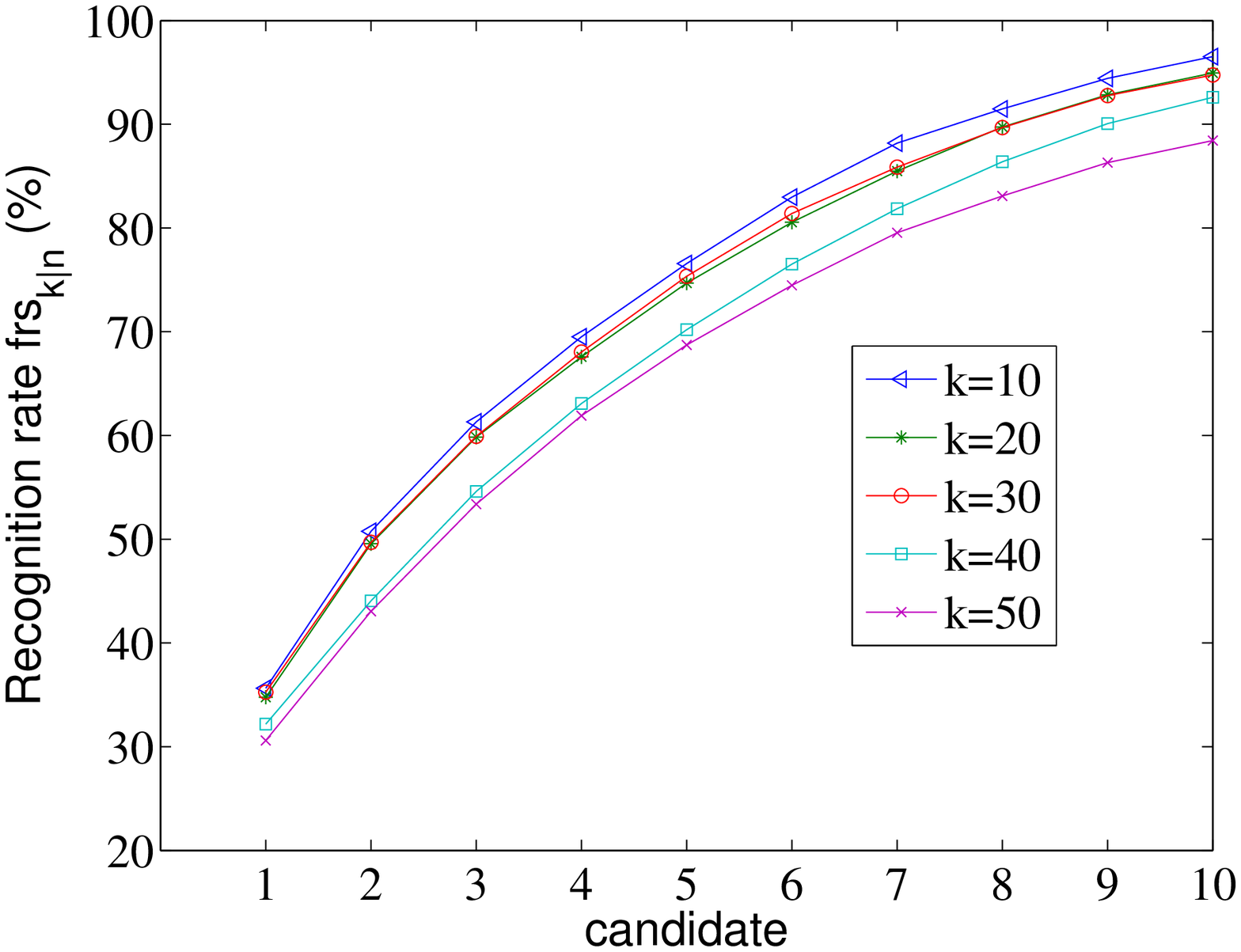}  &\ \ \includegraphics[width=4cm]{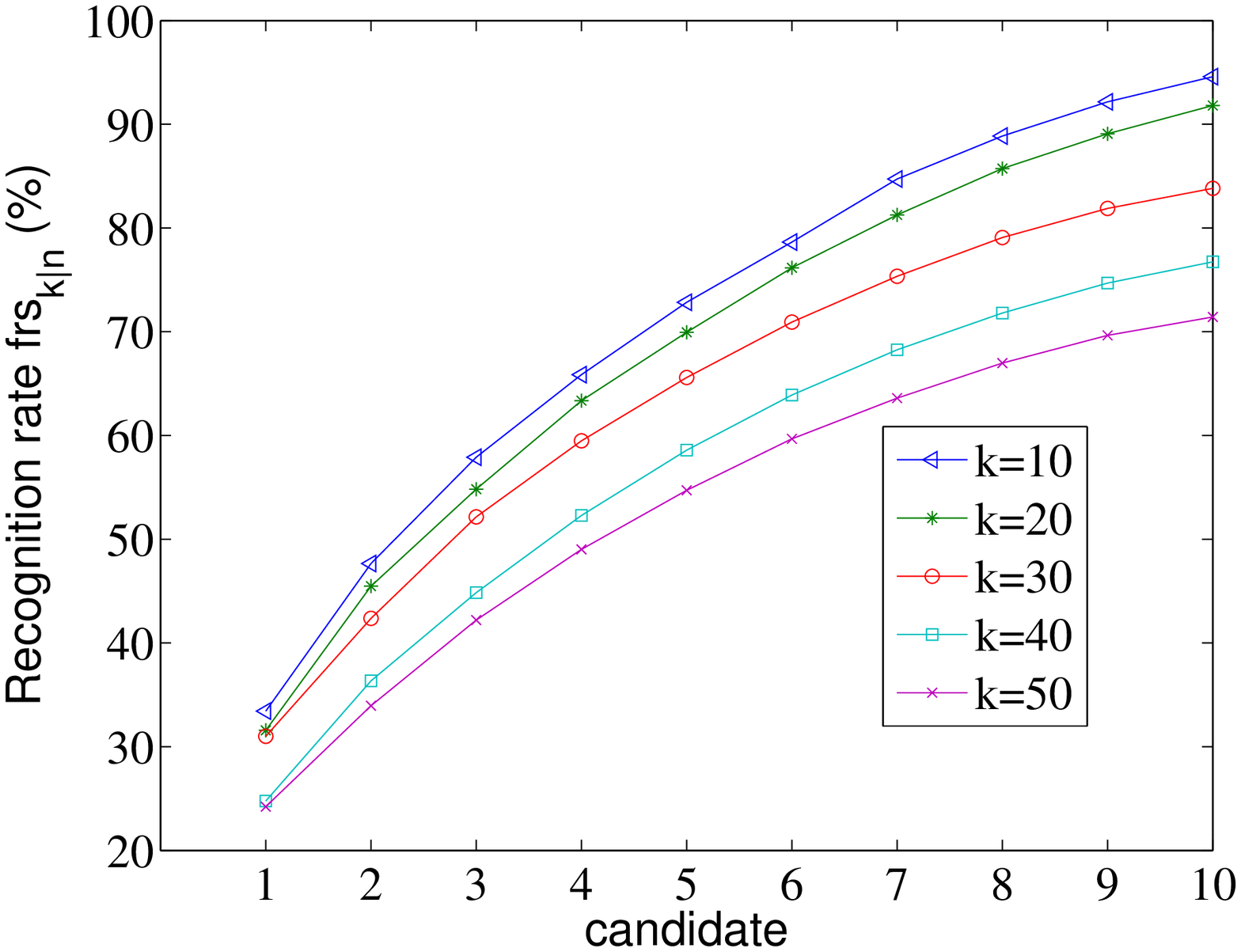} \\(e)&(f)\\
\end{tabular}
\caption{Sequence--by-sequence recognition rates on DI along multi--candidates. (a) $\mathscr{L}(*,0,100\%)$ (b) $\mathscr{L}(50,*,100\%)$ (c) $\mathscr{L}(*,0,60\%)$ (d) $\mathscr{L}(*,0,80\%)$ (e) $\mathscr{T}(*,0,40\%)$ (f)$\mathscr{T}(*,0,20\%)$. }
\end{center}
\end{figure}


Sixthly, as to the overlap problem on DI with split percent=60\% and 80\% for training and testing cases separatively, we give the discussion in the case of 50 k-mer size data sets.
In the dimension-by-dimension case, the performances on training sets and testing sets are different.
In training sets, the accuracy rates under each overlap percent increase along the top-n candidates (Fig.5(a)(b)) , and the accuracy rates under each top-n candidates decrease along the overlap percent (Supplementary Fig.3(a)(b)).
In testing sets, the accuracy rates under each overlap percent increase along the top-n candidates (Fig.5(c)(d)), and the accuracy rates under each top-n candidates are not precisely synchronized with the increasing of overlap percent (Supplementary Fig.3(c)(d)), in low top-n case, there are slightly enhancement, but do not hold in large top-n cases.

Seventhly, in sequence-by-sequence cases of 50 k-mer size data sets with split percent=60\% and 80\% for training and testing sets separatively on DI, for fixed overlap percent, the accuracy rates increases on both training (Fig.6(a)(b)) and testing data sets (Fig.6(c)(d)).
Given top-n candidates, the accuracy rates perform different along overlap percent.
In training sets, the accuracy rates decrease along low top-n candidates cases and increase in high top-n candidates cases (Supplementary Fig.4(a)(b)). But, in testing data, the accuracy rates increase along overlap percent (Supplementary Fig.4(c)(d)).
The above experimental results show that overlap  optimization is somewhat complicate in training and testing sets. The optimized overlap would be the balance between training and testing sets.

\begin{figure}[!htbp]
\begin{center}
\begin{tabular}{cccc}
\includegraphics[width=4cm]{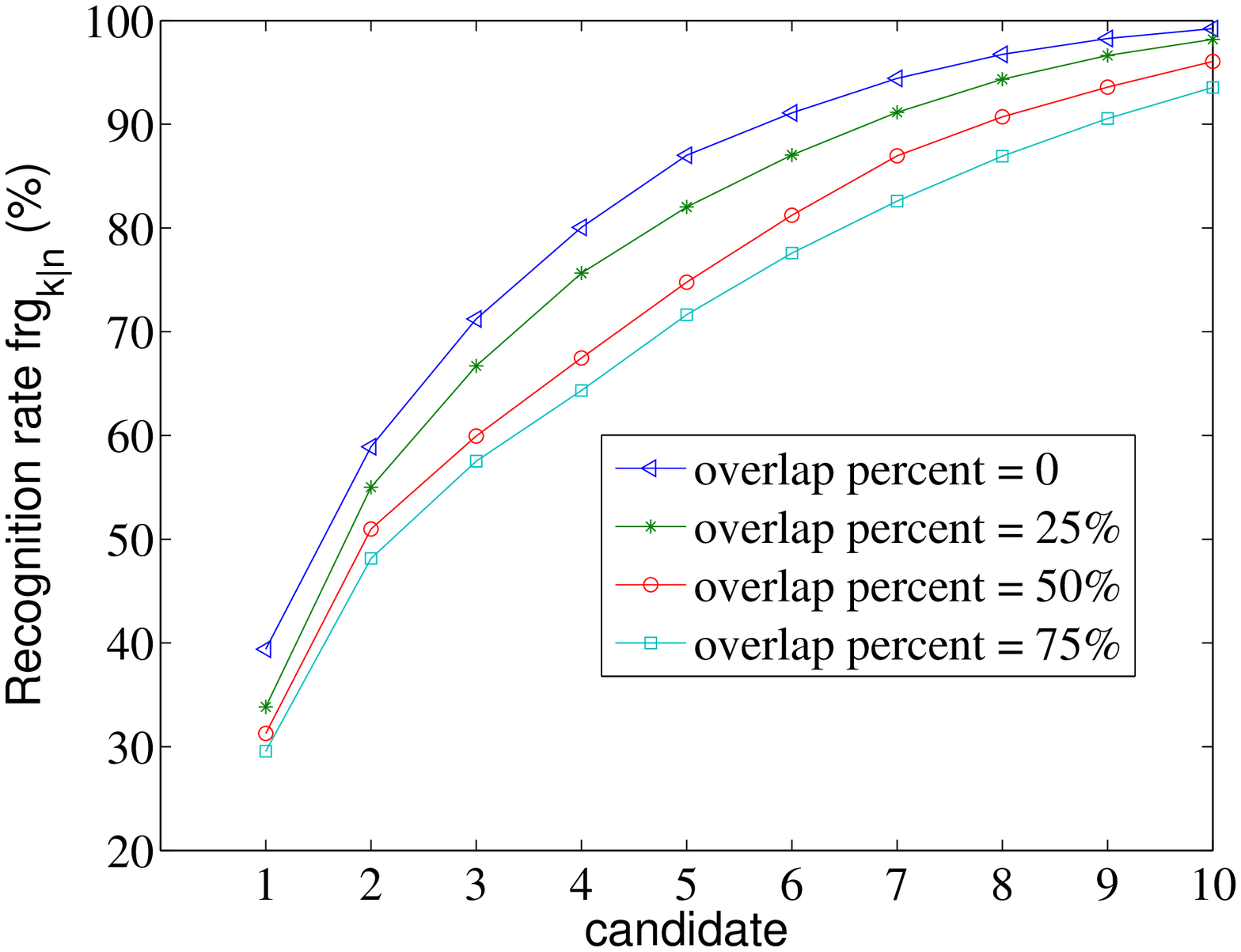} &\ \  \includegraphics[width=4cm]{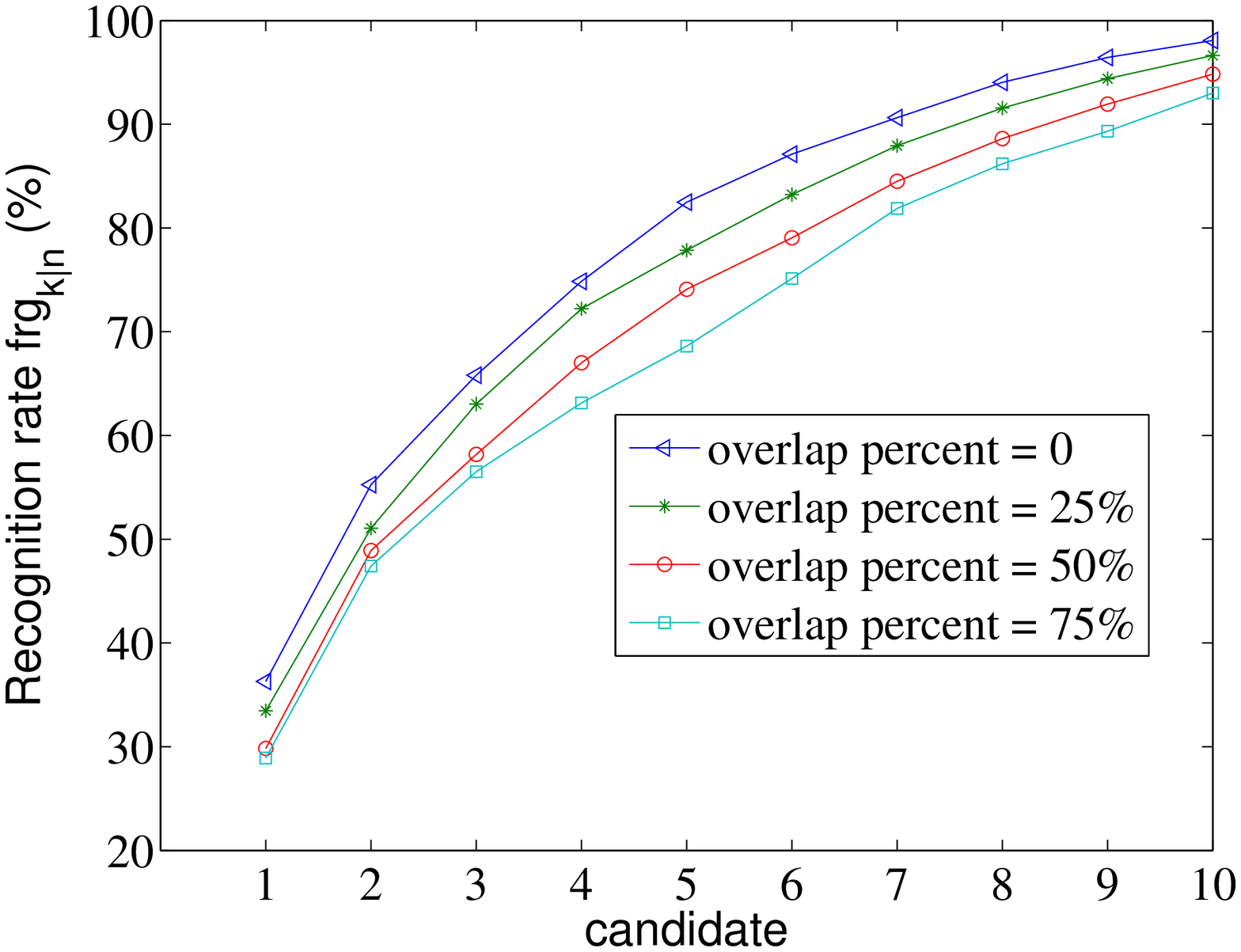} \\(a)&(b)\\
\includegraphics[width=4cm]{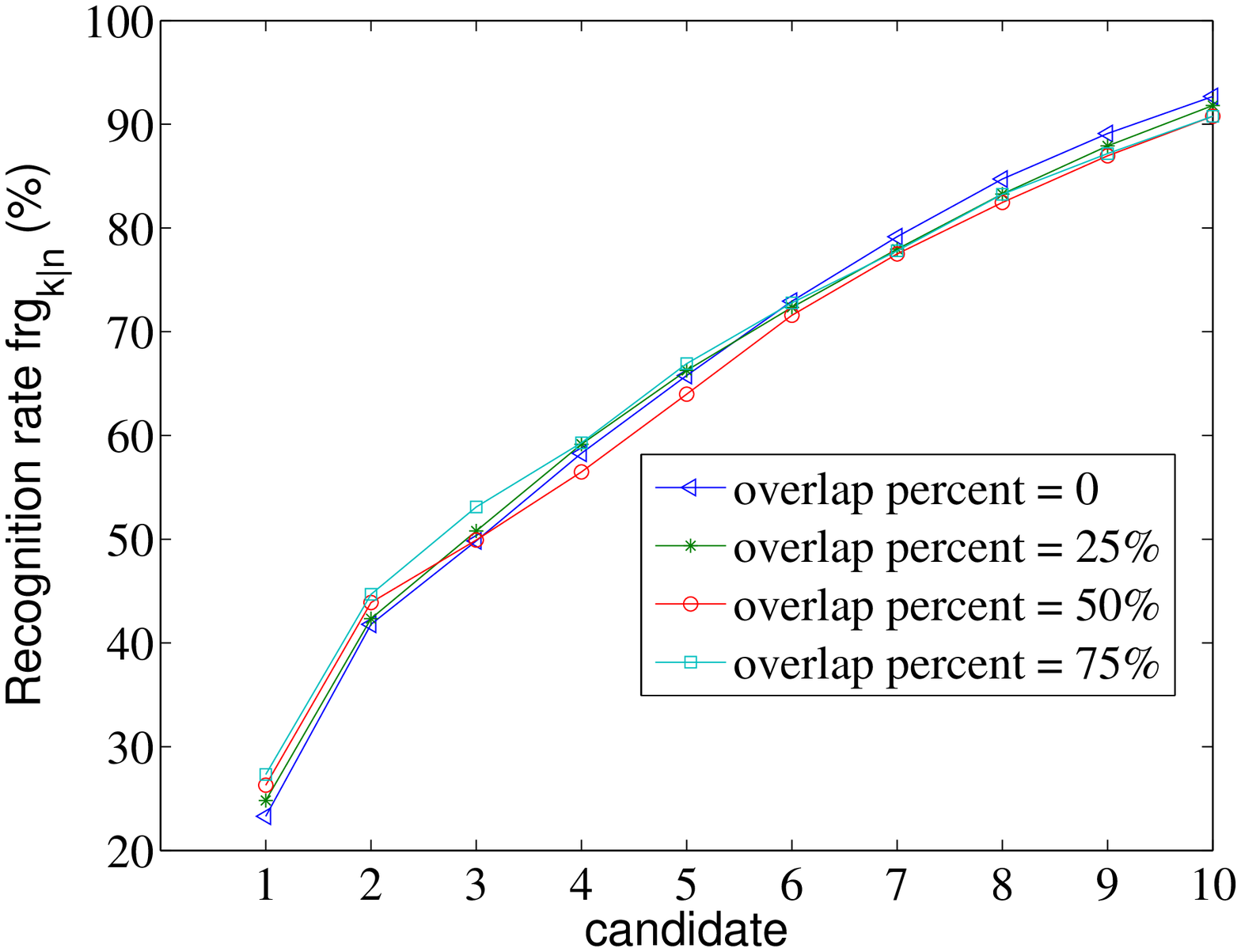} &\ \  \includegraphics[width=4cm]{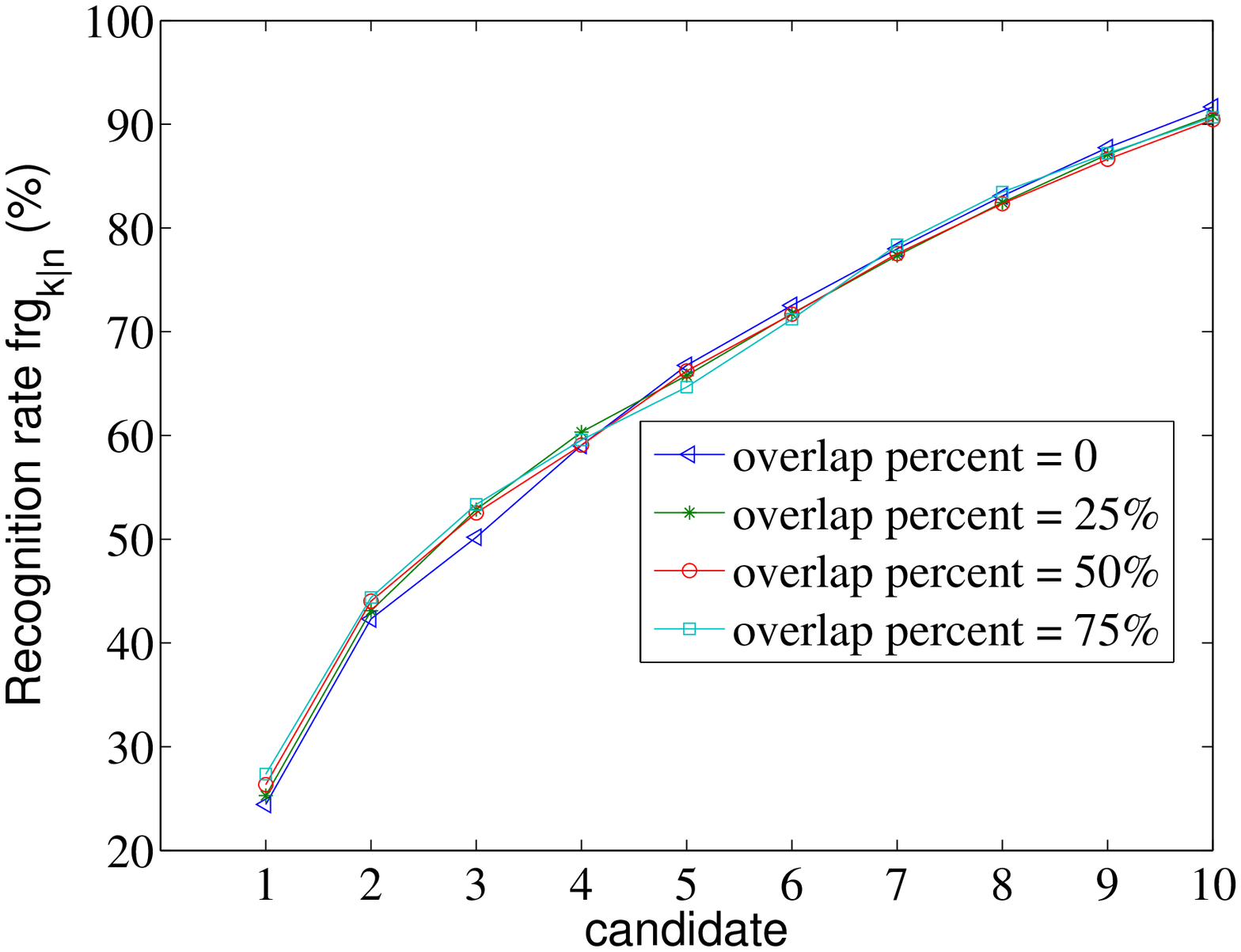} \\(c)&(d)\\
\end{tabular}
\caption{dimension-by-dimension recognition rates on DI. (a) $\mathscr{L}(50,*,60\%)$ (b) $\mathscr{L}(50,*,80\%)$ (c) $\mathscr{T}(50,*,40\%)$ (d) $\mathscr{T}(50,*,20\%)$.}
\end{center}
\end{figure}


\begin{figure}[!htbp]
\begin{center}
\begin{tabular}{cccc}
\includegraphics[width=4cm]{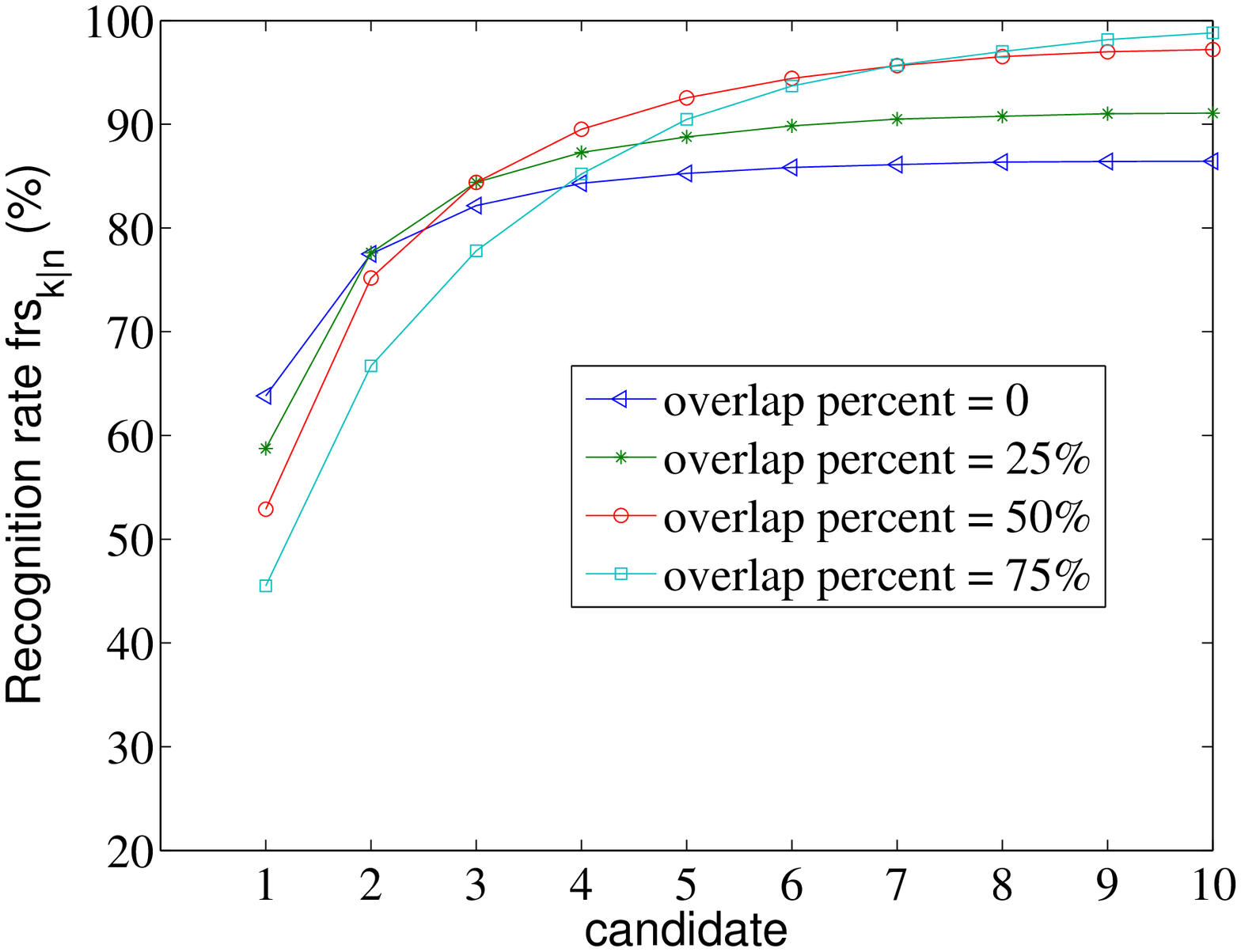} &\ \  \includegraphics[width=4cm]{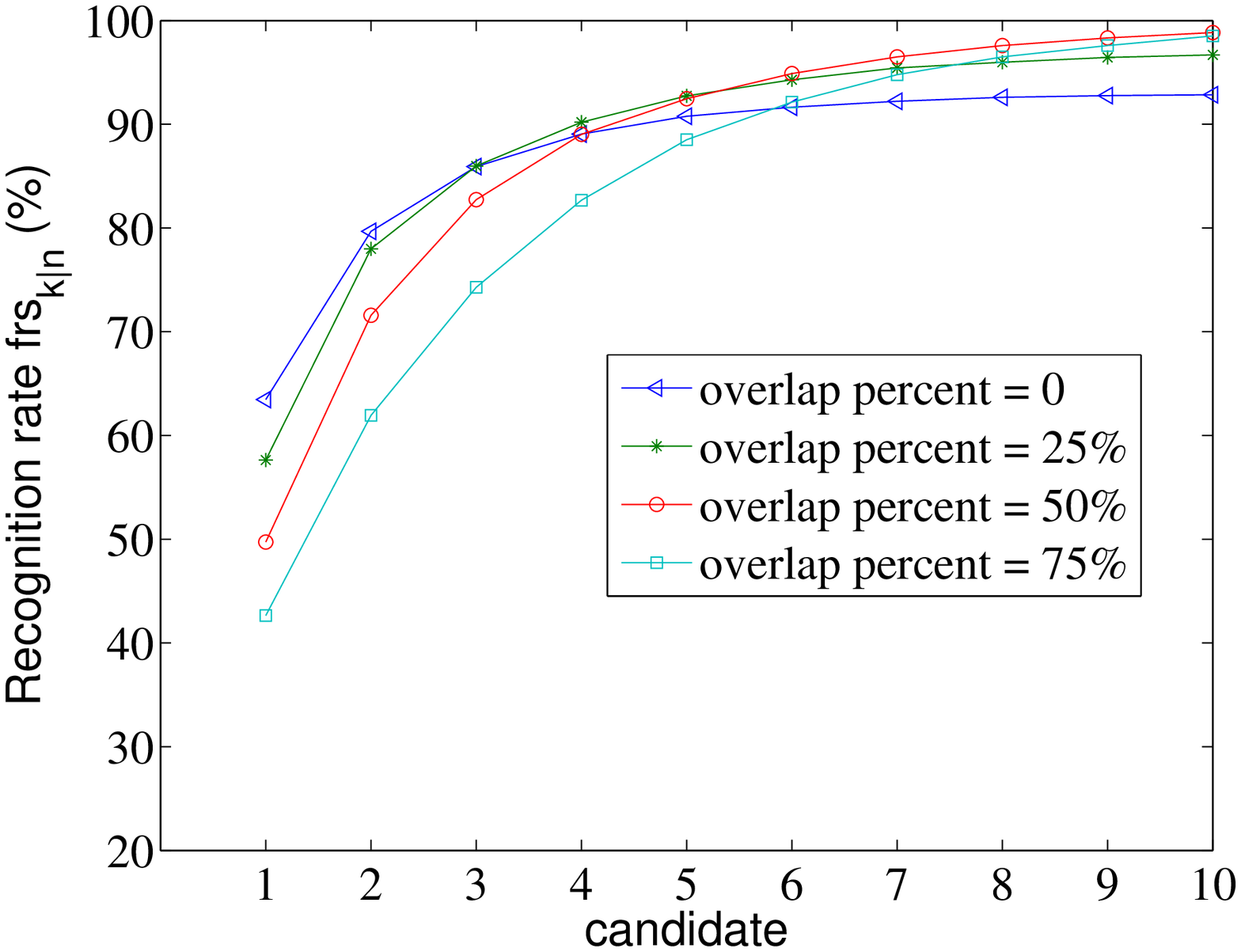} \\(a)&(b)\\
\includegraphics[width=4cm]{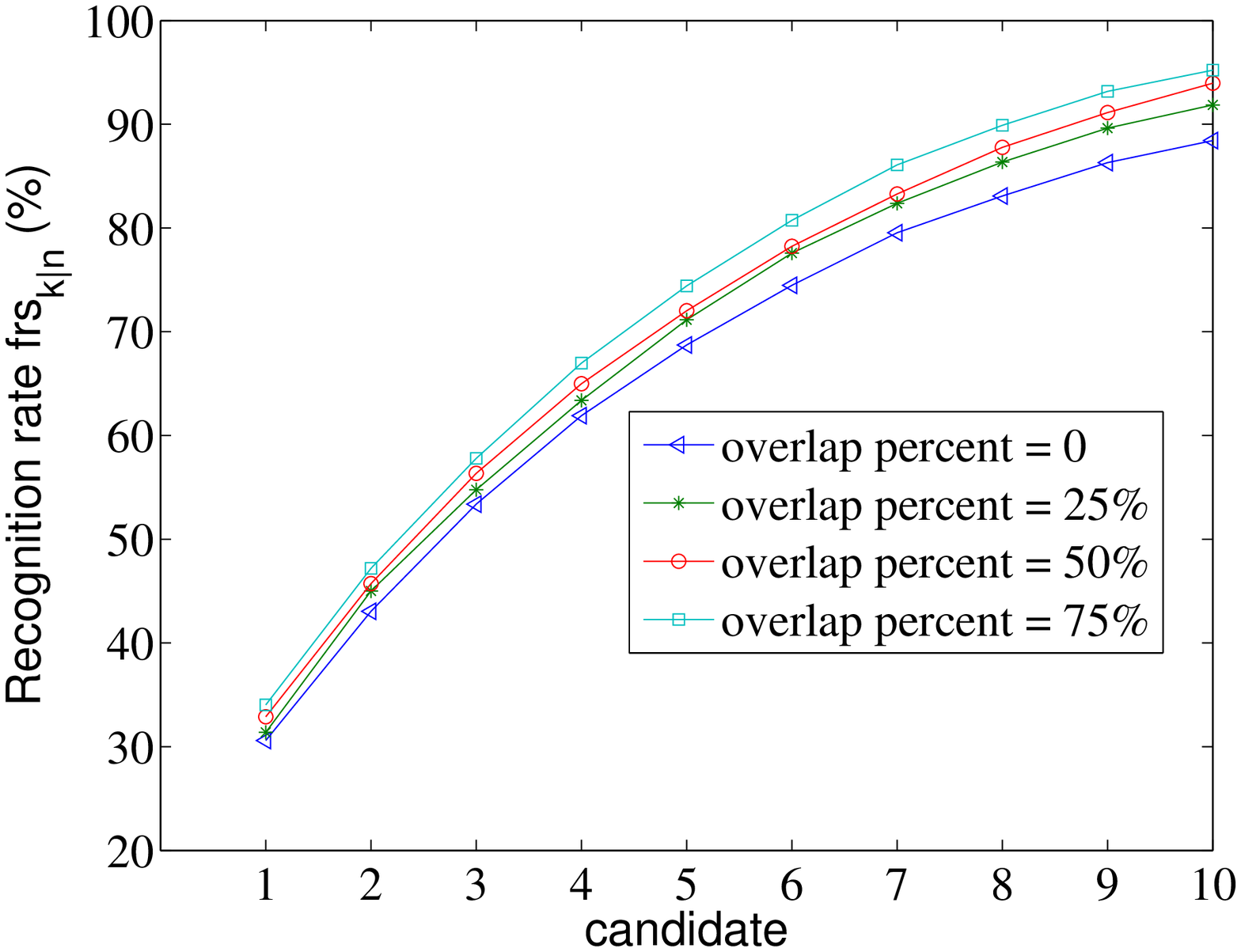} &\ \  \includegraphics[width=4cm]{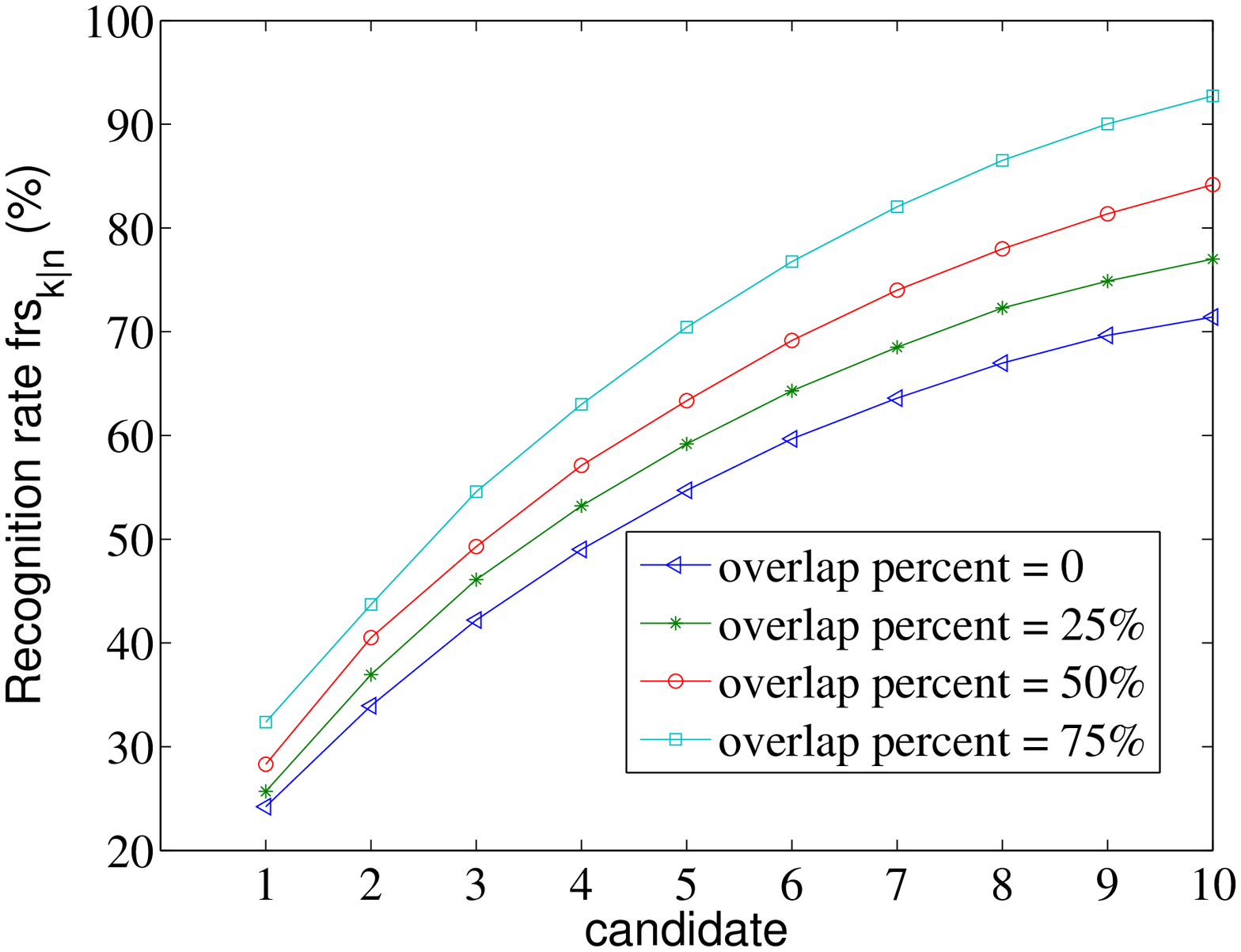} \\(c)&(d)\\
\end{tabular}
\caption{sequence-by-sequence recognition rates on DI. (a) $\mathscr{L}(50,*,60\%)$ (b) $\mathscr{L}(50,*,80\%)$ (c) $\mathscr{T}(50,*,40\%)$ (d) $\mathscr{T}(50,*,20\%)$.}
\end{center}
\end{figure}

The experimental results on DI above show that, on training sets, the high k-mer size is preferred, on testing sets, the relative low k-mer will obtain relative good performance. As to the overlap percent, the sequence-by-sequence recognition on testing data sets would prefer higher value of overlap. All of the experimental results of this section are listed in Supplementary Table 2 and Supplementary Table 3.



\subsection{Accuracy on DII with different k-mer size }

This section we will show the performance of multi-class SVM with N-best algorithm on DII, which is different from DI in biology information, where experiments on DI would be treated as a kind of gene prediction, and the experiments on DII would be treated as a kind of gene category.
The experimental results on data sets with split percents  60\% and 80\% are investigated.
All the experimental results  of dimension-by-dimension cases are listed in Supplementary Table 4, and sequence-by-sequence cases are listed in Supplementary Table 5.

Firstly, in dimension-by-dimension case,  the accuracy rates increase along top-n candidates and k-mer size (Fig.7(a)(b)(c)(d)) on training data sets,, while the accuracy rates on testing data sets increase along top-n candidates and slightly decrease along k-mer size in top-n candidates case, in other top-n case, the fluctuation of accuracy rate along k-mer size is not distinctive (Supplementary Fig.5(a)(b)(c)(d)).

\begin{figure}[!htbp]
\begin{center}
\begin{tabular}{cccc}
\includegraphics[width=4cm]{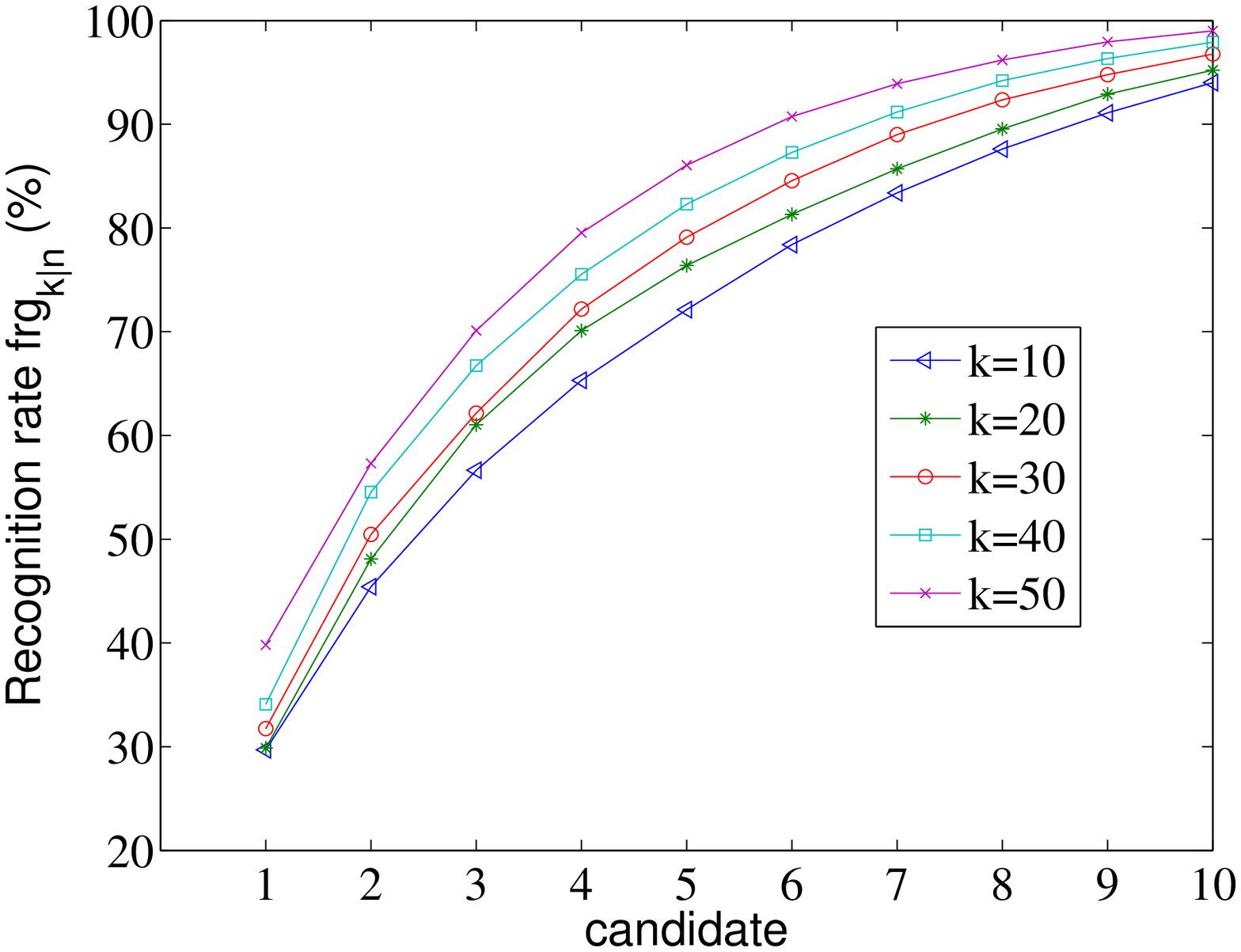} &\ \  \includegraphics[width=4cm]{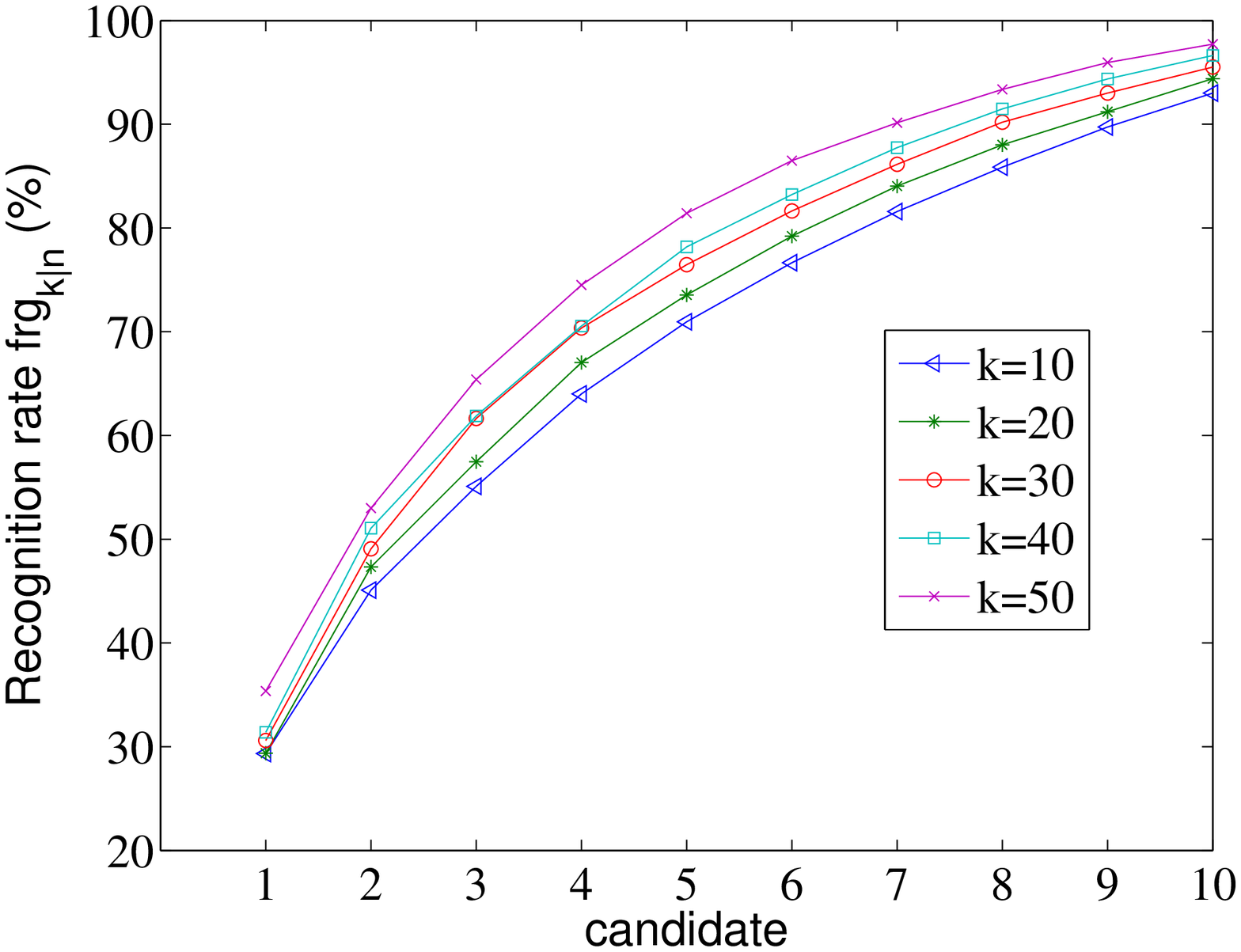} \\(a)&(b)\\
\includegraphics[width=4cm]{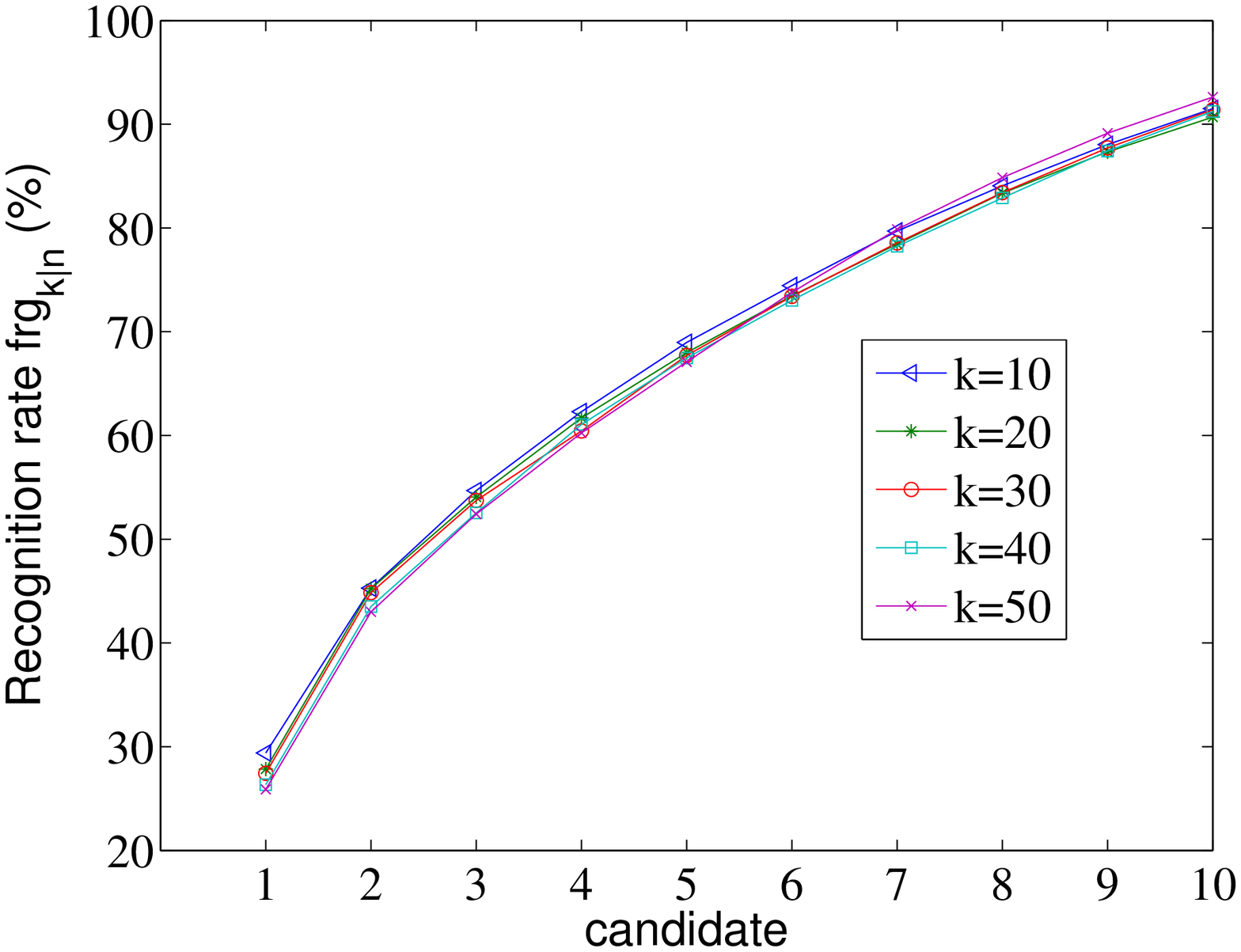} &\ \  \includegraphics[width=4cm]{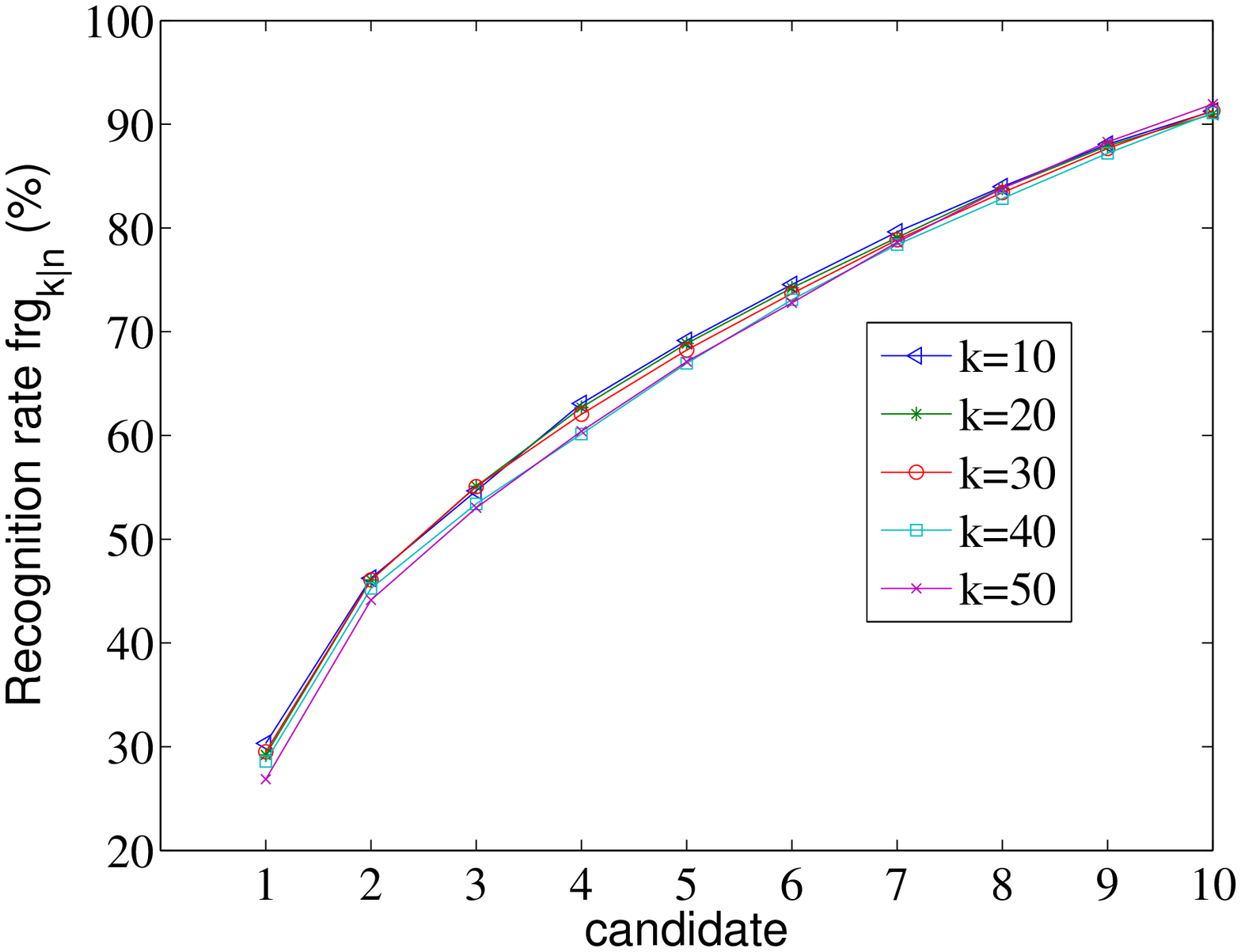} \\(c)&(d)\\
\end{tabular}
\caption{dimension-by-dimension recognition rates on DII. (a) $\mathscr{L}(*,0,60\%)$ (b) $\mathscr{L}(*,0,80\%)$  (c) $\mathscr{T}(*,0,40\%)$ (d) $\mathscr{T}(*,0,20\%)$. }
\end{center}
\end{figure}


Secondly, in sequence-by-sequence cases, the accuracy rates increase along both top-n candidates and k-mer size on training data sets (Fig.8(a)(b)(c)(d)), while the accuracy rates increase along top-n candidates and slightly decrease along k-mer size on testing data sets (Supplementary Fig.6(a)(b)(c)(d)).

Again, the appropriate selection of k-mer size on DII type data sets would be 10. This conclusion is also helpful to model metagenomic sequences with HMM though it is difficult for HMM framework to examine the high dimensional k-mer size cases.

\begin{figure}[!htbp]
\begin{center}
\begin{tabular}{cccc}
\includegraphics[width=4cm]{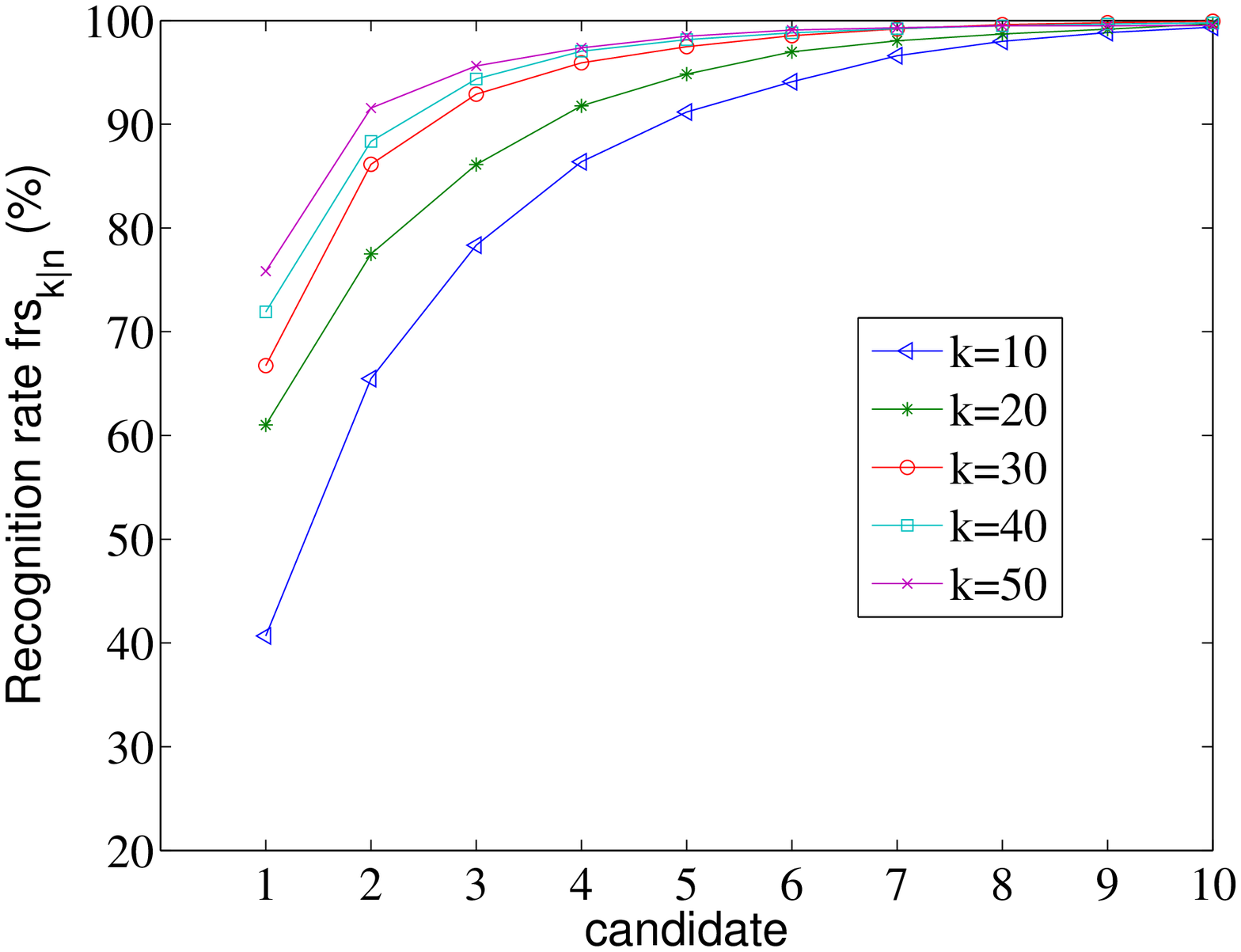} &\ \  \includegraphics[width=4cm]{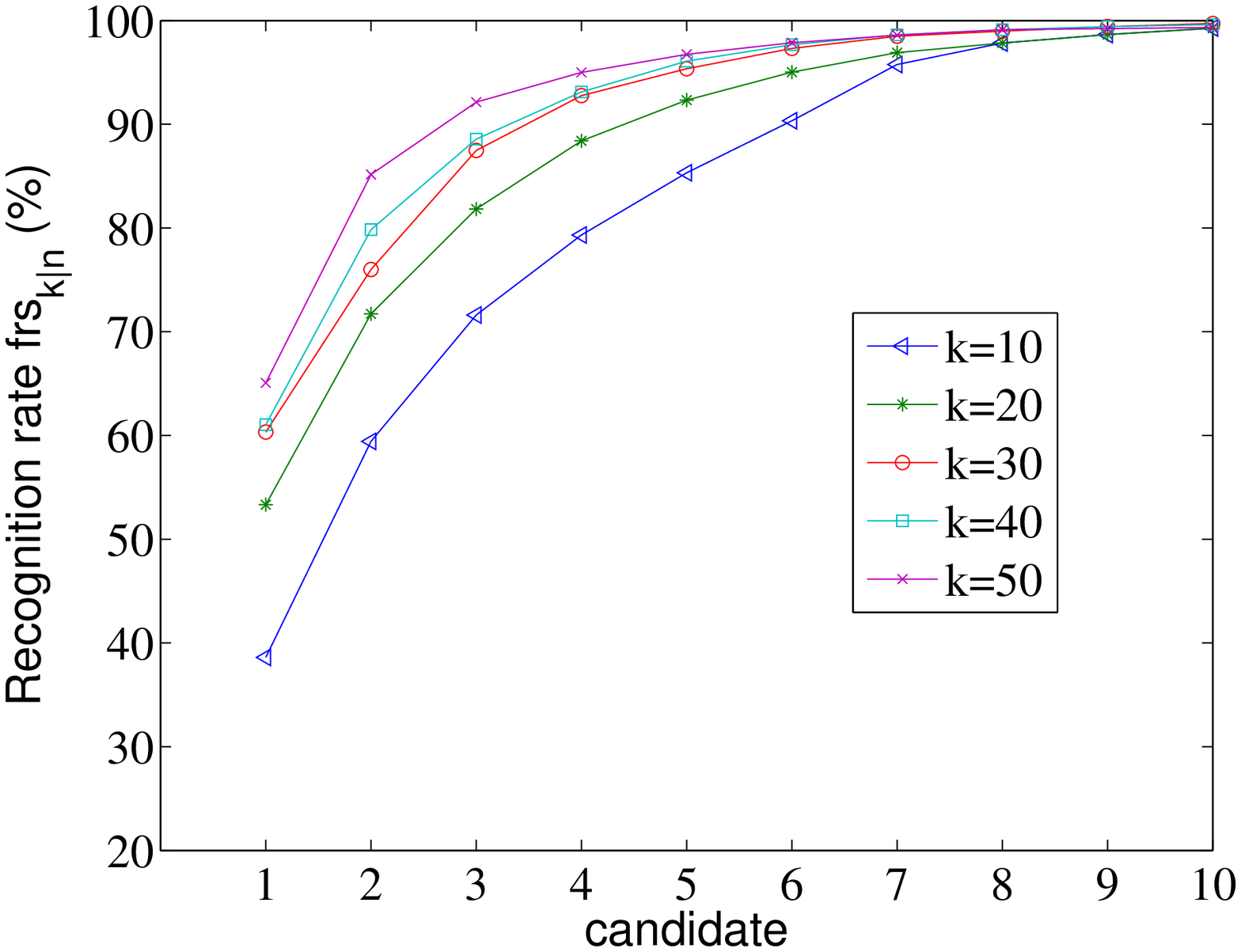} \\(a)&(b)\\
\includegraphics[width=4cm]{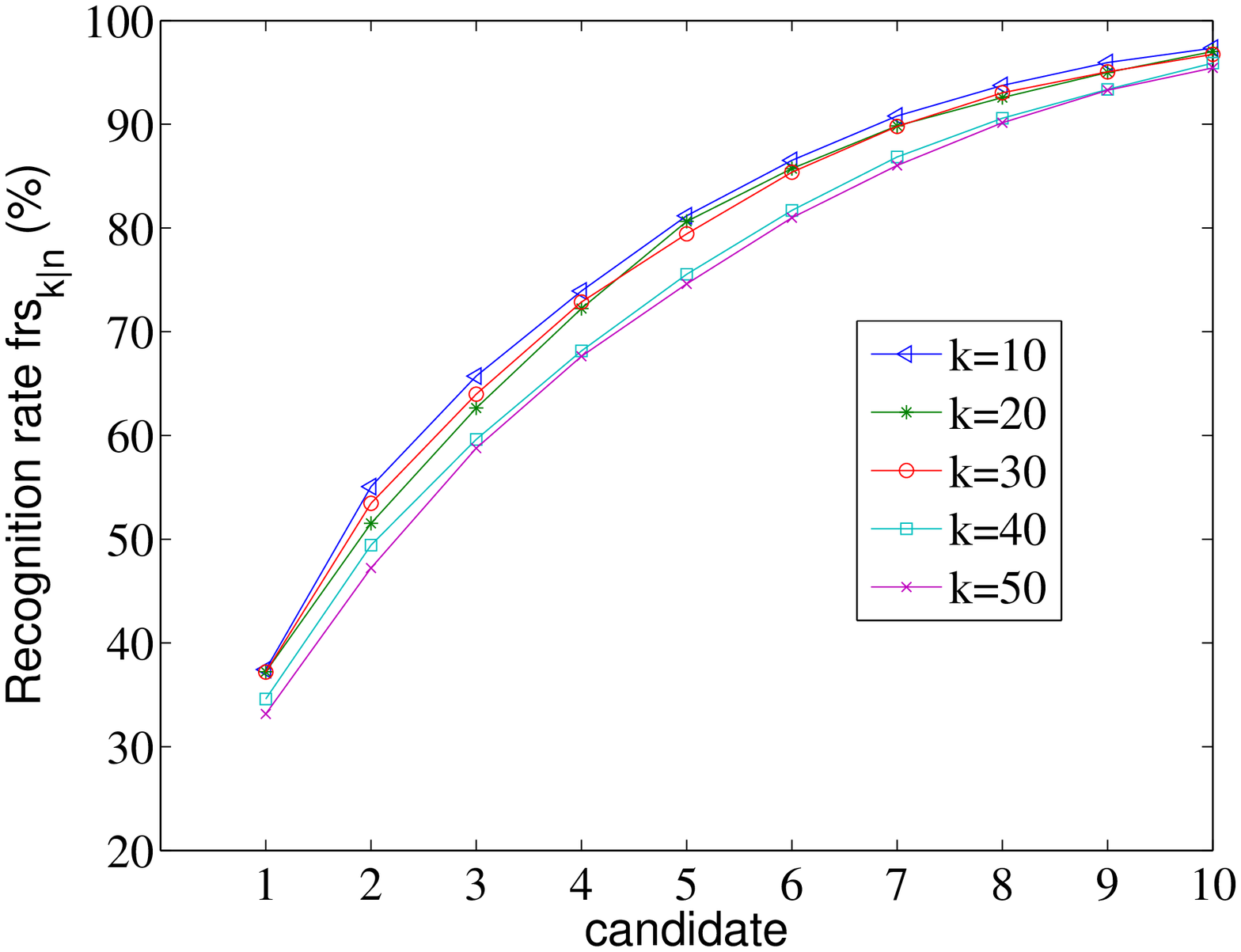} &\ \  \includegraphics[width=4cm]{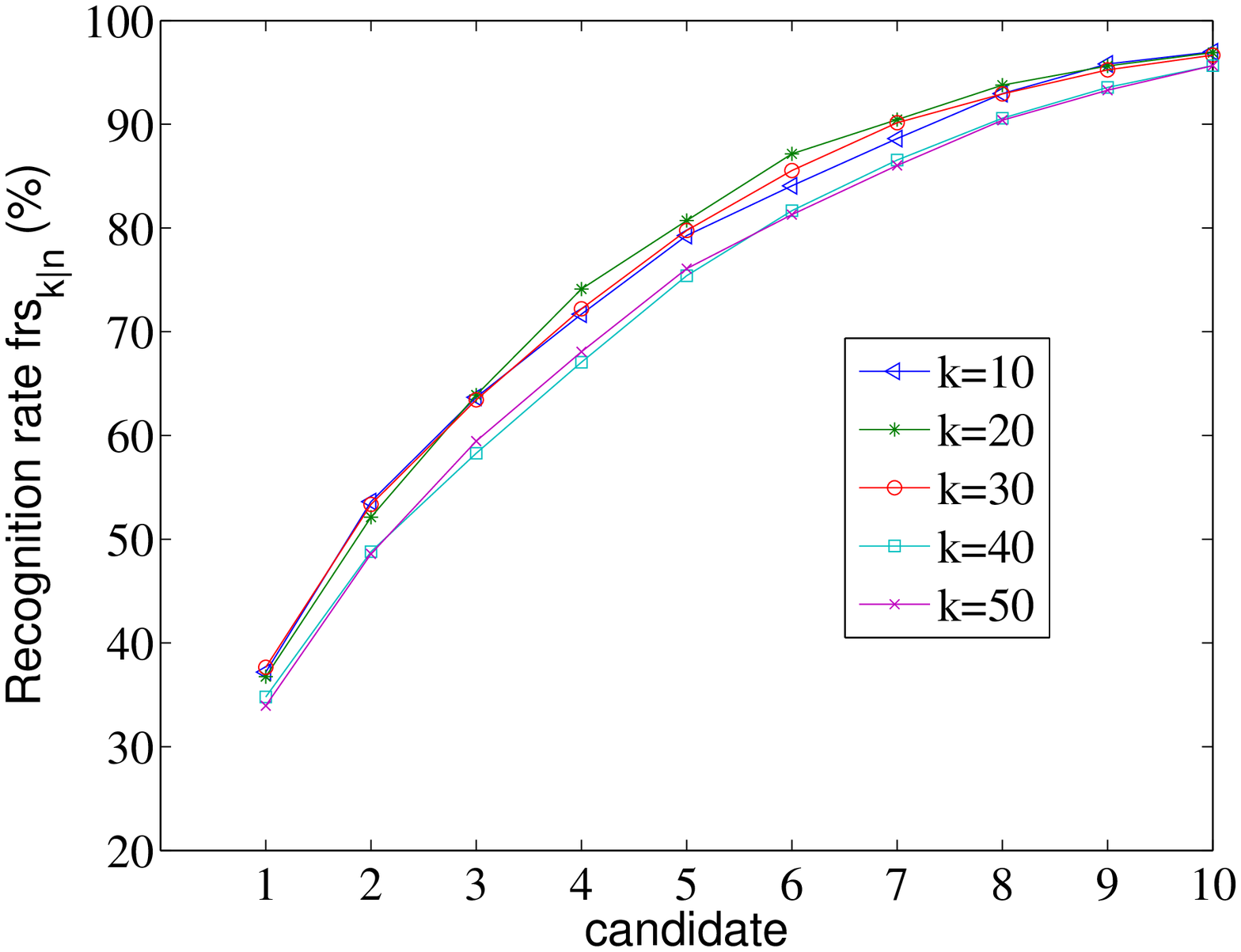} \\(c)&(d)\\
\end{tabular}
\caption{sequence-by-sequence recognition rates on DII. (a) $\mathscr{L}(*,0,60\%)$ (b) $\mathscr{L}(*,0,80\%)$  (c) $\mathscr{T}(*,0,40\%)$ (d) $\mathscr{T}(*,0,20\%)$. }
\end{center}
\end{figure}

Moreover, from all the experimental results on split percent 60\% and 80\%, the principles along parameters of overlap and k-mer size have almost the same tendency, which is also provide the reasonable explanations of our models and performance.
Regardless of DI and DII, the fact that 10 k-mer size with top-10 candidates can achieve high recognition rates demonstrates that the centroid marker family genomic sequences of all microbial would be highly discriminated with the toleration of confusion of 10 species among the 17 species within the multi-class SVM with N-best algorithm framework, the confusion is mainly caused by the similarities among species measured by multi-class SVM model in the view of evolution[55,56].

\section{DISCUSSION}

We develop the multi-class SVM based N-best algorithm on microbial species classification, realize the true sense of fragment classification of genomic sequences without preprocessing, and discuss the effectiveness of k-mer size,overlap and split percent. The genome sequences of maker family from all microbial on MetaRef are treated as time series and employed in experiments with different lengths. Generally, the experimental results on the given range of SVM parameters grids show that the larger the k-mer size from 10 to 50, the higher the accuracy rates on training sets. The conclusions on testing sets are reverse. The overlap of fragments also takes effect on different data sets. In the view of prediction, the low overlap is recommended. Furthermore, N-best algorithm shows that top-10 candidates would provide good experimental results on both dimension-by-dimension and sequence-by-sequence recognition tasks. The experiments are designed only based on the \{A,C,T,G\} information along genomic sequences, if the detail 3-D genomic structure information or biology information are available, the prediction accuracy rates would be higher.
Though the multi-class SVM with N-best algorithm and feature selection are constructed for challenging fragment classification, it would be applied to cancer diagnose of genomic fragments in future. And, how to overcome the time--consumption and memory storage of SVM and propose new kernel SVM are our future research interests. Finally, great improvement of top-1 classification accuracy is also our research goal in microbial genomic analysis.


\section{ACKNOWLEDGEMENTS}

 This paper is partially supported by CHINA SCHOLARSHIP COUNCIL (No. 201303070216), 863 Project of China (2008AA02Z306), it was finished while I visited University of Southern California from March.2014 to Feb.2015.

\subsubsection{Conflict of interest statement.} None declared.


\onecolumn
\clearpage

\begin{flushleft}
{\Large
\textbf{Supplementary Material : Random Fragments Classification of Microbial Marker Clades with Multi-class SVM and N-Best Algorithm}}
\\
Jingwei Liu
\\
{\small School of Mathematics and System Sciences,Beihang University,Beijing,100191,P.R. China}
\\
E-mail: liujingwei03@tsinghua.org.cn
\end{flushleft}


\vspace{1cm}

\begin{table}[!htbp]
\caption{Vector number of marker gene database with different dimension (k-bp).}
\label{table:1}
\renewcommand{\tabcolsep}{1pc} 
\renewcommand{\arraystretch}{1}
\begin{center}
\begin{tabular}{ccccc}
\toprule
DI& & & & \\
\hline
\multirow{2}{*}{Dimension}& \multicolumn{4}{c}{Overlap percent} \\
\cmidrule{2-5}
 & 0 & 25\% & 50\% & 75\% \\
\midrule
10  &921139	&1149824&1835989&3055903 \\
\rowcolor{mygray}
20	&457482	&608681	&908749	&1811209 \\
30	&302566	&393194	&598807	&1118850 \\
\rowcolor{mygray}
40	&225659	&298783	&445092	&883969  \\
50	&179205	&233895	&352308	&671869  \\

\hline

DII& & & & \\
\hline
\multirow{2}{*}{Dimension}& \multicolumn{4}{c}{Split percent (training/testing)} \\
\cline{2-3}
\cline{4-5}
 & 60\% & 40\% & 80\% & 20\% \\
\hline
10&552213 &368926 &739042 & 182097\\
\rowcolor{mygray}
20&274266 &183216 &367056 & 90426\\
30&181389 &121177 &242767 & 59799\\
\rowcolor{mygray}
40&135302 &90357  &181062 & 44597 \\
50&107420 &71785  &143791 & 35414 \\
\bottomrule
\end{tabular}
\end{center}
\end{table}

\setcounter{figure}{1}
\begin{figure}[!htbp]
\begin{center}
\begin{tabular}{cccc}
\includegraphics[width=7cm]{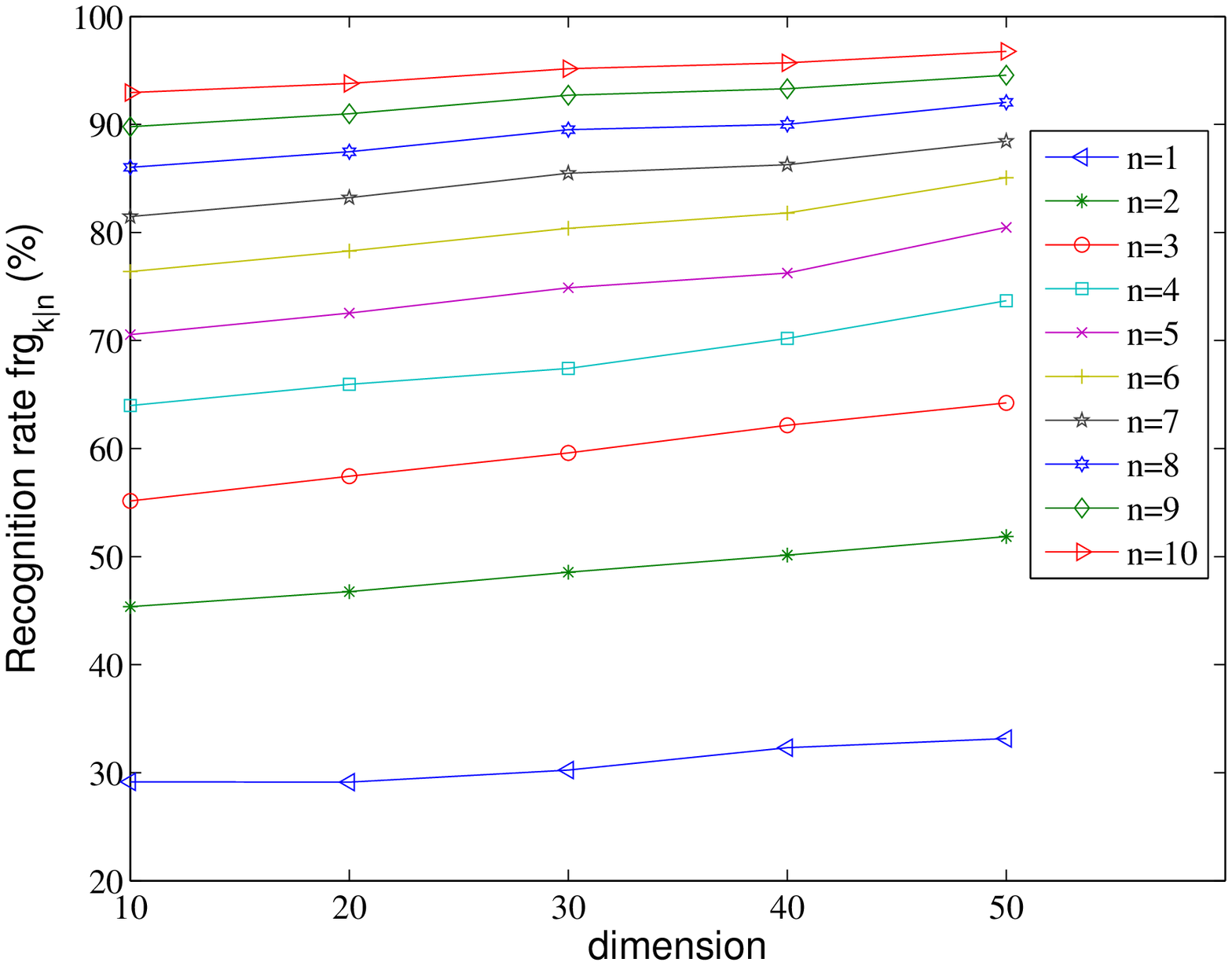} &\ \  \includegraphics[width=7cm]{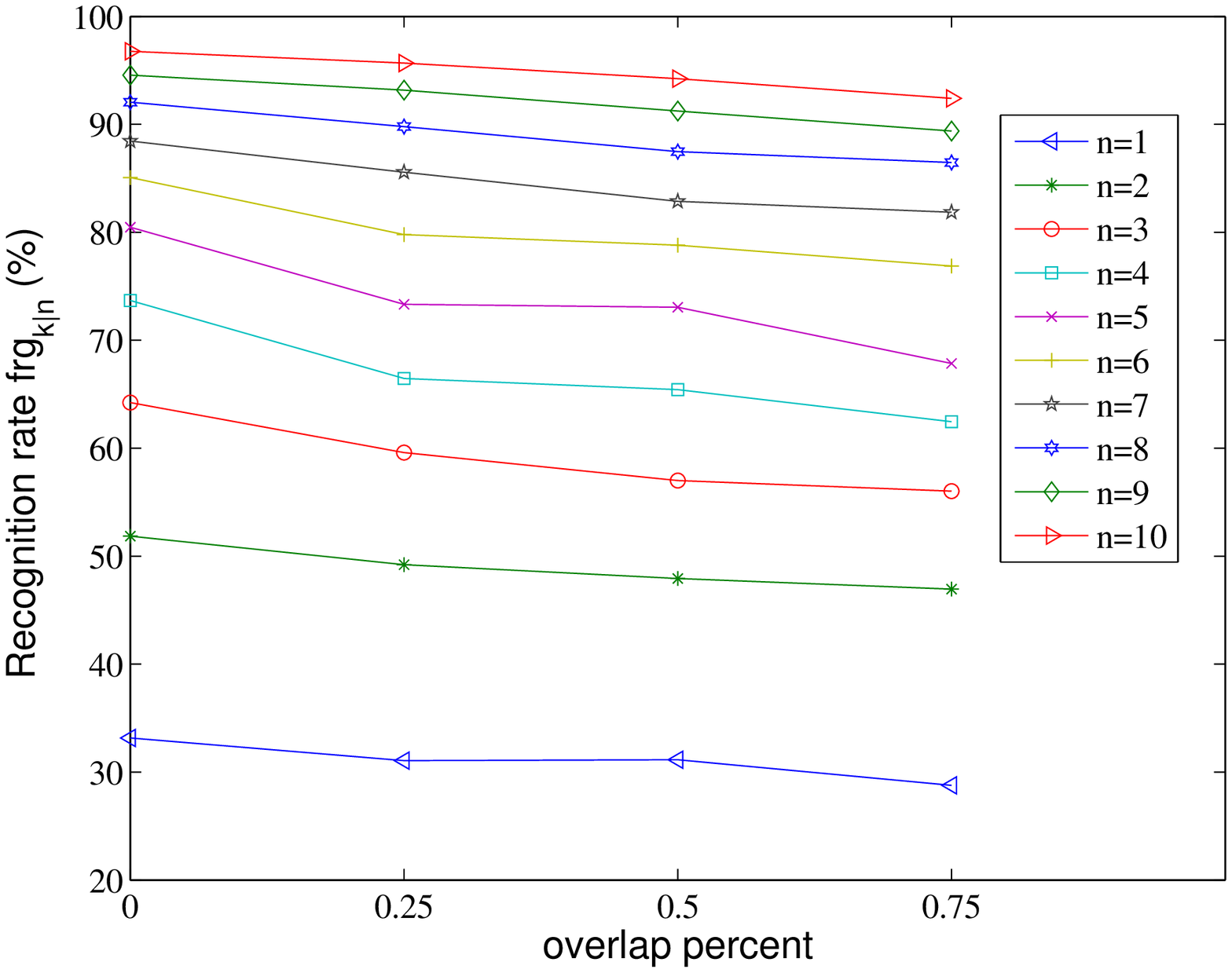} \\(a)&(b)\\
\includegraphics[width=7cm]{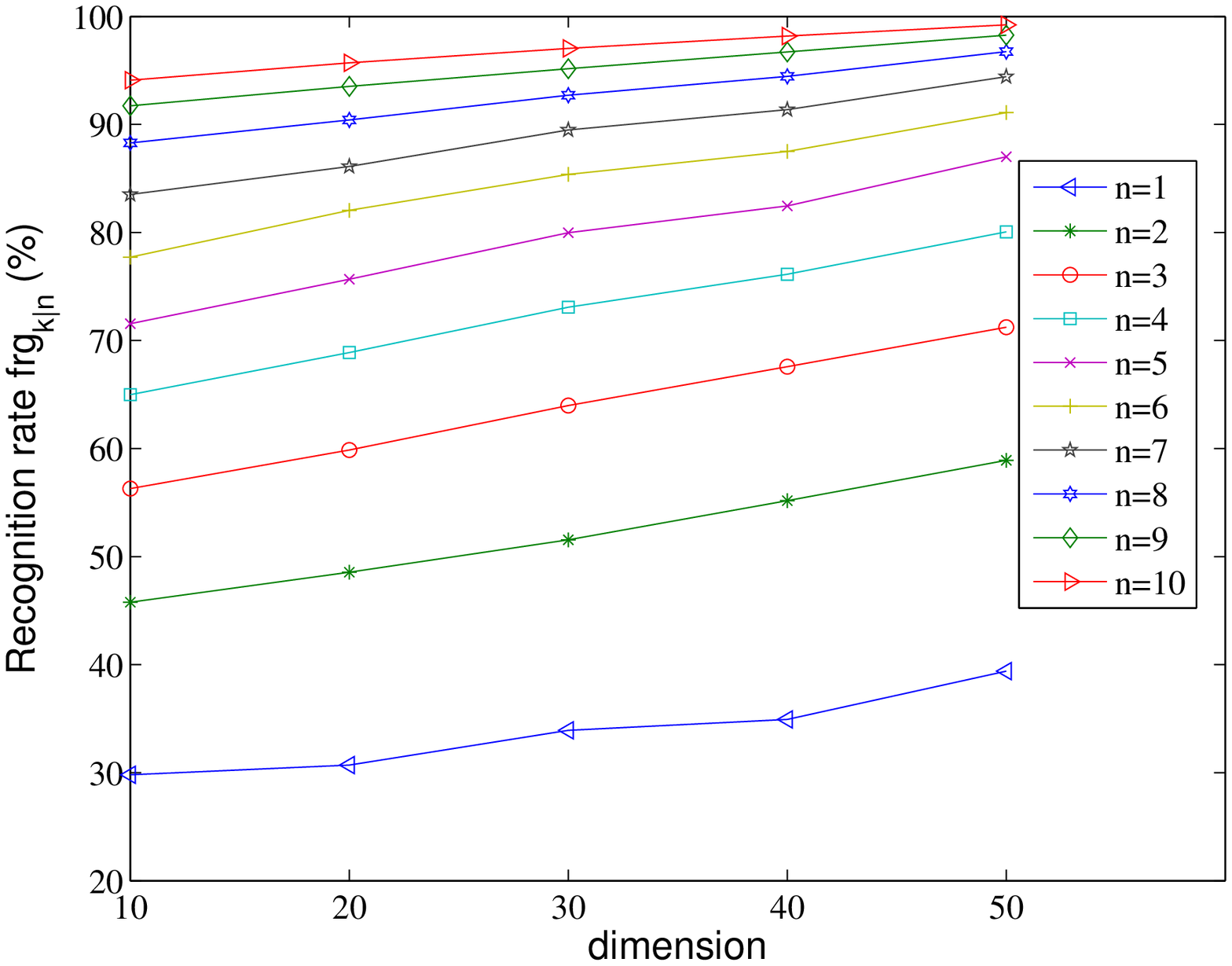} &\ \  \includegraphics[width=7cm]{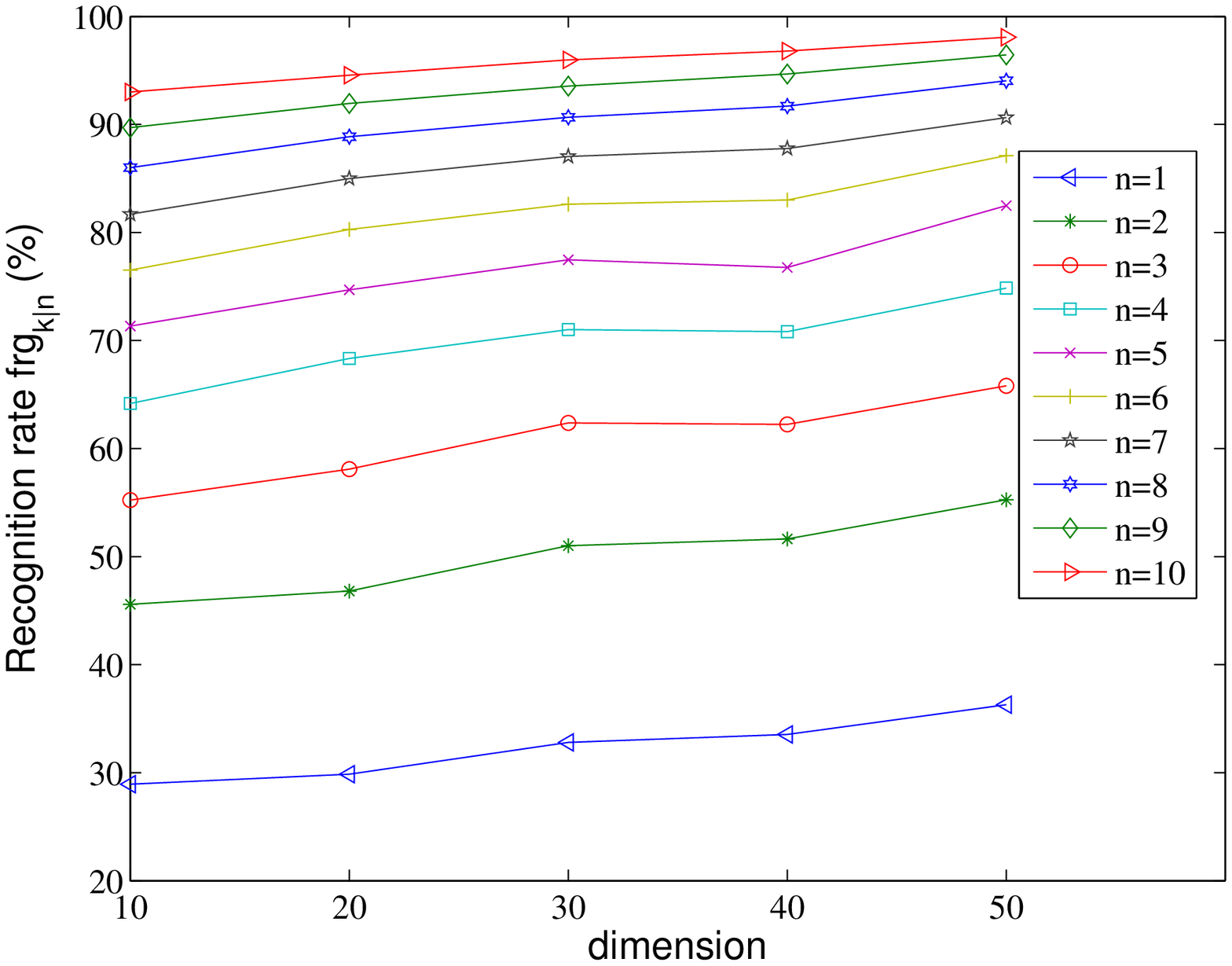} \\(c)&(d)\\
\includegraphics[width=7cm]{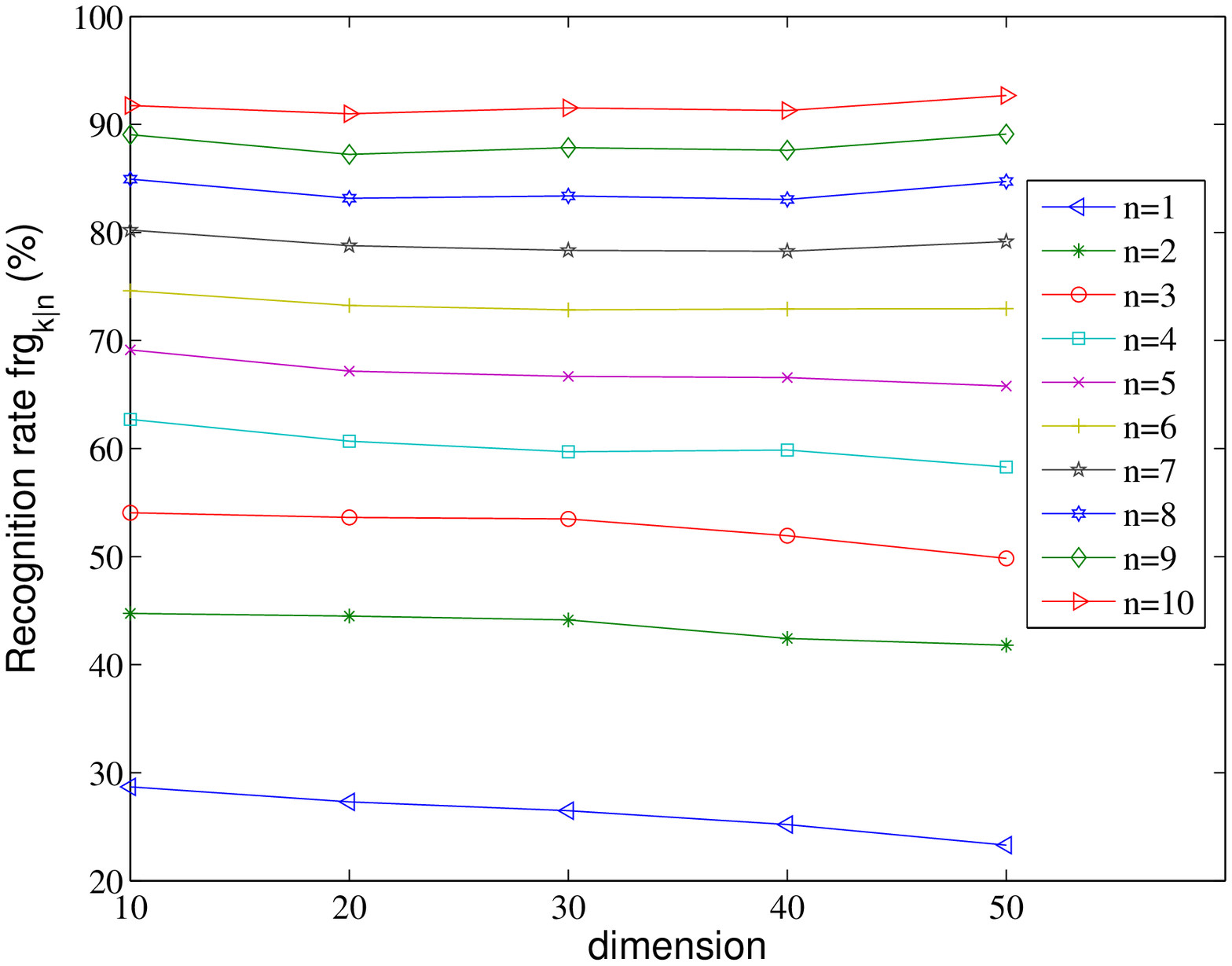} &\ \  \includegraphics[width=7cm]{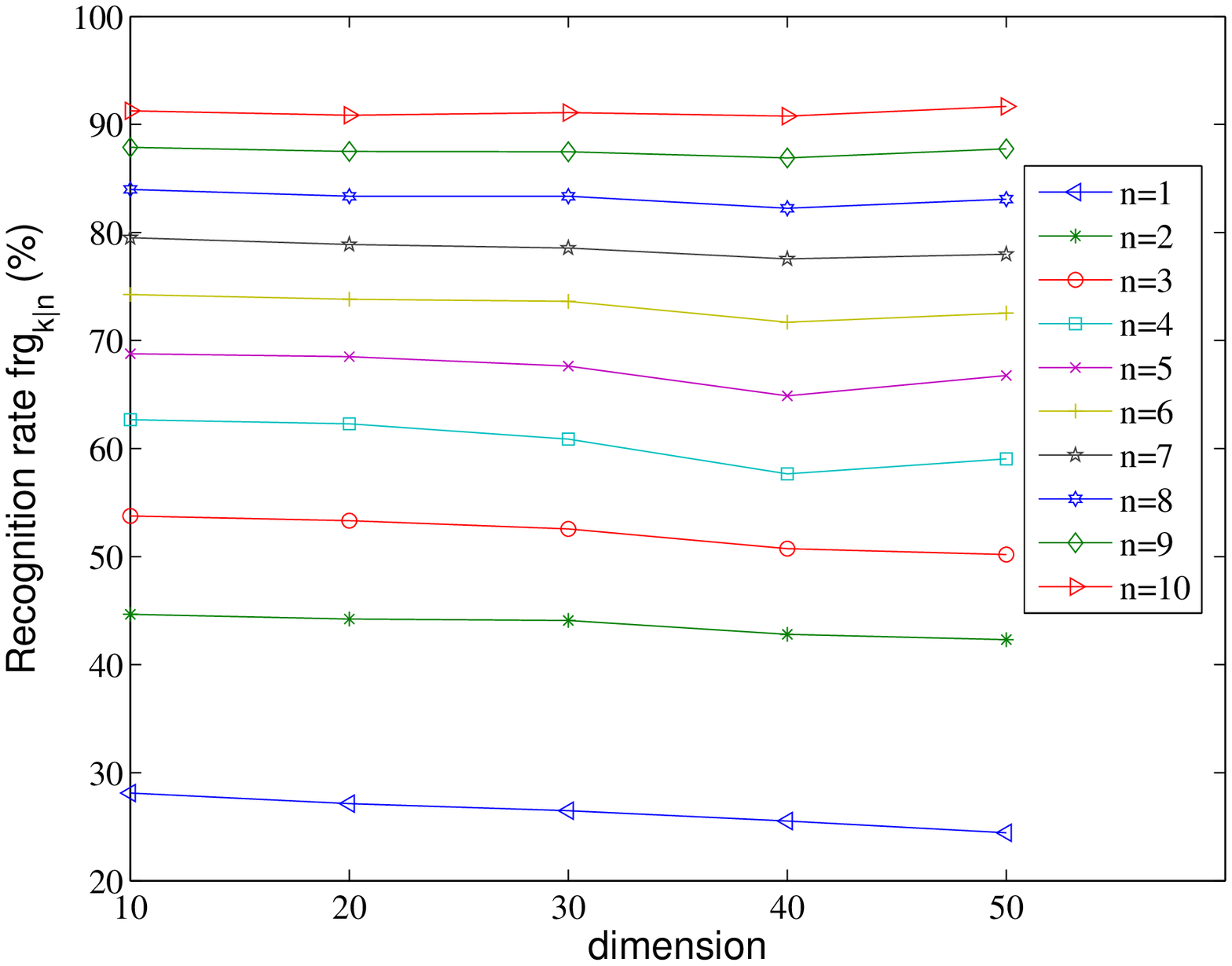} \\(e)&(f)\\
\end{tabular}
\caption*{Dimension-by-Dimension recognition rates on DI. (a) $\mathscr{L}(*,0,100\%)$ (b) $\mathscr{L}(50,*,100\%)$ (c) $\mathscr{L}(*,0,60\%)$ (d) $\mathscr{L}(*,0,80\%)$  (e) $\mathscr{T}(*,0,40\%)$ (f)$\mathscr{T}(*,0,20\%)$. }
\end{center}
\end{figure}

\setcounter{figure}{2}
\begin{figure}[!htbp]
\begin{center}
\begin{tabular}{cccc}
\includegraphics[width=7cm]{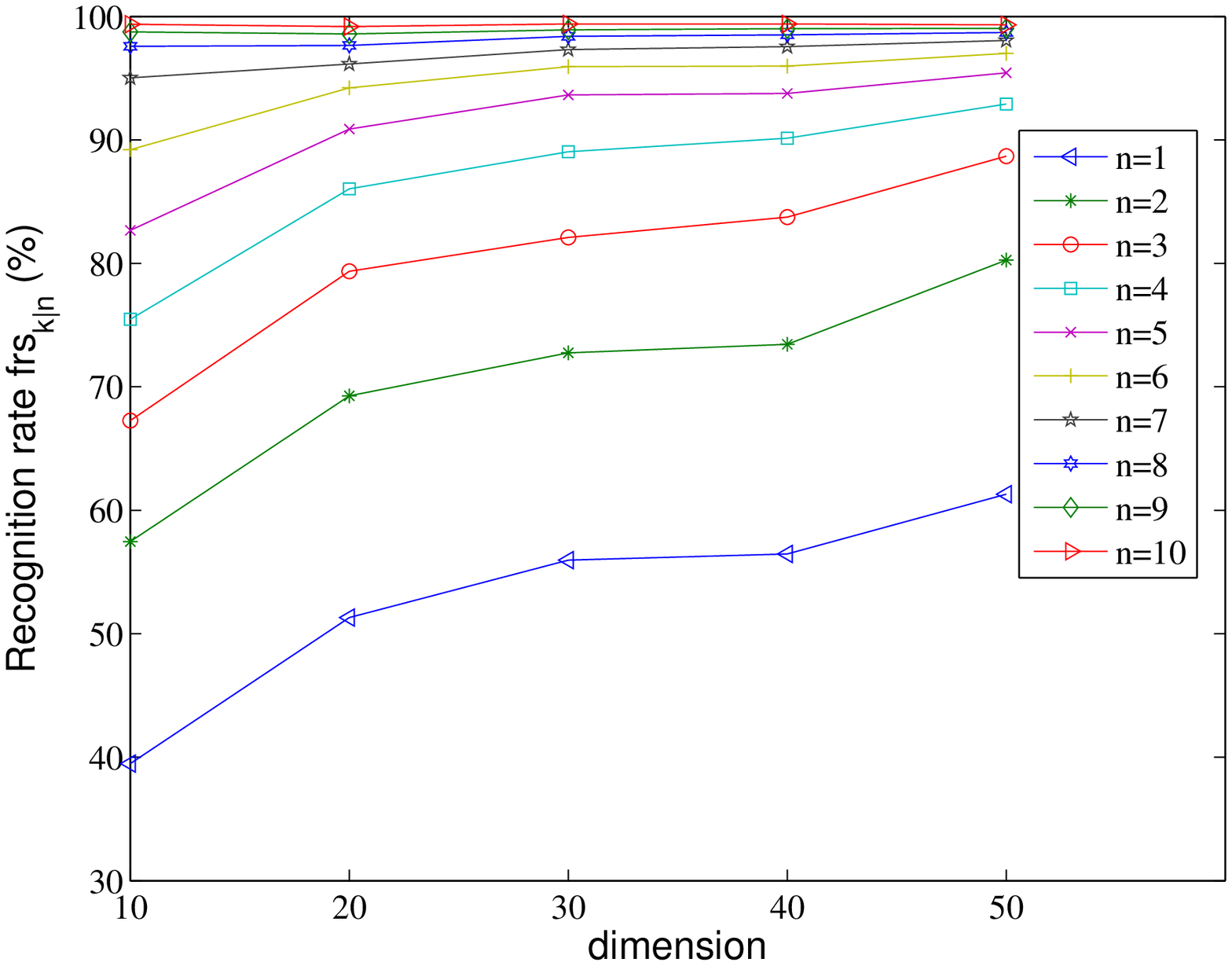} &\ \  \includegraphics[width=7cm]{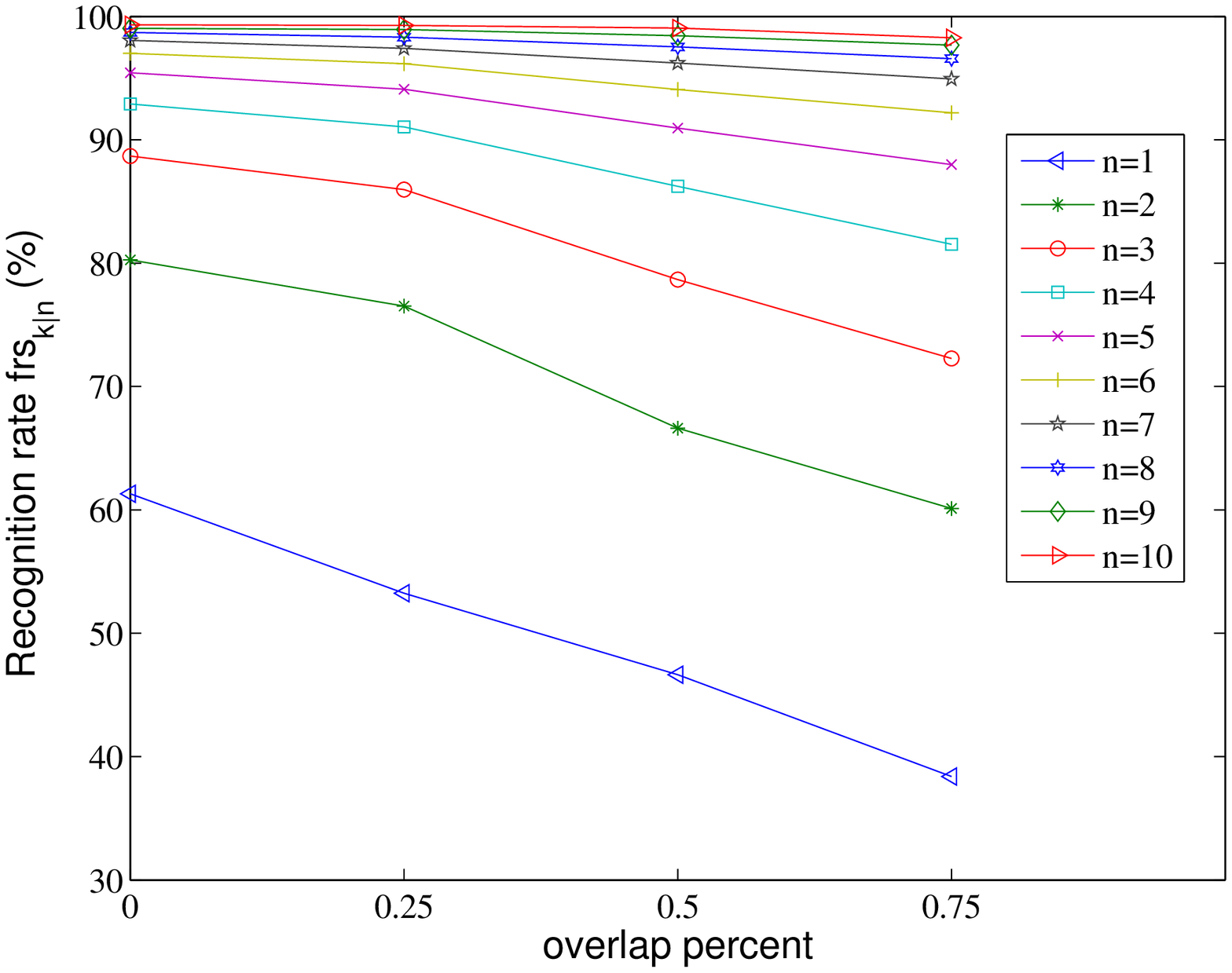} \\(a)&(b)\\
\includegraphics[width=7cm]{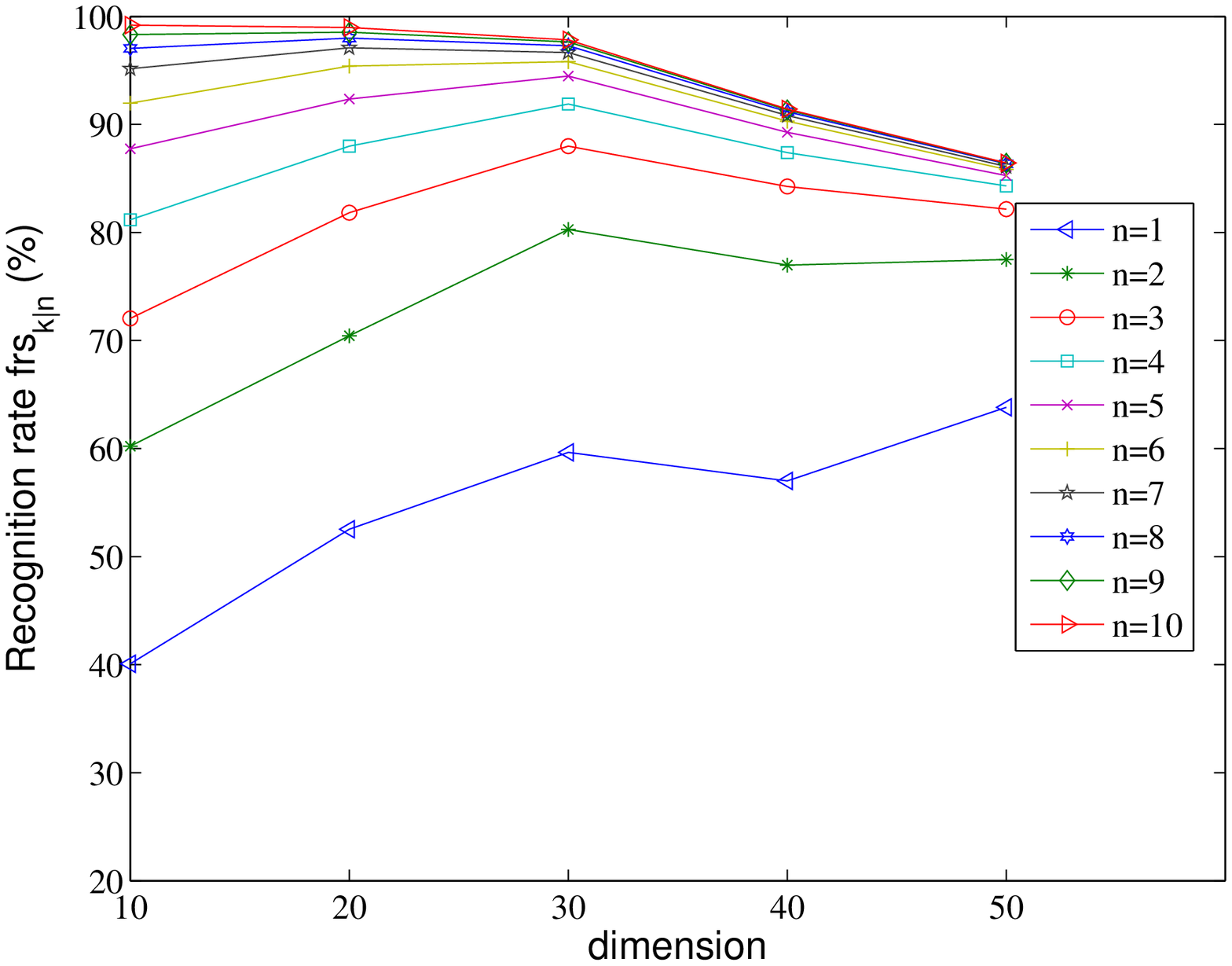}  &\ \ \includegraphics[width=7cm]{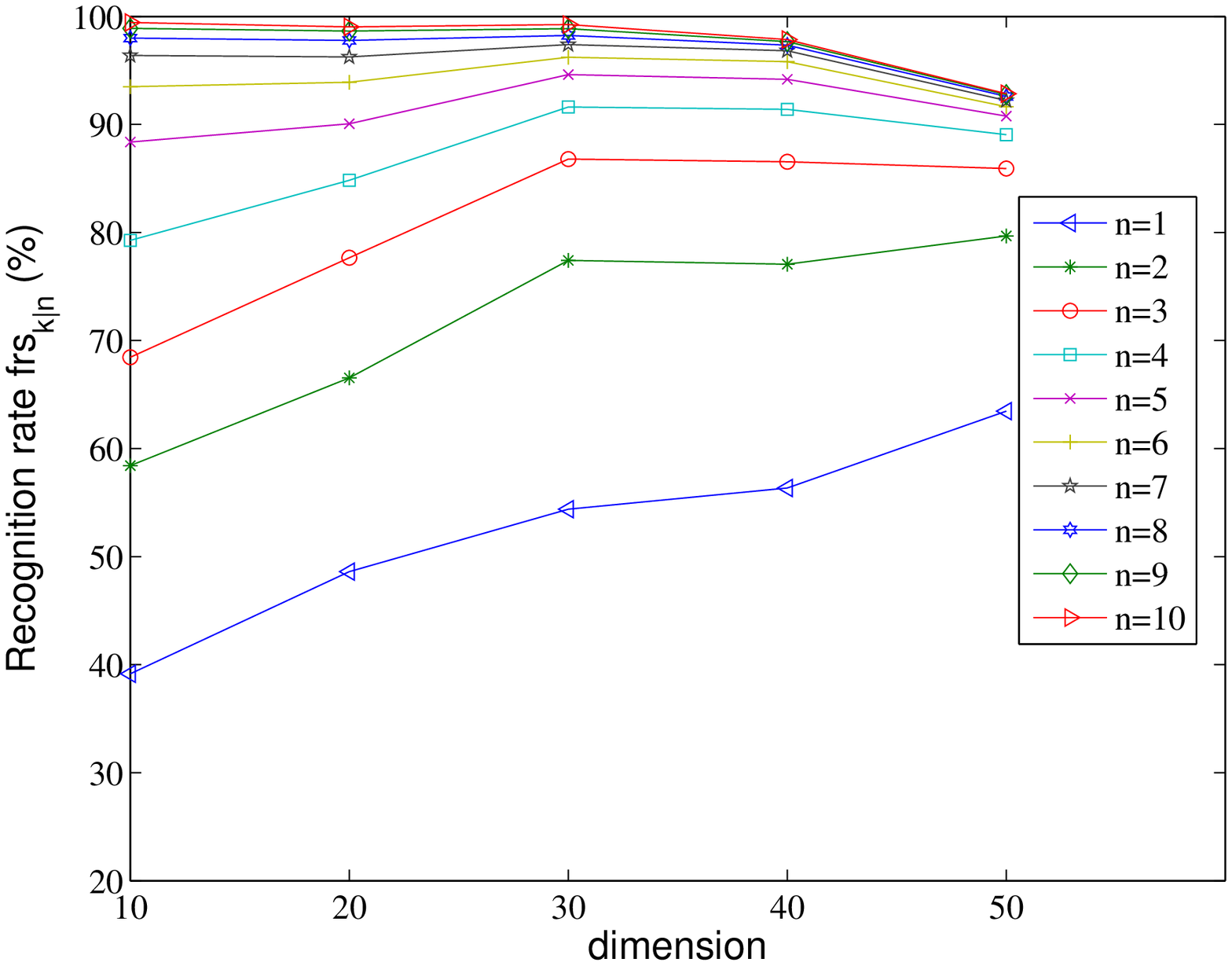} \\(c)&(d)\\
\includegraphics[width=7cm]{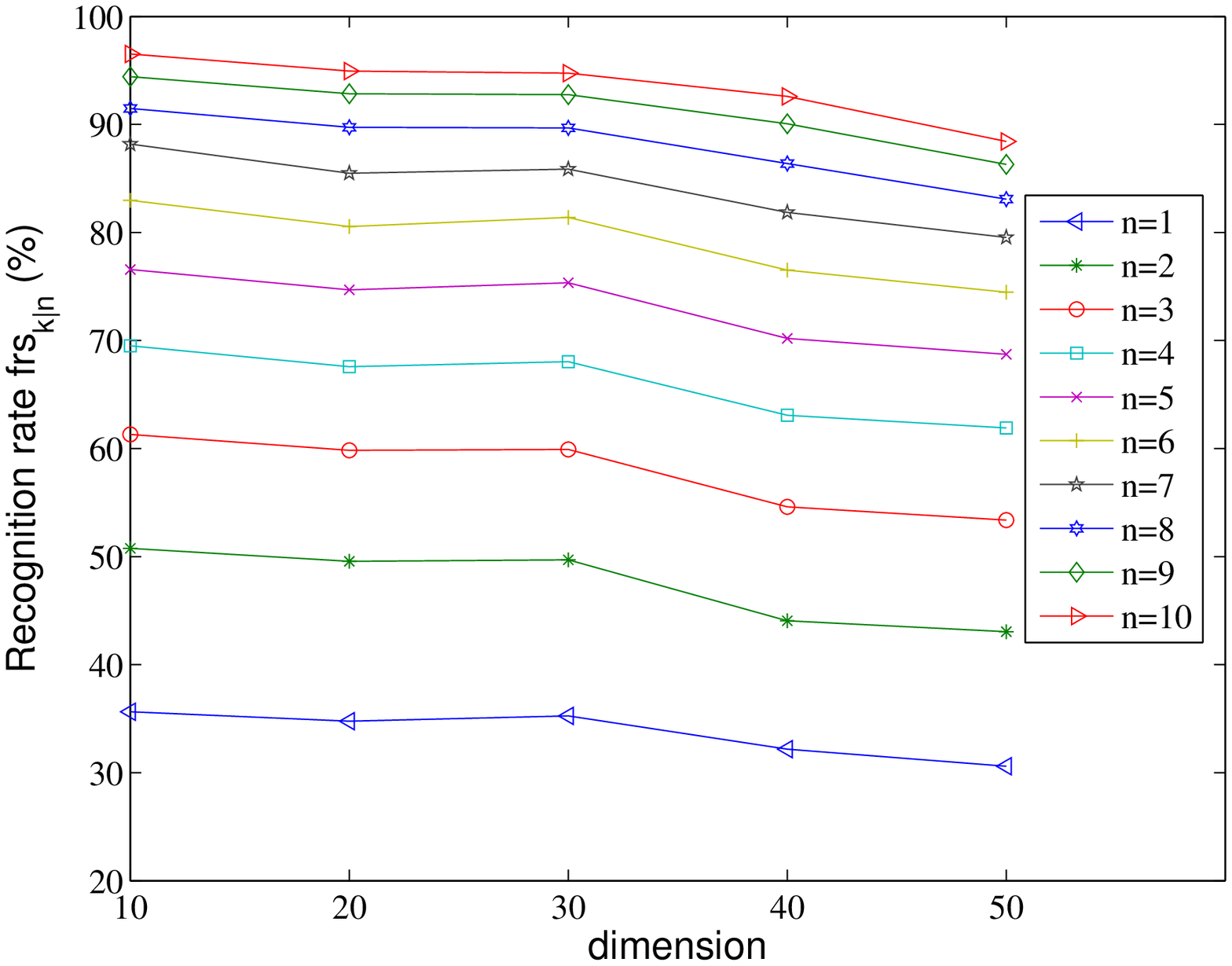}  &\ \ \includegraphics[width=7cm]{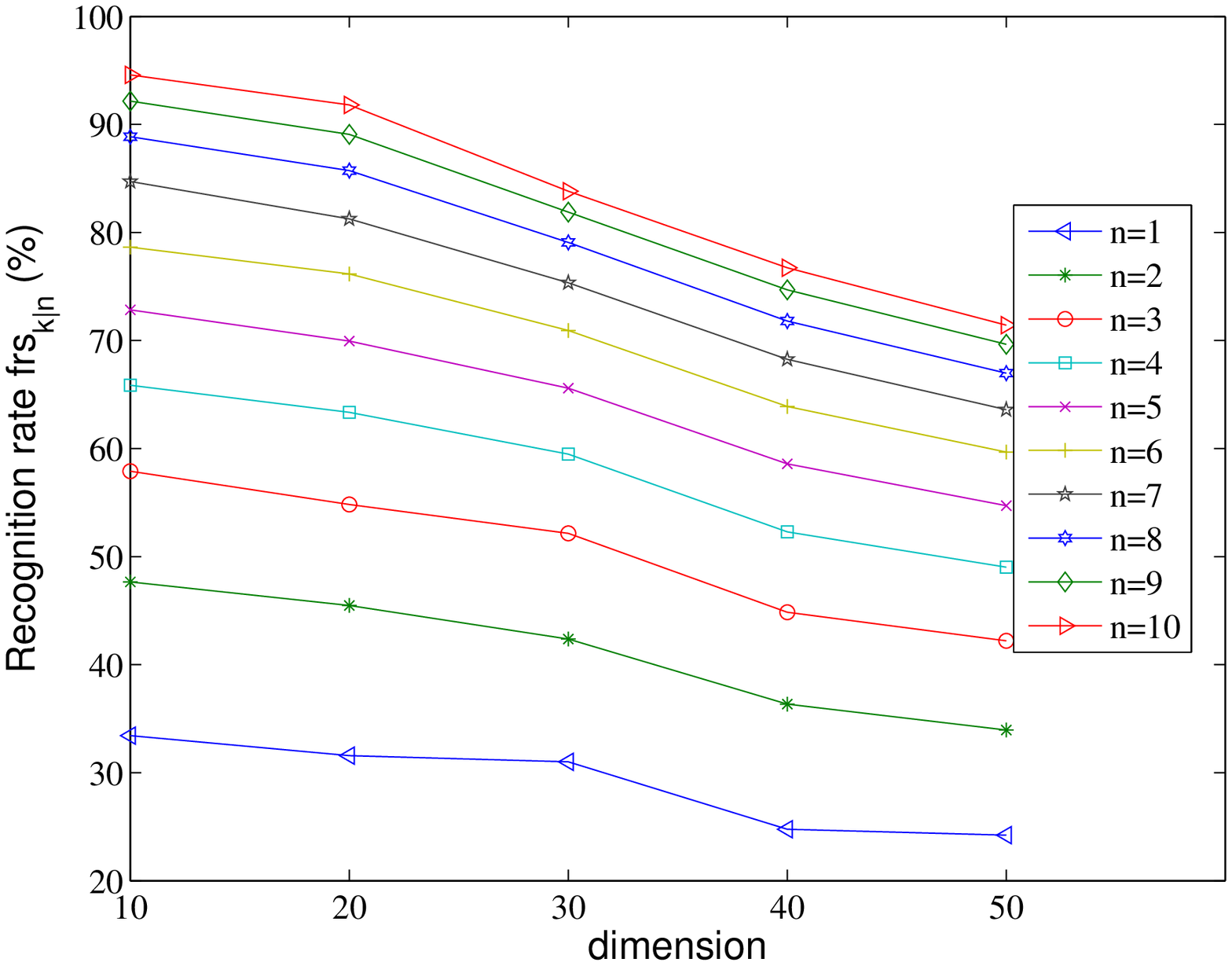} \\(e)&(f)\\
\end{tabular}
\caption*{ Sequence--by--sequence recognition rates on DI. (a) $\mathscr{L}(*,0,100\%)$ (b) $\mathscr{L}(50,*,100\%)$ (c) $\mathscr{L}(*,0,60\%)$ (d) $\mathscr{L}(*,0,80\%)$ (e) $\mathscr{T}(*,0,40\%)$ (f)$\mathscr{T}(*,0,20\%)$.}
\end{center}
\end{figure}

\setcounter{figure}{3}
\begin{figure}[!htbp]
\begin{center}
\begin{tabular}{cccc}
\includegraphics[width=7cm]{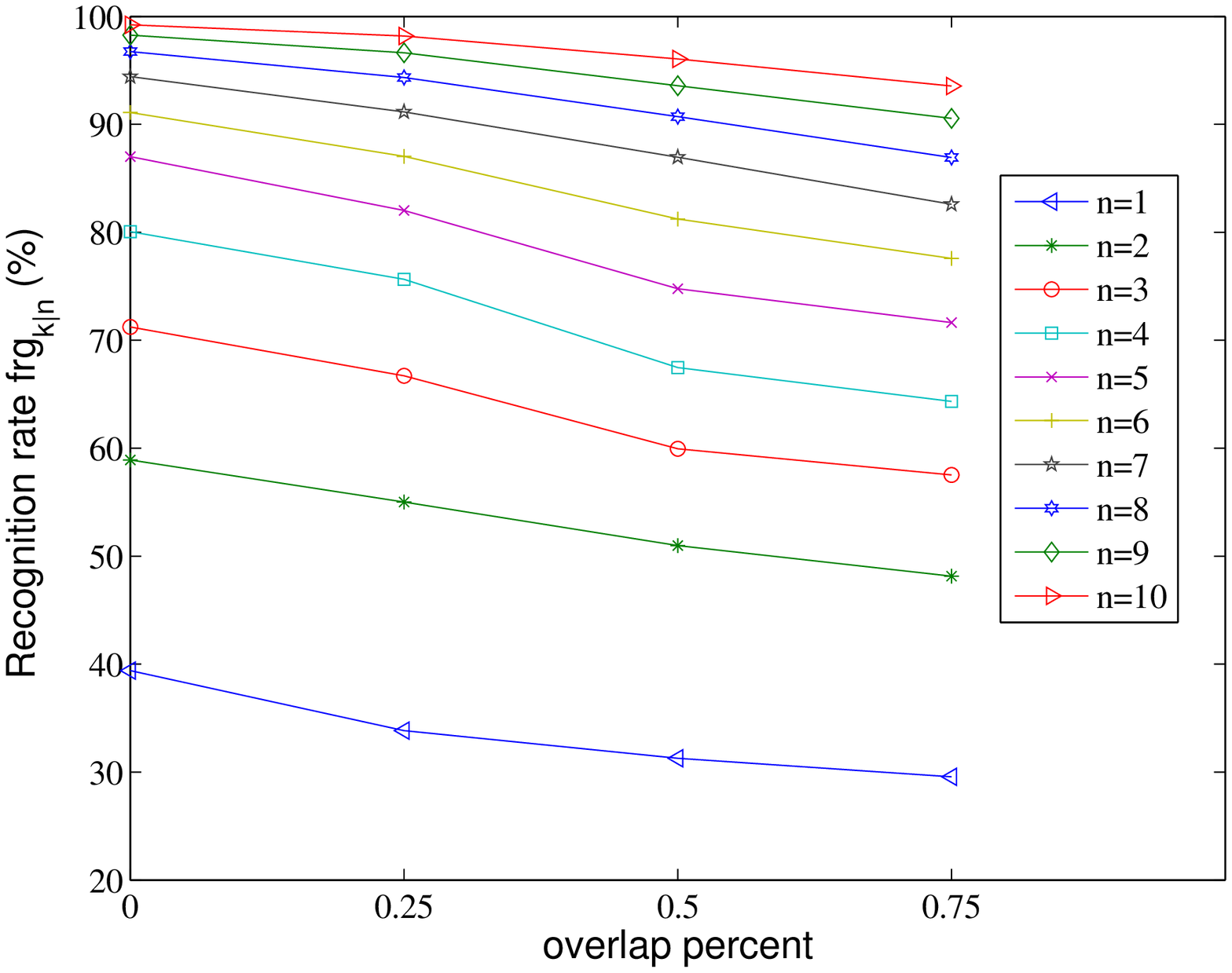} &\ \  \includegraphics[width=7cm]{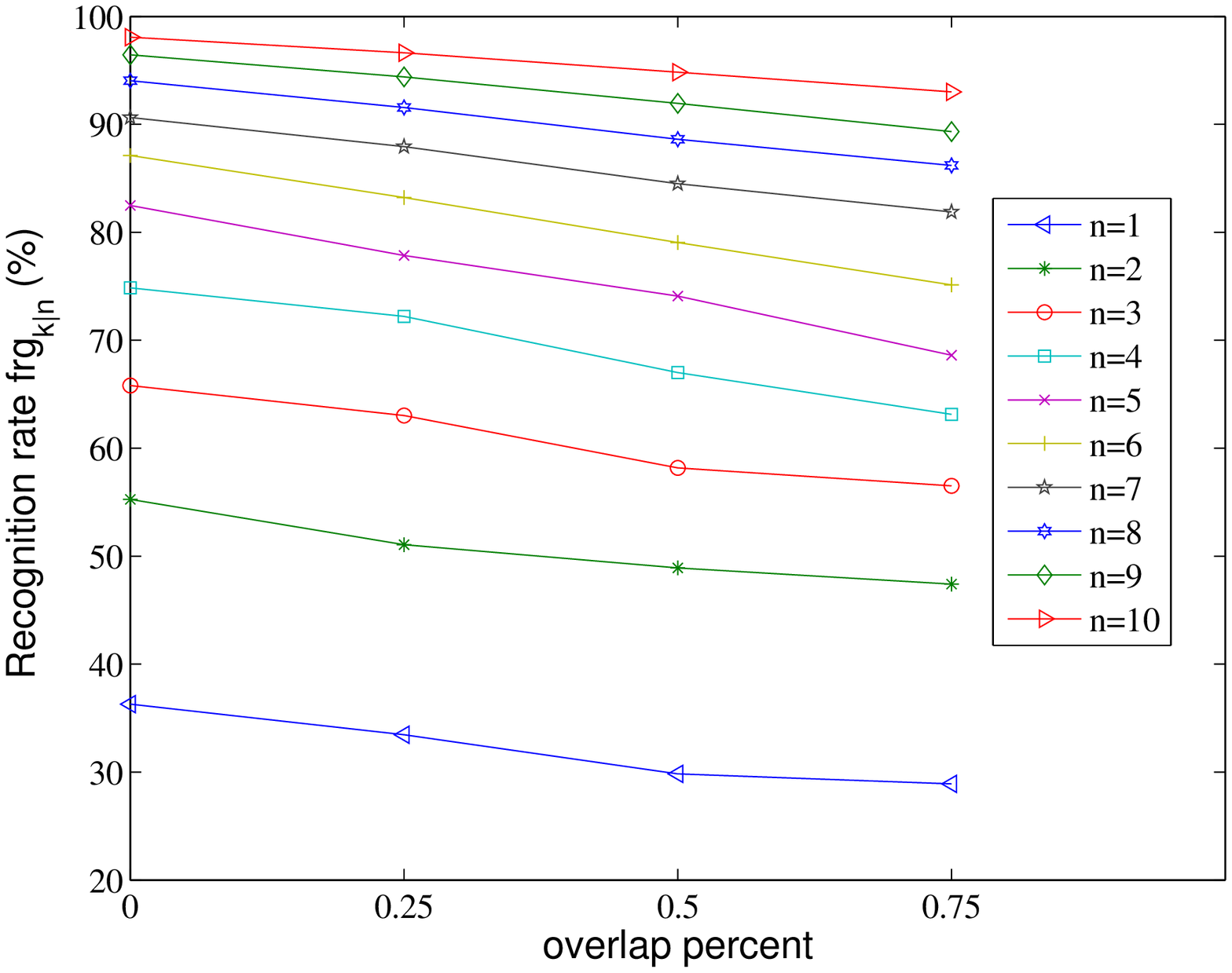} \\(a)&(b)\\
\includegraphics[width=7cm]{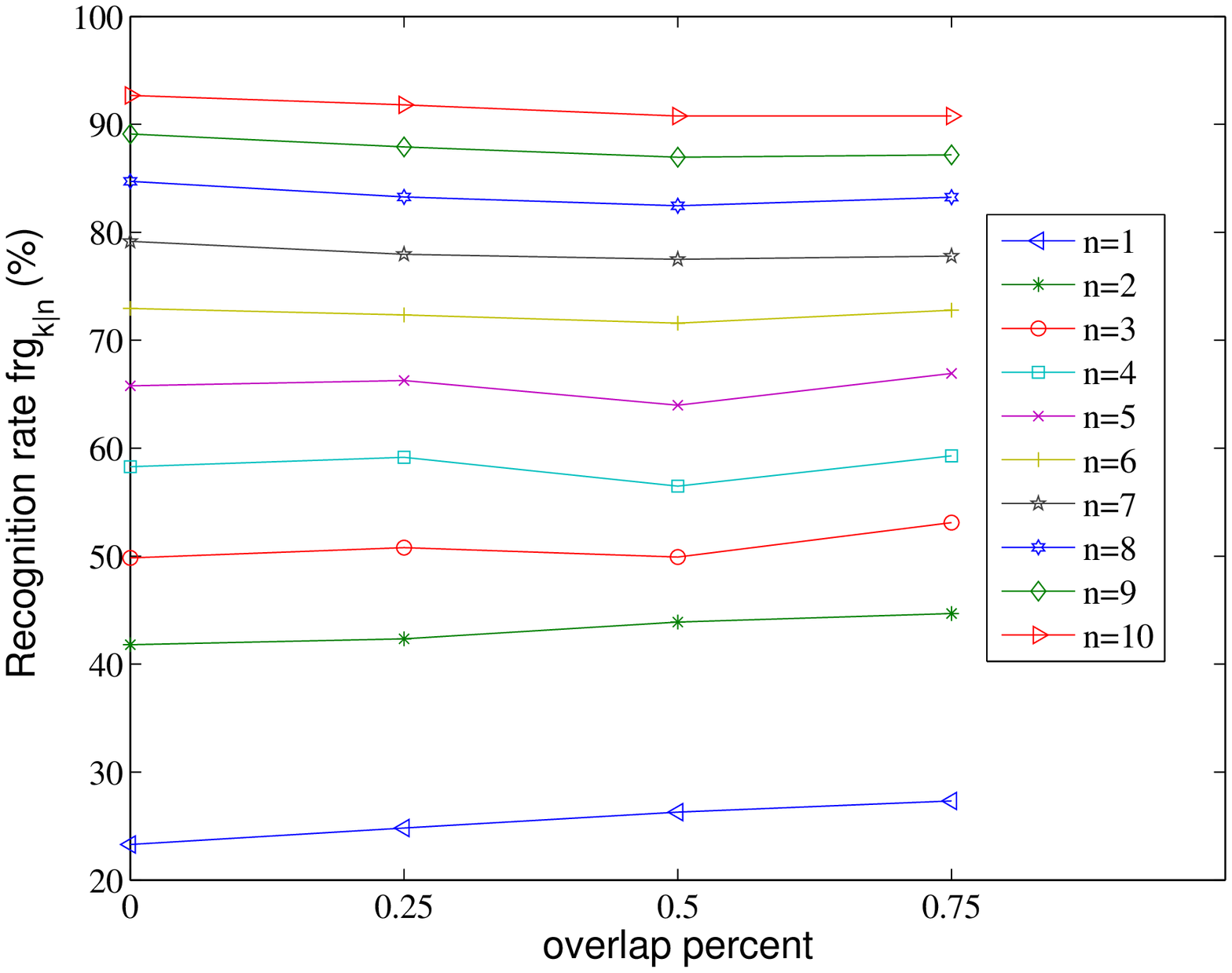} &\ \  \includegraphics[width=7cm]{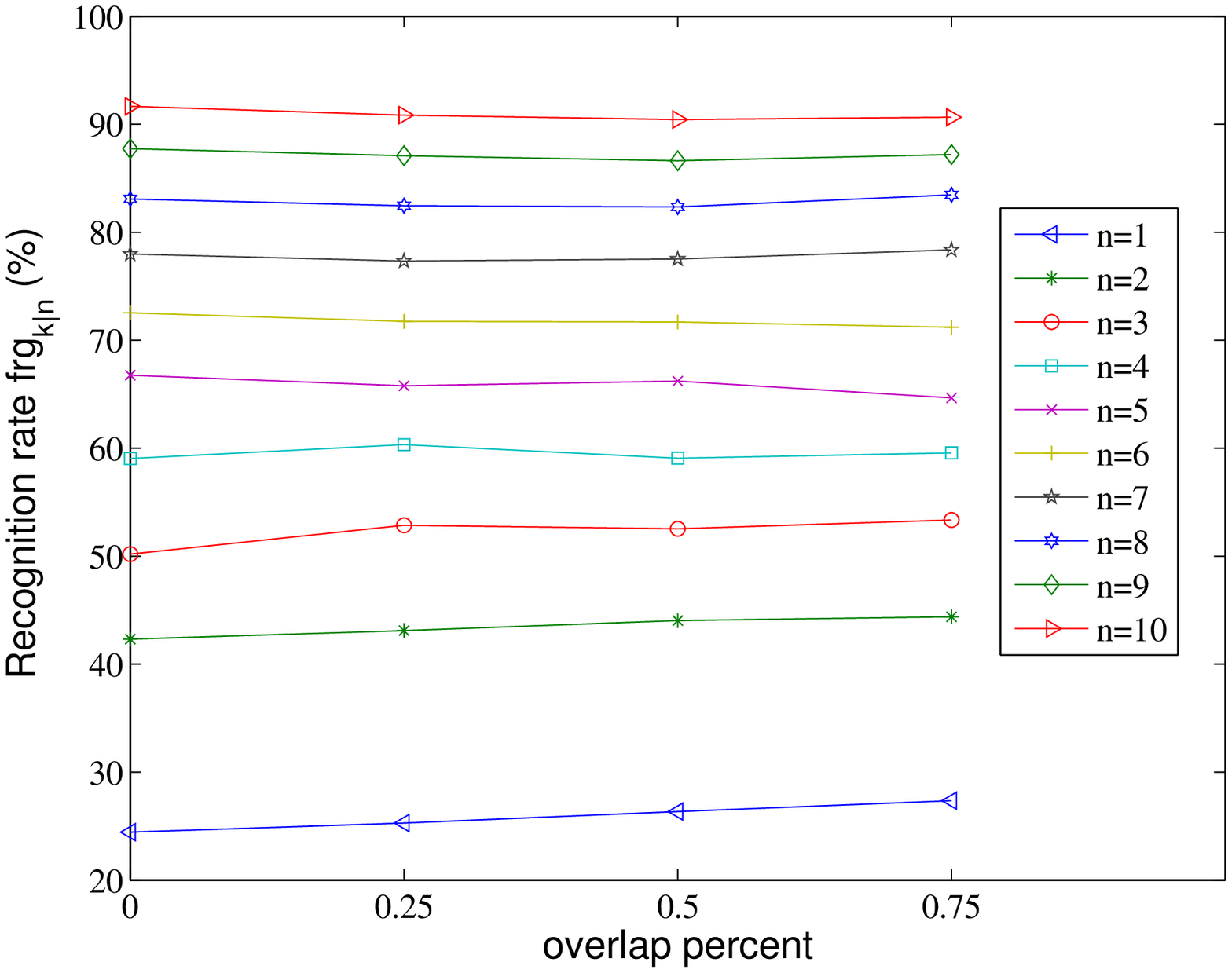} \\(c)&(d)\\
\end{tabular}
\caption*{ Dimension--by--dimension recognition rates on DI. (a) $\mathscr{L}(50,*,60\%)$ (b) $\mathscr{L}(50,*,80\%)$ (c) $\mathscr{T}(50,*,40\%)$ (d) $\mathscr{T}(50,*,20\%)$.}
\end{center}
\end{figure}

\setcounter{figure}{4}
\begin{figure}[!htbp]
\begin{center}
\begin{tabular}{cccc}
\includegraphics[width=7cm]{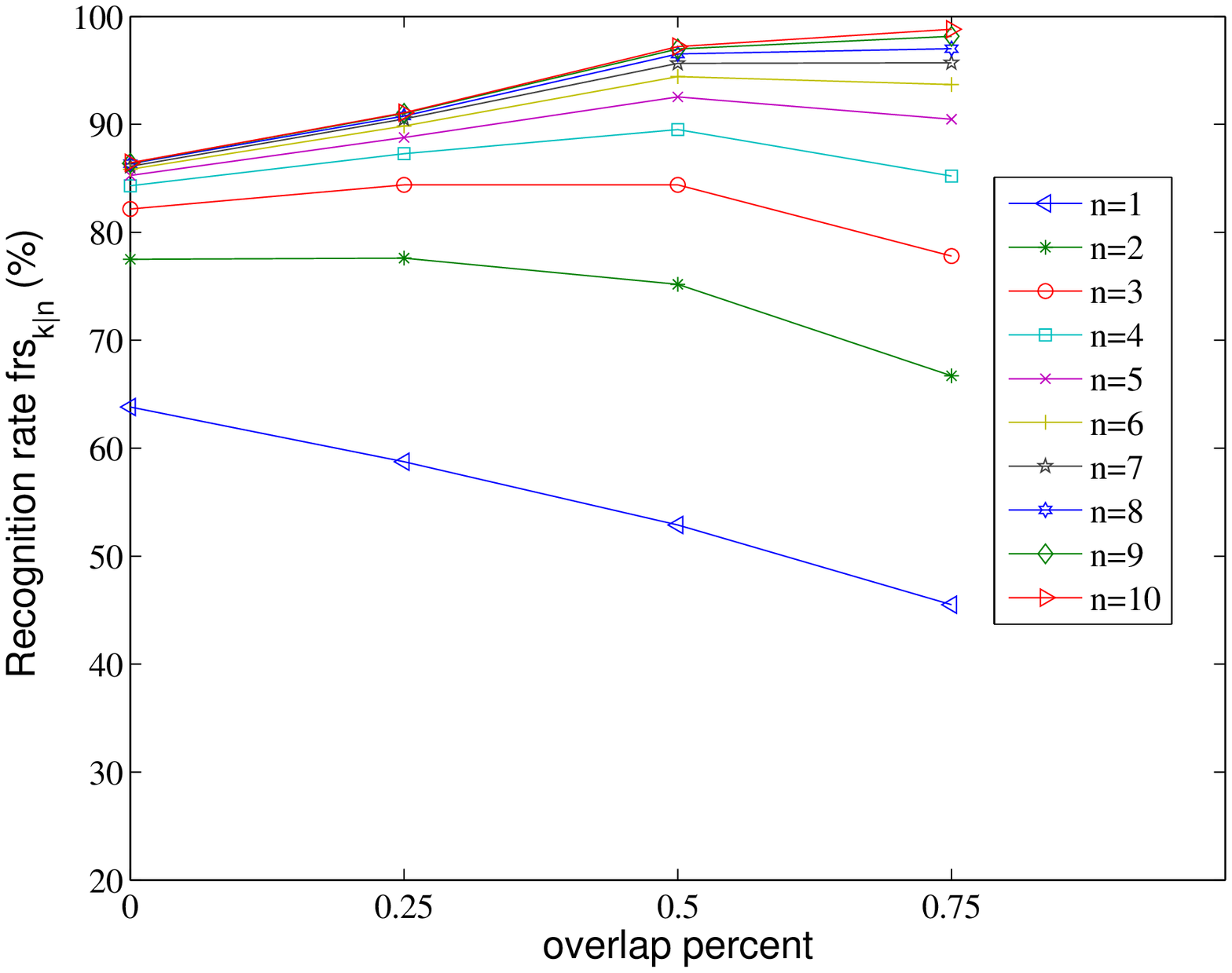} &\ \  \includegraphics[width=7cm]{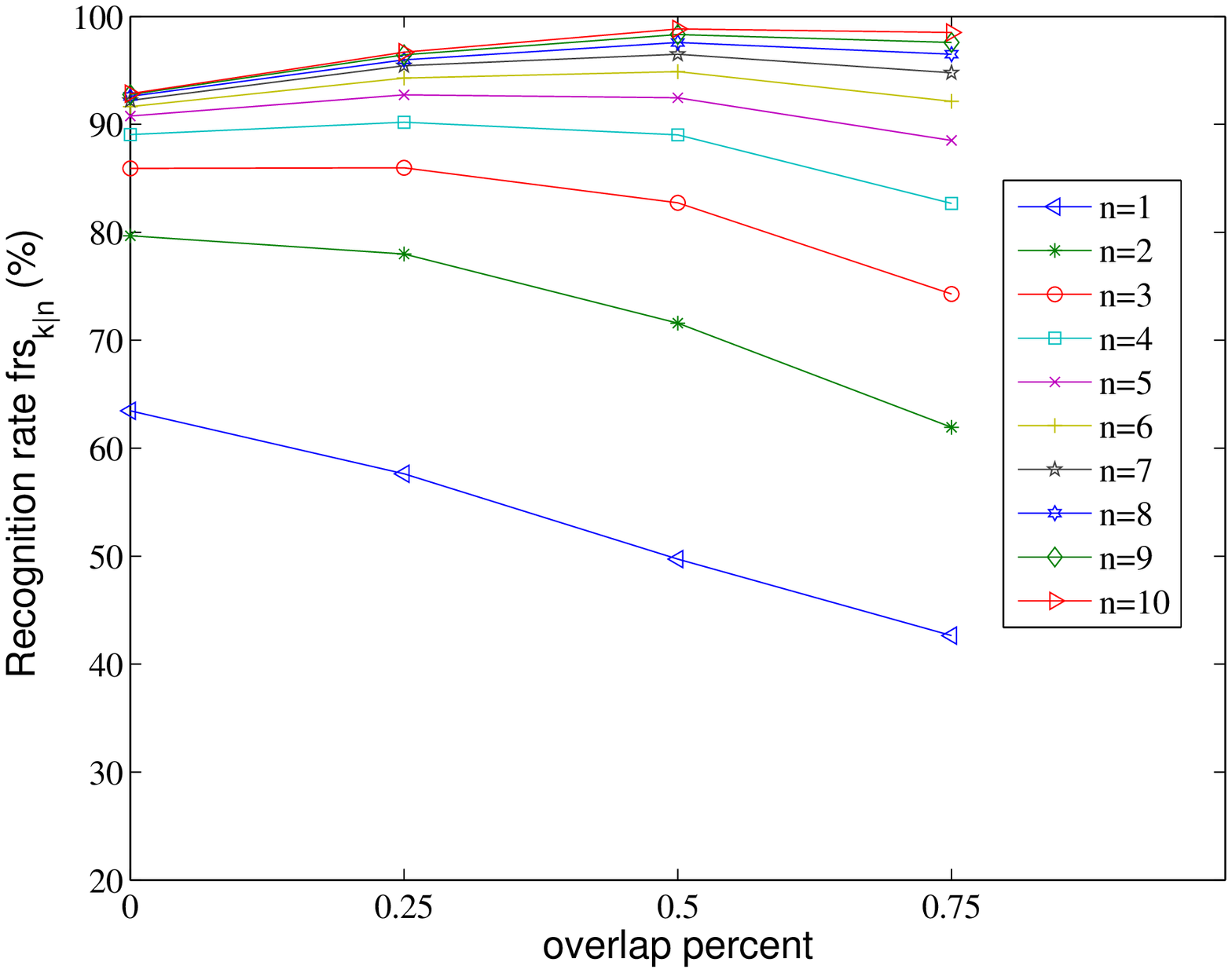} \\(a)&(b)\\
\includegraphics[width=7cm]{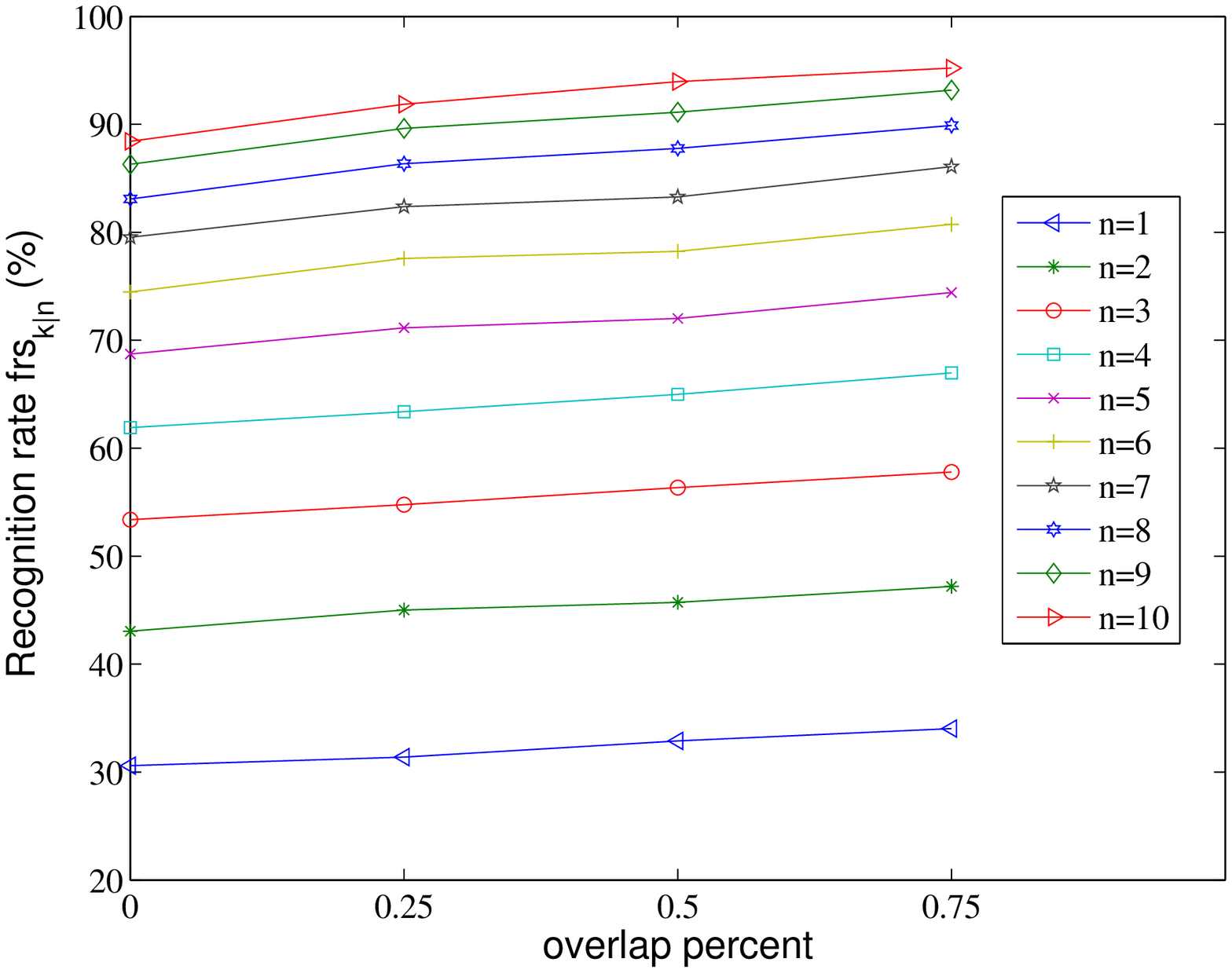} &\ \  \includegraphics[width=7cm]{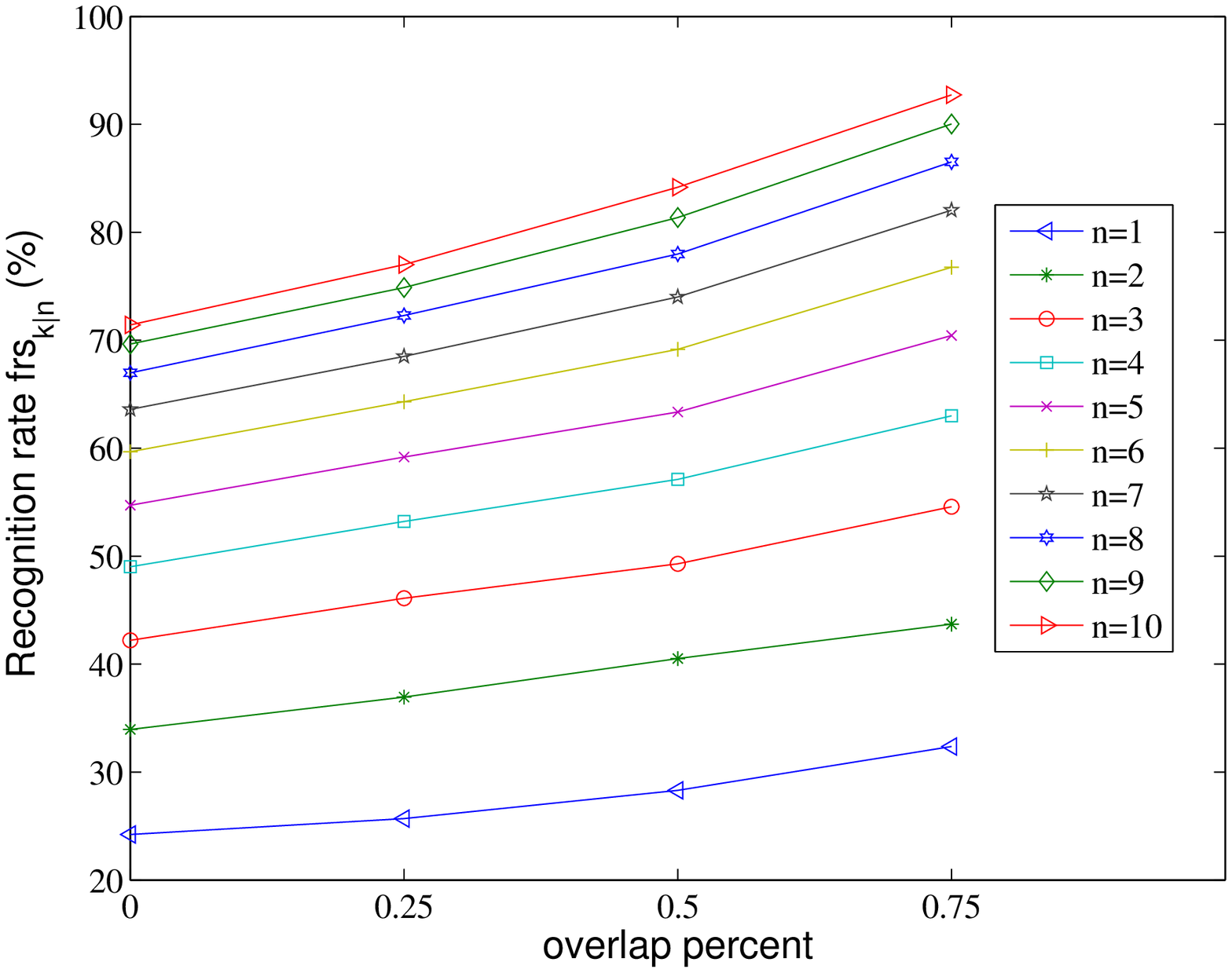} \\(c)&(d)\\
\end{tabular}
\caption*{ Sequence--by--sequence recognition rates on DI. (a) $\mathscr{L}(50,*,60\%)$ (b) $\mathscr{L}(50,*,80\%)$ (c) $\mathscr{T}(50,*,40\%)$ (d) $\mathscr{T}(50,*,20\%)$. }
\end{center}
\end{figure}

\setcounter{figure}{5}
\begin{figure}[!htbp]
\begin{center}
\begin{tabular}{cccc}
\includegraphics[width=7cm]{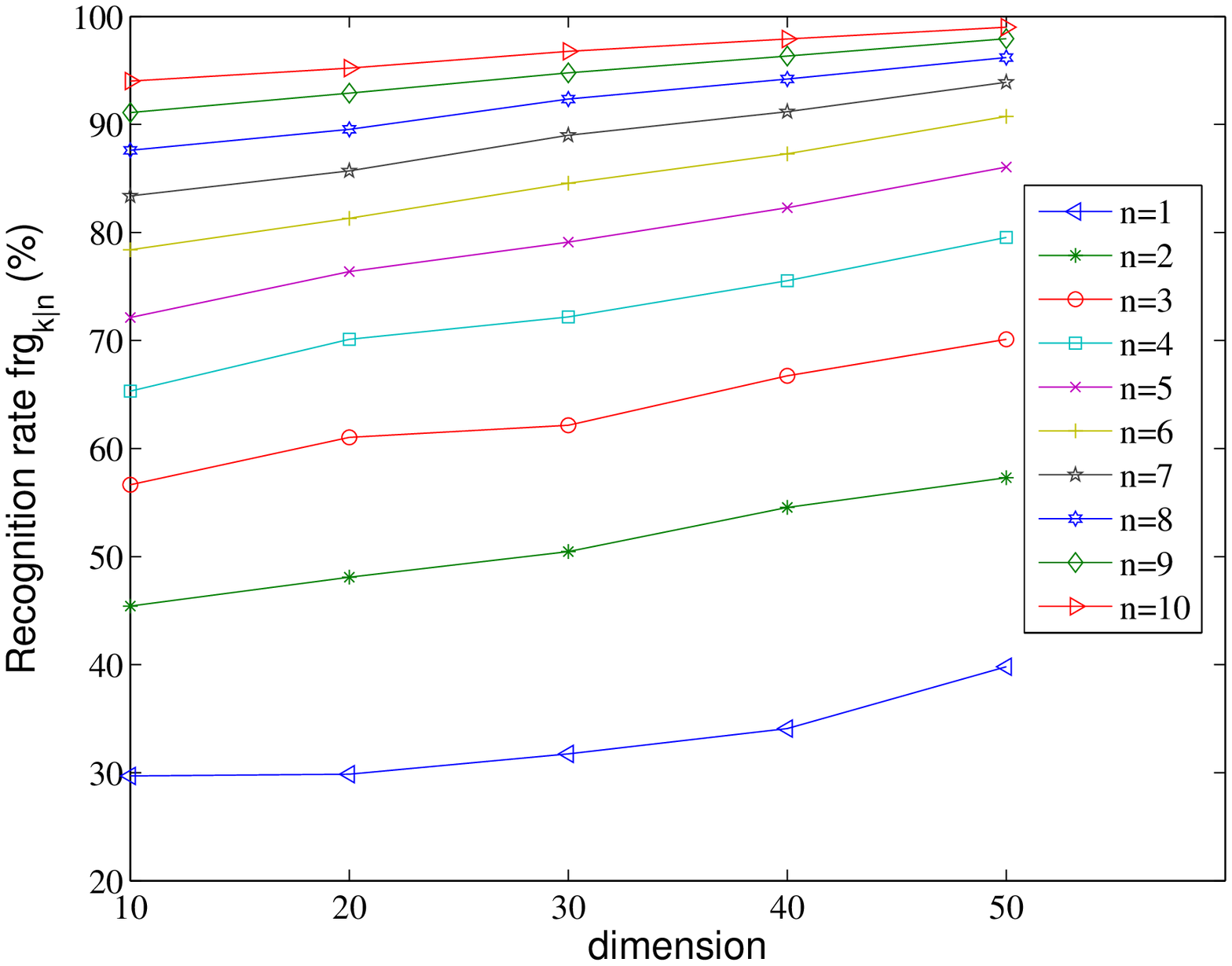} &\ \  \includegraphics[width=7cm]{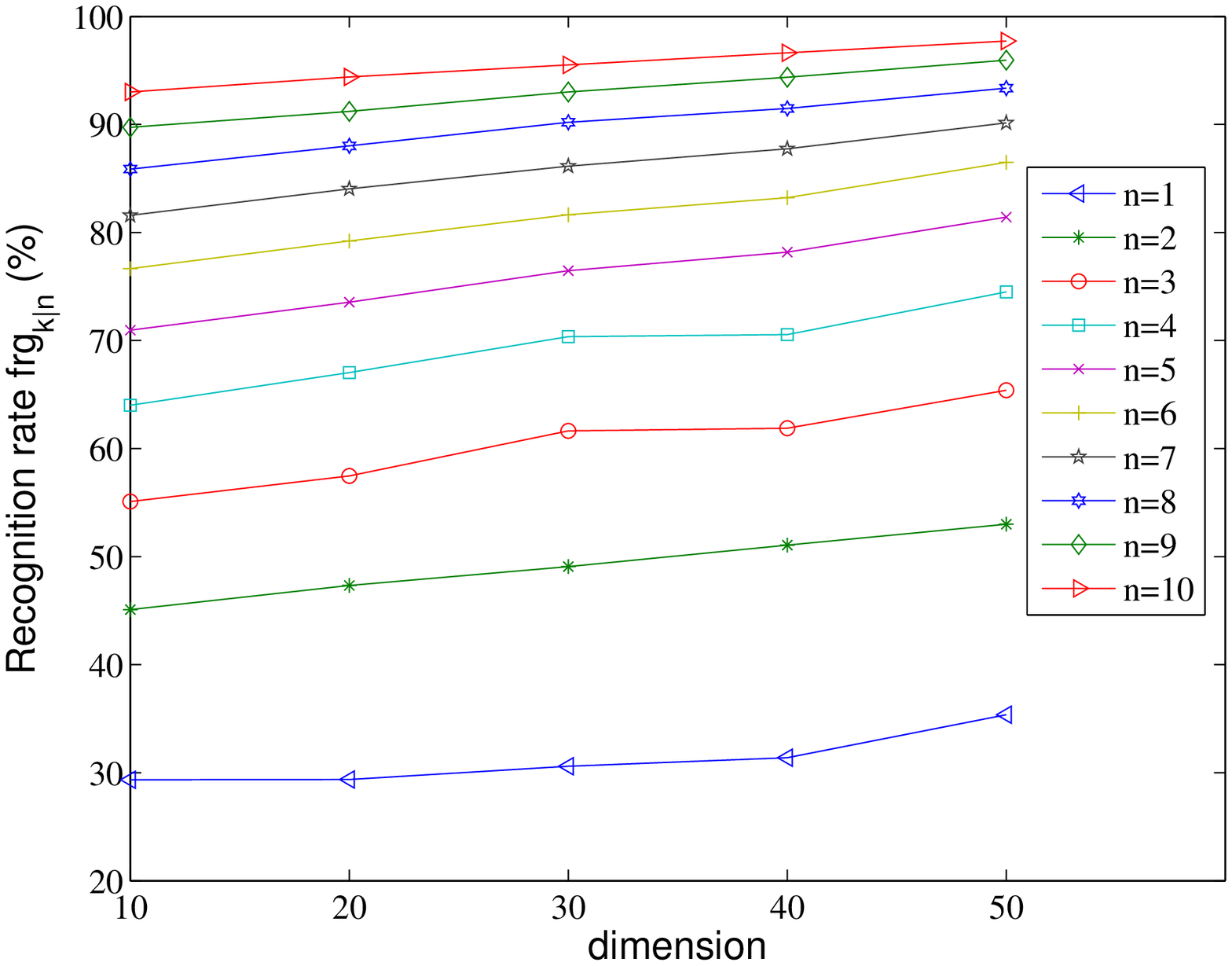} \\(a)&(b)\\
\includegraphics[width=7cm]{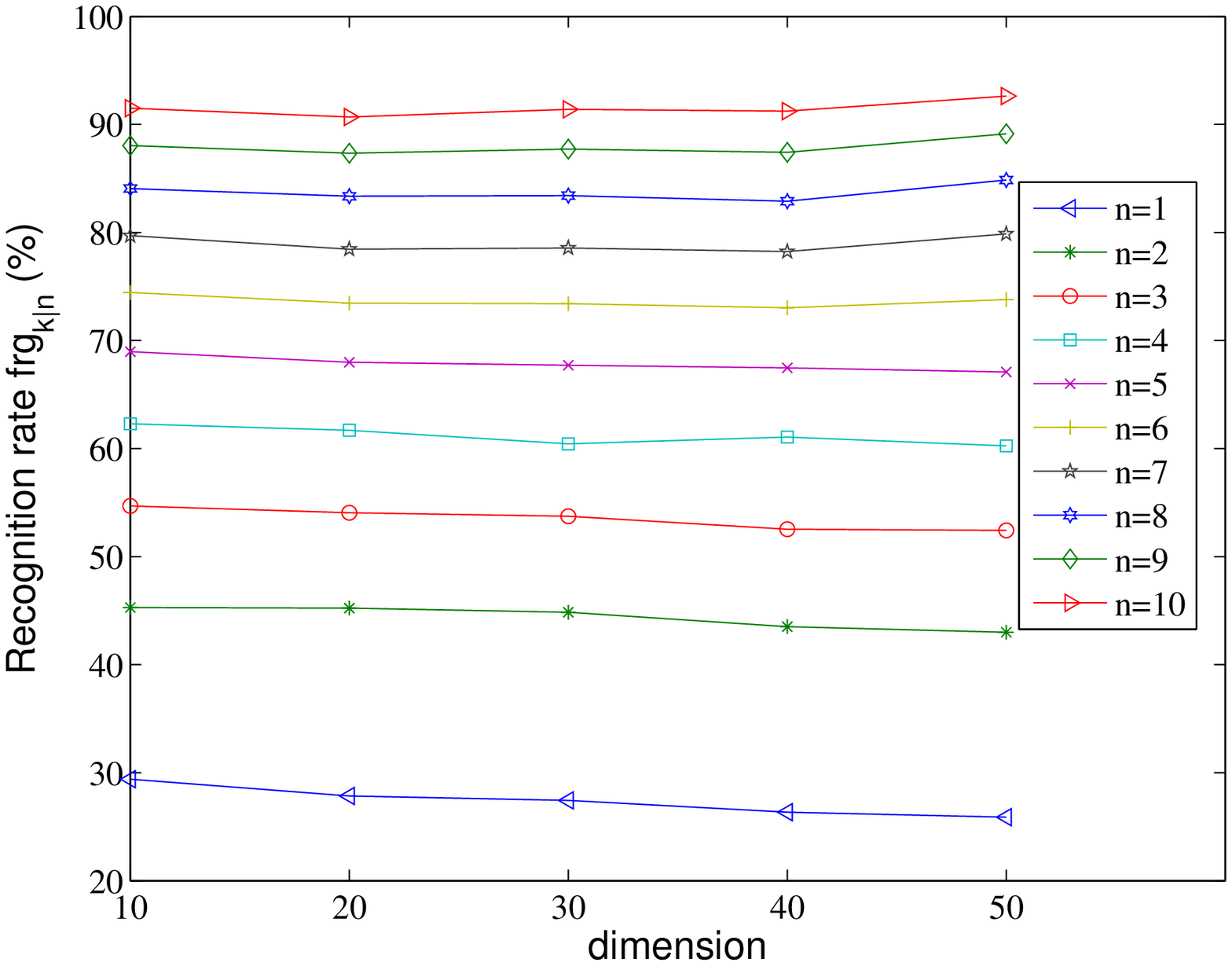} &\ \  \includegraphics[width=7cm]{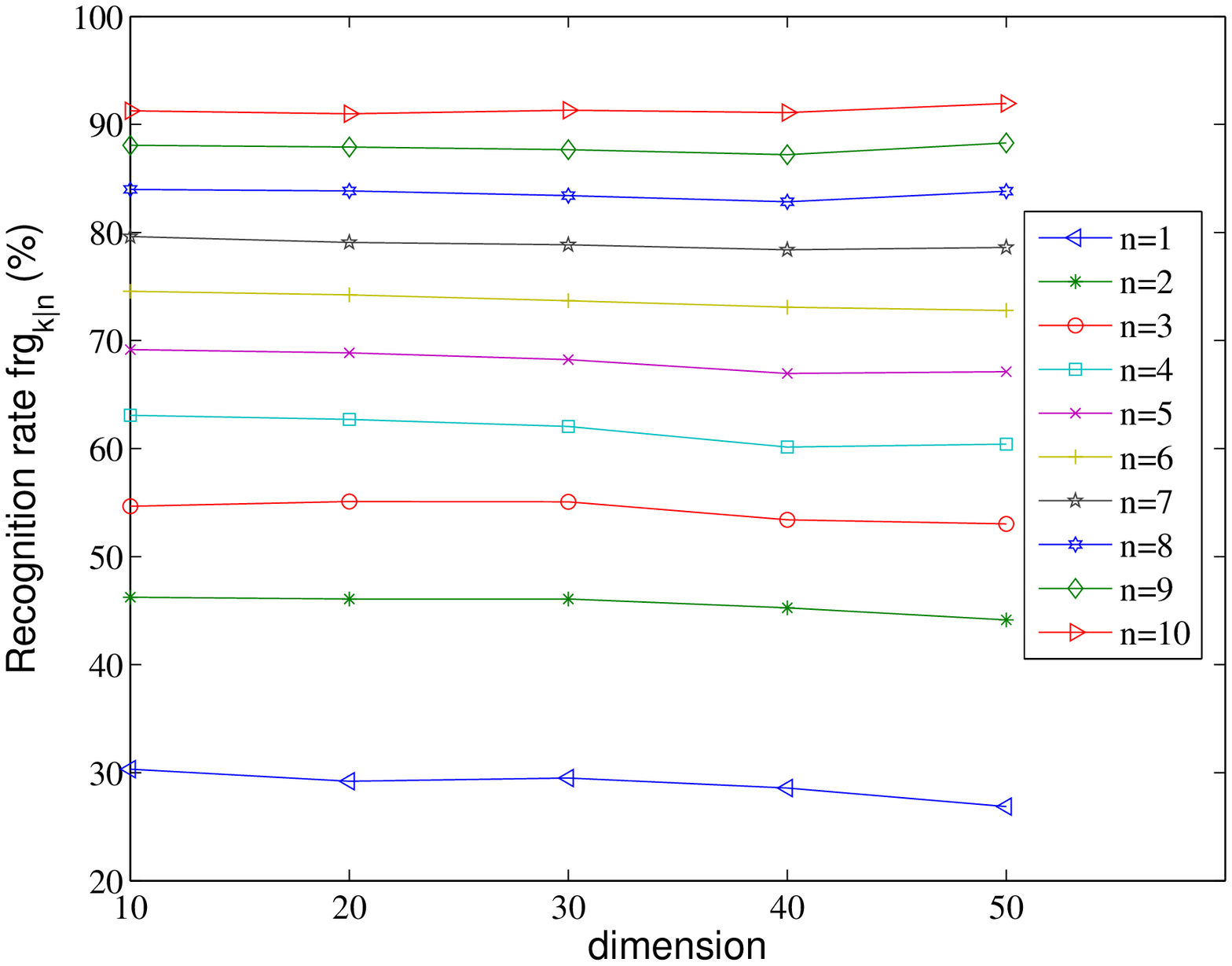} \\(c)&(d)\\
\end{tabular}
\caption*{ Dimension--by--dimension recognition rates on DII . (a) $\mathscr{L}(*,0,60\%)$ (b) $\mathscr{L}(*,0,80\%)$  (c) $\mathscr{T}(*,0,40\%)$ (d) $\mathscr{T}(*,0,20\%)$. }
\end{center}
\end{figure}

\setcounter{figure}{6}
\begin{figure}[!htbp]
\begin{center}
\begin{tabular}{cccc}
\includegraphics[width=7cm]{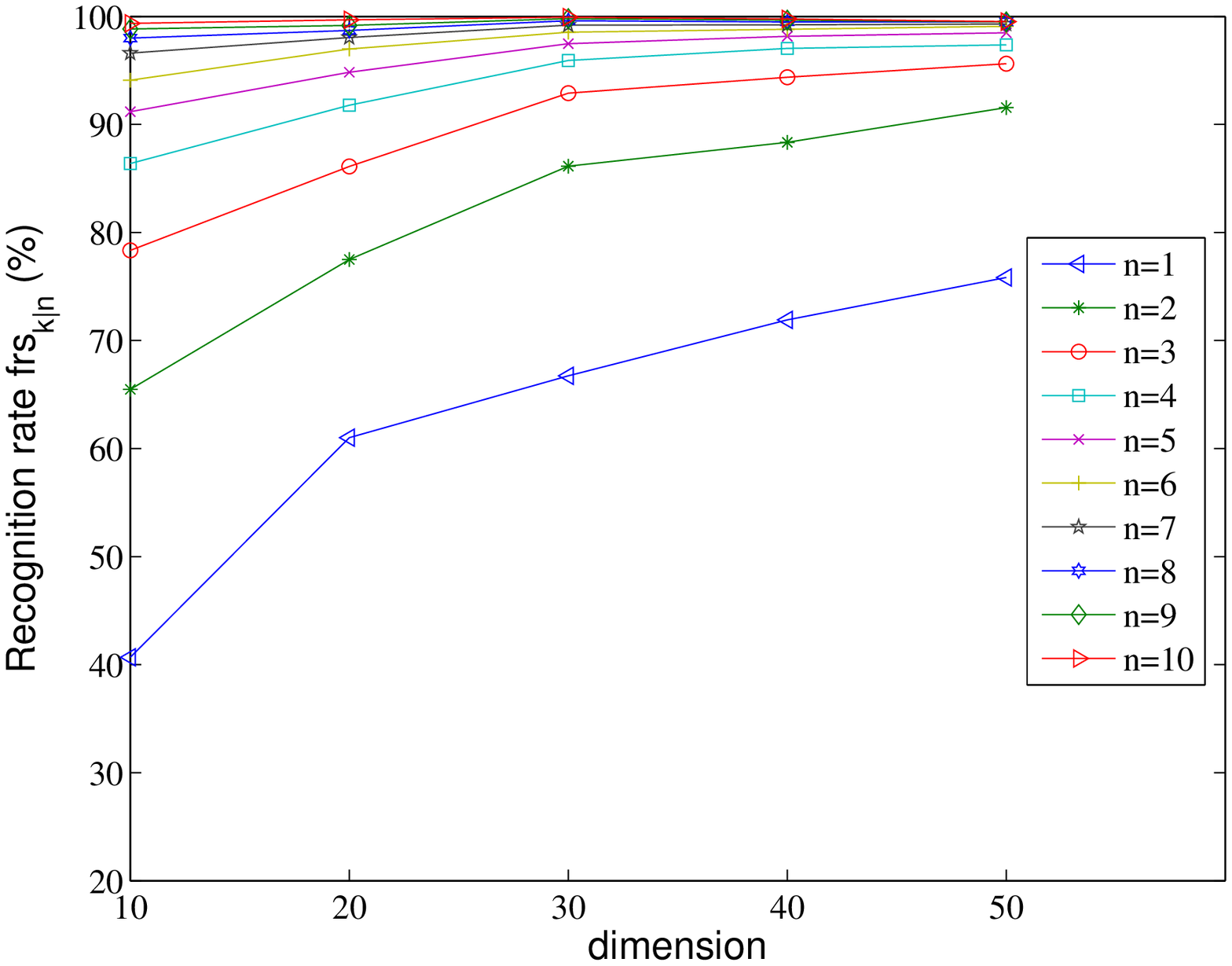} &\ \  \includegraphics[width=7cm]{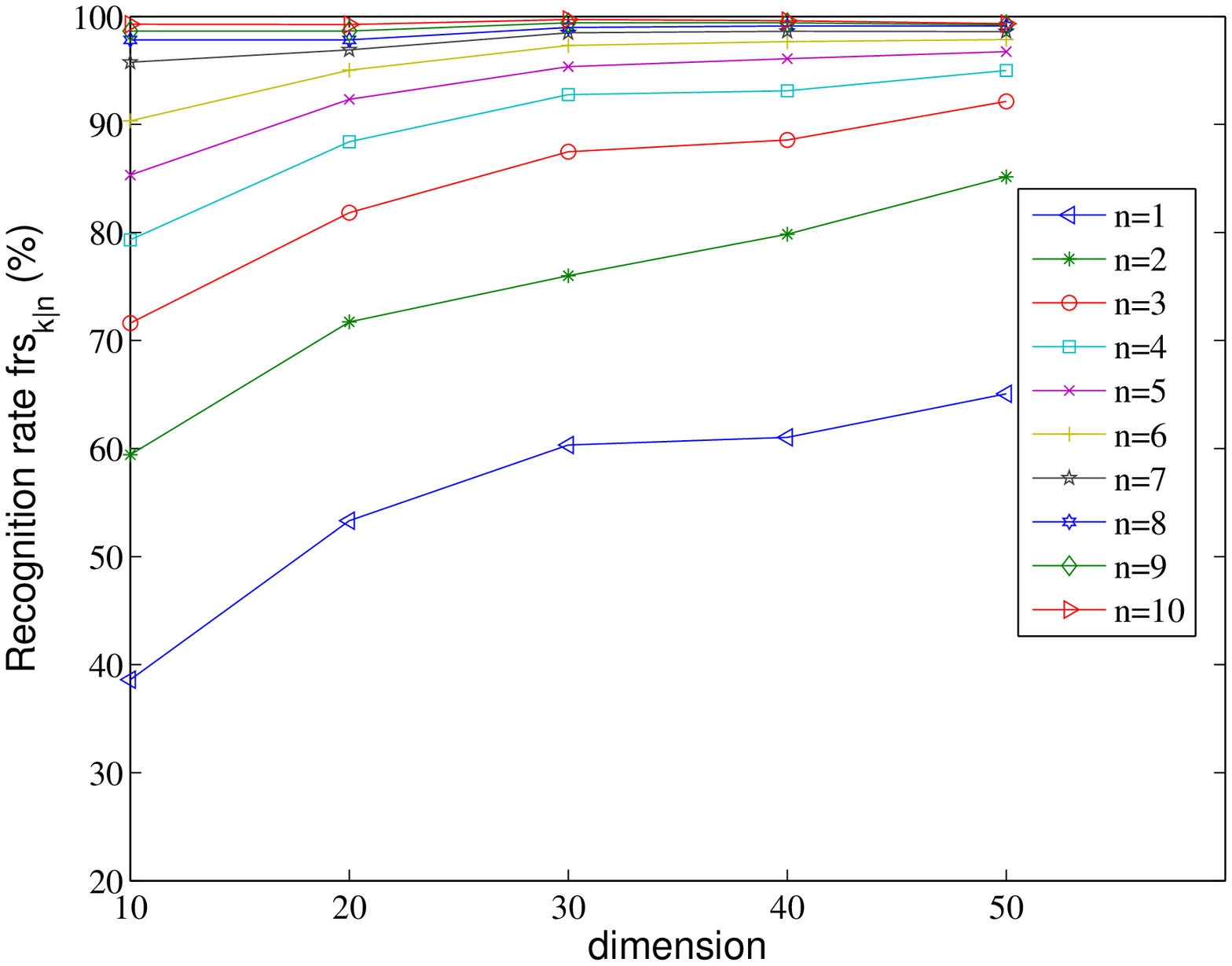} \\(a)&(b)\\
\includegraphics[width=7cm]{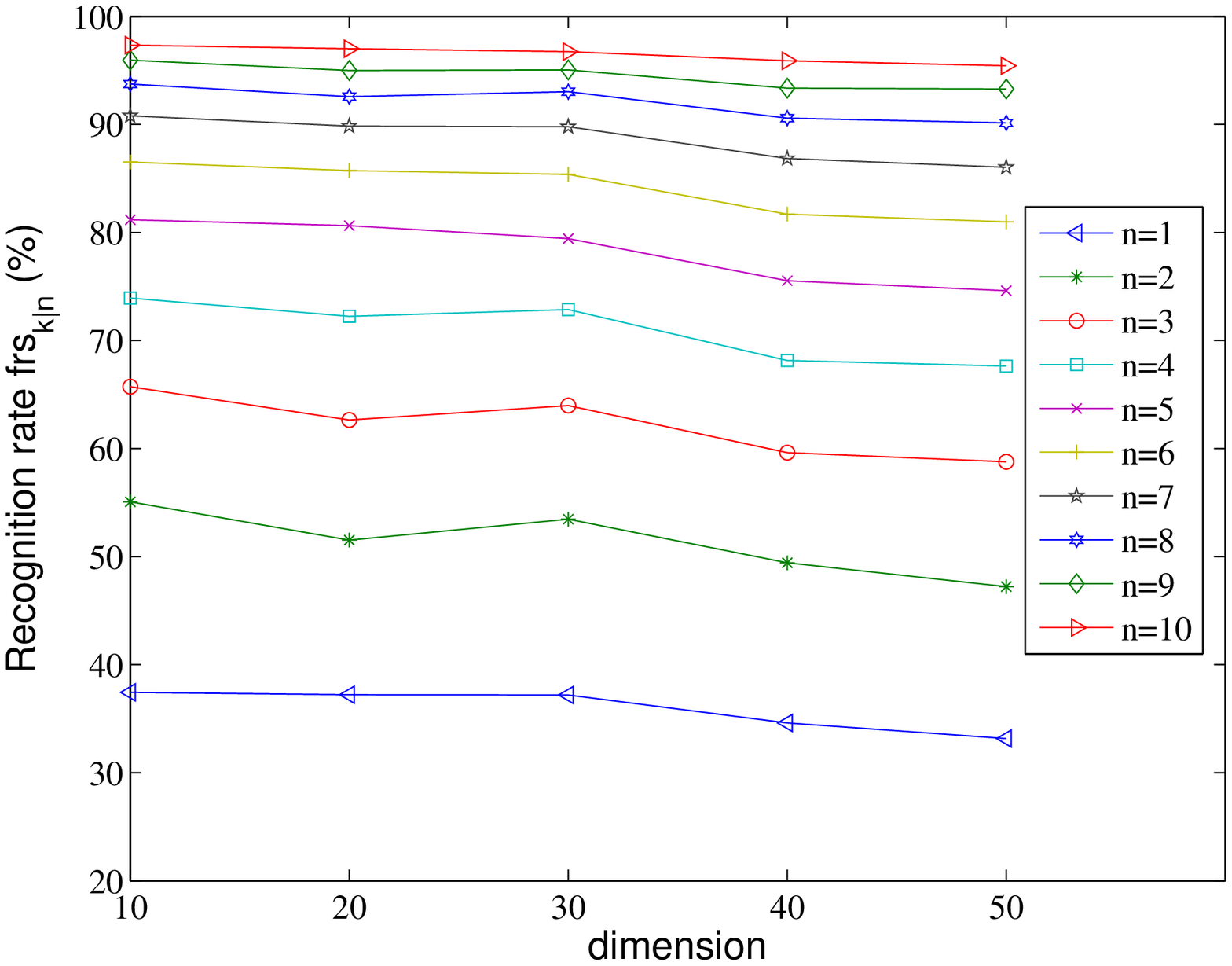} &\ \  \includegraphics[width=7cm]{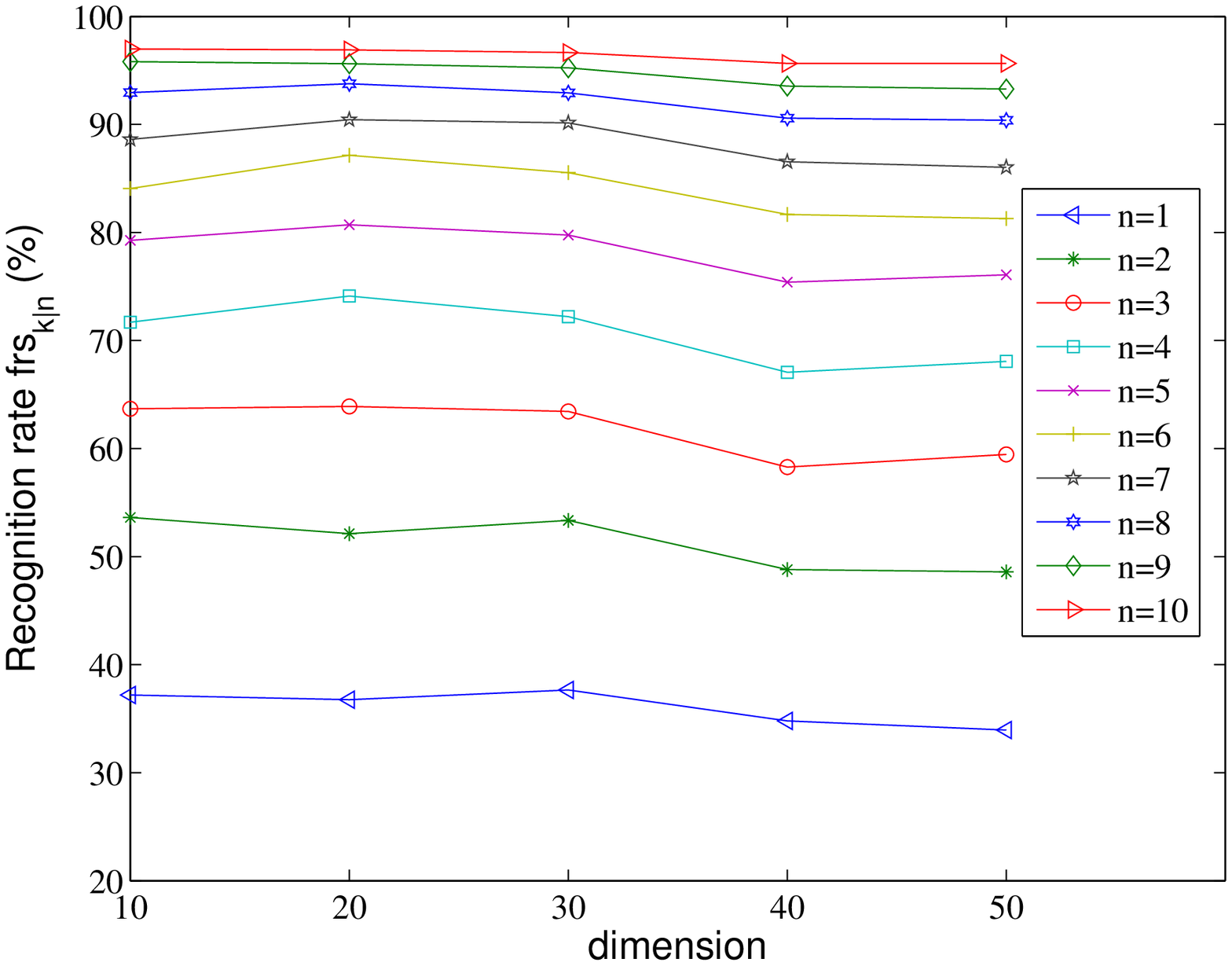} \\(c)&(d)\\
\end{tabular}
\caption*{ Sequence--by--sequence recognition rates on DII. (a) $\mathscr{L}(*,0,60\%)$ (b) $\mathscr{L}(*,0,80\%)$  (c) $\mathscr{T}(*,0,40\%)$ (d) $\mathscr{T}(*,0,20\%)$. }
\end{center}
\end{figure}


\clearpage

\begin{table}[!htbp]
\caption{Maximum accuracy rates of dimension-by-dimension recognition with SVM and top N-Best on DI.}
\label{table:2}
\renewcommand{\tabcolsep}{0.2pc} 
\renewcommand{\arraystretch}{1}
\begin{center}
\begin{tabular}{cllllllllll}
\toprule
& & & & & & & & & &\\
\hline
\multirow{2}{*}{Data set} & \multicolumn{10}{c}{N--candidate} \\
\cmidrule{2-11}
&1 &2 &3 &4 &5 &6 &7 &8 &9 &10\\
\midrule
$\mathscr{L}(10,0,100\%)$  & 29.1524&  45.3499&  55.1381&  63.9802&  70.5578&  76.3853&  81.4770&  86.0207&  89.7962&  92.9417 \\
$\mathscr{L}(20,0,100\%)$	&29.1290&  46.7605&  57.4512&  65.9442&  72.5504&  78.3019&  83.2321&  87.4885&  90.9828&  93.8046  \\
$\mathscr{L}(30,0,100\%)$	&30.2417&  48.5501&  59.5906&  67.4187&  74.8785&  80.3749&  85.4805&  89.5094&  92.7047&  95.1597 \\
$\mathscr{L}(40,0,100\%)$	&32.3072&  50.1270&  62.1664&  70.1953&  76.2544&  81.7982&  86.2868&  89.9982&  93.2974&  95.6939 \\
$\mathscr{L}(50,0,100\%)$	&33.1408&  51.8390&  64.2131&  73.6849&  80.4710&  85.0886&  88.4445&  92.0622&  94.5749&  96.7707  \\
\rowcolor{mygray}
$\mathscr{L}(50,0.25,100\%)$&31.0661& 49.1964&  59.6015&  66.4508&  73.3321&  79.8003&  85.5709&  89.7882&  93.1794&  95.6673 \\
$\mathscr{L}(50,0.5,100\%)$	&31.1293&  47.9385&  57.0064&  65.4158&  73.0545&  78.7950&  82.8590&  87.4695&  91.2446&  94.2269 \\
$\mathscr{L}(50,0.75,100\%)$&28.7537&  46.9359&  56.0094&  62.4396&  67.8412&  76.8814&  81.8509&  86.4579&  89.3755&  92.4086 \\

\hline
\hline

$\mathscr{L}(10,0,80\%)$	&28.9219 & 45.5890 & 55.2286&  64.1692 & 71.3453 & 76.5182 & 81.6939 & 86.0141 & 89.7028 & 93.0153 \\
$\mathscr{L}(20,0,80\%)$	&29.8525&  46.8040&  58.0903 & 68.3442&  74.6969 & 80.2725 & 85.0012&  88.8769 & 91.9582 & 94.5587 \\
$\mathscr{L}(30,0,80\%)$	&32.8037 & 50.9962 & 62.3726 & 71.0052 & 77.4849 & 82.6210 & 87.0528 & 90.6602 & 93.5443 & 95.9695 \\
$\mathscr{L}(40,0,80\%)$	&33.5453 & 51.6201 & 62.2330 & 70.8199 & 76.7607 & 82.9937 & 87.7732 & 91.6892 & 94.6637 & 96.8024 \\
$\mathscr{L}(50,0,80\%)$	&36.2889 & 55.2549 & 65.8190 & 74.8511 & 82.4900 & 87.1315 & 90.6297 & 94.0369 & 96.4533 & 98.0879 \\
\rowcolor{mygray}
$\mathscr{L}(50,0.25,80\%)$	&33.4600 & 51.0611 & 63.0256 & 72.2075 & 77.8655 & 83.2365 & 87.9323 & 91.5702 & 94.4105&  96.6420 \\
$\mathscr{L}(50,0.50,80\%)$	&29.8364  &48.9057 & 58.1688 & 67.0047 & 74.0874  &79.0414&  84.4936 & 88.6204 & 91.9367 & 94.8254 \\
$\mathscr{L}(50,0.75,80\%)$	&28.9125 & 47.4129 & 56.5023 & 63.1323 & 68.6258 & 75.1206 & 81.8824 & 86.2010 & 89.3410 & 93.0019 \\
\hline
$\mathscr{T}(10,0,20\%)$	& 28.1011 & 44.6525 & 53.7493&  62.6669 & 68.7712 & 74.2429 & 79.5064  & 83.9734 & 87.8928 & 91.2690\\
$\mathscr{T}(20,0,20\%)$	& 27.1238 & 44.2305 & 53.3159&  62.3020 & 68.5094 & 73.8141 & 78.8983&  83.3520 & 87.5116 & 90.8618\\
$\mathscr{T}(30,0,20\%)$	& 26.4781 & 44.0914 & 52.5489&  60.8670 & 67.6340 & 73.6352 & 78.5645 & 83.3591 & 87.4677 & 91.0892\\
$\mathscr{T}(40,0,20\%)$	& 25.5126 & 42.7923 & 50.7391 & 57.6458 & 64.8765 & 71.6992 & 77.5538 & 82.2442 & 86.8907 & 90.7711\\
$\mathscr{T}(50,0,20\%)$	& 24.4188 & 42.2969 & 50.1801 & 59.0578 & 66.7598 & 72.5435 & 77.9893 & 83.0920 & 87.7441 & 91.6783  \\
\rowcolor{mygray}
$\mathscr{T}(50,0.25,20\%)$	& 25.2634 & 43.1037 & 52.8523 & 60.3333 & 65.7691 & 71.7519 & 77.3482 & 82.4728 & 87.0851 & 90.8430\\
$\mathscr{T}(50,0.50,20\%)$	& 26.3453 & 44.0362 & 52.5299 & 59.0685 & 66.2043 & 71.7062 & 77.5346 & 82.3529 & 86.6402 & 90.4444\\
$\mathscr{T}(50,0.75,20\%)$	& 27.3373 & 44.3945 & 53.3592 & 59.5510 & 64.6706 & 71.2024 & 78.3791 & 83.4635 & 87.2125 & 90.6695 \\
\hline
\hline
$\mathscr{L}(10,0,60\%)$	& 29.8095 & 45.7840&  56.3059 & 64.9796 & 71.5621 & 77.7058 & 83.5297 & 88.3057 & 91.7147 & 94.1014\\
$\mathscr{L}(20,0,60\%)$	& 30.6973 & 48.5402&  59.8552 & 68.8960 & 75.6688 & 82.0383 & 86.1187 & 90.4233&  93.5281&  95.6958 \\
$\mathscr{L}(30,0,60\%)$	& 33.9040 & 51.5536&  63.9864 & 73.0764 & 79.9770 &85.3750 & 89.5019 & 92.7104 & 95.1563 & 97.0519\\
$\mathscr{L}(40,0,60\%)$	& 34.9225&  55.1855&  67.5837 & 76.1416 & 82.4520 & 87.4890&  91.3761 & 94.4557 & 96.7062 & 98.1885 \\
$\mathscr{L}(50,0,60\%)$	& 39.3829 & 58.9222&  71.2244 & 80.0714 & 87.0090 & 91.0924 & 94.4168 & 96.7459&  98.2642 & 99.2233\\
\rowcolor{mygray}
$\mathscr{L}(50,0.25,60\%)$	& 33.8457 & 55.0146 & 66.6992 & 75.6584 & 82.0342 & 87.0414 & 91.1657 & 94.3555 & 96.6417 & 98.1922 \\
$\mathscr{L}(50,0.50,60\%)$	& 31.2826 & 50.9855 & 59.9542 & 67.4814 & 74.7800 & 81.2390 & 86.9697 & 90.7208 & 93.5934 & 96.0711\\
$\mathscr{L}(50,0.75,60\%)$	& 29.5668 & 48.1559 & 57.5237 & 64.3312 & 71.6415 & 77.5680 & 82.5913 & 86.9179&  90.5514 & 93.5667\\
\hline
$\mathscr{T}(10,0,40\%)$	&28.6860 & 44.7459 & 54.0663 & 62.7004 & 69.1209 & 74.6020 & 80.2309 & 84.9403 & 89.0569 & 91.7479  \\
$\mathscr{T}(20,0,40\%)$	&27.2974 & 44.4829 & 53.6333 & 60.6738 & 67.1645 & 73.2349 & 78.7768 & 83.1719 & 87.2416 & 90.9932  \\
$\mathscr{T}(30,0,40\%)$	&26.4721 & 44.1347 & 53.4889 & 59.7083 & 66.6915 & 72.8363 &  78.3563&  83.3982 & 87.8438 & 91.5343  \\
$\mathscr{T}(40,0,40\%)$	&25.2008 & 42.4063 & 51.9196 & 59.8656 & 66.5702 & 72.9194 & 78.2758 & 83.0555 & 87.6245 & 91.2929 \\
$\mathscr{T}(50,0,40\%)$	&23.2928 & 41.7797 & 49.8286 & 58.2750 & 65.7790 & 72.9532 & 79.1590 & 84.7130 & 89.1239 & 92.6863 \\
\rowcolor{mygray}
$\mathscr{T}(50,0.25,40\%)$	&24.8183 & 42.3461 & 50.7740 & 59.1511 & 66.2693 & 72.3535 & 77.9521 & 83.2761 & 87.9005 & 91.8124 \\
$\mathscr{T}(50,0.50,40\%)$	&26.2837 & 43.8785 & 49.9088 & 56.4861 & 63.9817 & 71.5918 & 77.5077 & 82.4559 & 86.9606 & 90.7753 \\
$\mathscr{T}(50,0.75,40\%)$	&27.3154 & 44.6684 & 53.1084 & 59.2998 & 66.9273 & 72.7711 & 77.8036 & 83.2448 & 87.1742 & 90.7759 \\

\bottomrule
\end{tabular}
\end{center}
\end{table}


\begin{table}[!htbp]
\caption{Maximum accuracy rates of sequence-by-sequence recognition with SVM and top N-Best on DI.}
\label{table:3}
\renewcommand{\tabcolsep}{0.2pc} 
\renewcommand{\arraystretch}{1}
\begin{center}
\begin{tabular}{cllllllllll}
\toprule
& & & & & & & & & &\\
\hline
\multirow{2}{*}{Data set} & \multicolumn{10}{c}{N--candidate} \\
\cmidrule{2-11}
&1 &2 &3 &4 &5 &6 &7 &8 &9 &10\\
\midrule
$\mathscr{L}(10,0,100\%)$   &39.4915 & 57.4576 & 67.2639 & 75.4722 & 82.6634 & 89.2010 & 95.0202 & 97.5706 & 98.7571&  99.3624\\
$\mathscr{L}(20,0,100\%)$	&51.3156 & 69.2575 & 79.3462 & 86.0371 & 90.8878 & 94.2050 & 96.1582 & 97.6513 & 98.5876&  99.1768\\
$\mathscr{L}(30,0,100\%)$	&55.9726 & 72.7441 & 82.1065 & 89.0395 & 93.6481 & 95.9241 & 97.3204 & 98.3858 & 98.9266&  99.3866\\
$\mathscr{L}(40,0,100\%)$	&56.4568 & 73.4544 & 83.7369 & 90.1291 & 93.7611 & 95.9887 & 97.5545 & 98.5069 & 99.0073 & 99.3947\\
$\mathscr{L}(50,0,100\%)$	&61.2994 & 80.2663 & 88.6844 & 92.8975 & 95.4237 & 96.9976 & 98.0468 & 98.7006 & 99.0395 & 99.3140\\
\rowcolor{mygray}
$\mathscr{L}(50,0.25,100\%)$&53.2526 & 76.5214 & 85.9564 & 91.0412 & 94.1001 & 96.1824 & 97.4173 & 98.3293 & 98.9346 & 99.2655\\
$\mathscr{L}(50,0.5,100\%)$	&46.6344 & 66.6182 & 78.6683 & 86.2308 & 90.9524 & 94.0678 & 96.2228 & 97.5383 & 98.4262 & 99.0557\\
$\mathscr{L}(50,0.75,100\%)$&38.3939 & 60.1049 & 72.2680 & 81.5335 & 87.9984 & 92.1872 & 94.9314 & 96.5698 & 97.6755 & 98.2728\\

\hline
\hline

$\mathscr{L}(10,0,80\%)$	&39.1364 & 58.4262 & 68.4504 & 79.2575 & 88.3777 & 93.5109 & 96.3842 & 97.9903 & 98.8943 & 99.4350 \\
$\mathscr{L}(20,0,80\%)$	&48.6037 & 66.5456 & 77.6513 & 84.8265 & 90.0646 & 93.8983 & 96.2550 & 97.7805 & 98.6602 & 99.0234  \\
$\mathscr{L}(30,0,80\%)$	&54.3906 & 77.4173 & 86.8039 & 91.6303 & 94.6086 & 96.2308 & 97.4011 & 98.2405 & 98.8701 & 99.2494 \\
$\mathscr{L}(40,0,80\%)$	&56.3600 & 77.0541 & 86.5456 & 91.3882 & 94.1727 & 95.8031 & 96.8119 & 97.3527 & 97.6755 & 97.8692 \\
$\mathscr{L}(50,0,80\%)$	&63.4544 & 79.6772 & 85.9161 & 89.0638 & 90.7667 & 91.6546 & 92.2195 & 92.5989 & 92.7603 & 92.8571 \\
\rowcolor{mygray}
$\mathscr{L}(50,0.25,80\%)$	&57.6352 & 77.9903 & 85.9887 & 90.2018 & 92.7441 & 94.2776 & 95.4479 & 95.9726 & 96.4487 & 96.6747 \\
$\mathscr{L}(50,0.50,80\%)$	&49.7256 & 71.5900 & 82.7199 & 89.0315 & 92.4697 & 94.8910 & 96.4972 & 97.5868 & 98.3212 & 98.8378 \\
$\mathscr{L}(50,0.75,80\%)$	&42.6312 & 61.9290 & 74.2938 & 82.6877 & 88.5149 & 92.1388 & 94.7700 & 96.5052 & 97.5868 & 98.5149 \\
\hline
$\mathscr{T}(10,0,20\%)$	&33.4221 & 47.6433 & 57.8935 & 65.8596 & 72.8491 & 78.6441 & 84.7215 & 88.8539 & 92.1711 & 94.5924 \\
$\mathscr{T}(20,0,20\%)$	&31.5658 & 45.4641 & 54.8265 & 63.3495 & 69.9596 & 76.1501 & 81.2510 & 85.7304 & 89.0799 & 91.7998  \\
$\mathscr{T}(30,0,20\%)$	&30.9847 & 42.3648 & 52.1630 & 59.4754 & 65.6013 & 70.9201 & 75.3511 & 79.0718 & 81.8967 & 83.8337 \\
$\mathscr{T}(40,0,20\%)$	&24.7700 & 36.3519 & 44.8426 & 52.2841 & 58.5714 & 63.8902 & 68.2647 & 71.7918 & 74.6893 & 76.7232  \\
$\mathscr{T}(50,0,20\%)$	&24.2131 & 33.9387 & 42.2115 & 49.0073 & 54.7054 & 59.6691 & 63.6077 & 66.9814 & 69.6529 & 71.4205 \\
\rowcolor{mygray}
$\mathscr{T}(50,0.25,20\%)$	&25.6739 & 36.9492 & 46.0856 & 53.2203 & 59.1687 & 64.2938 & 68.5069 & 72.2841 & 74.8830 & 76.9976  \\
$\mathscr{T}(50,0.50,20\%)$	&28.3132 & 40.5004 & 49.2978 & 57.1186 & 63.3656 & 69.1525 & 74.0032 & 77.9903 & 81.3801 & 84.1727  \\
$\mathscr{T}(50,0.75,20\%)$	&32.3648 & 43.7046 & 54.5763 & 62.9863 & 70.4358 & 76.7635 & 82.0420 & 86.5214 & 90.0242 & 92.7441 \\
\hline
\hline
$\mathscr{L}(10,0,60\%)$	& 40.0726 & 60.2098 & 72.0420 & 81.1784 & 87.7401 & 91.9774 & 95.1574 & 97.0299 & 98.3132 & 99.2010\\
$\mathscr{L}(20,0,60\%)$	& 52.5424 & 70.4439 & 81.8321 & 87.9984 & 92.3487 & 95.4076 & 97.0944 & 97.9903 & 98.5391 & 98.9911 \\
$\mathscr{L}(30,0,60\%)$	& 59.6449 & 80.2906 & 87.9822 & 91.8886 & 94.4794 & 95.8192 & 96.6586 & 97.2801 & 97.6513 & 97.8370\\
$\mathscr{L}(40,0,60\%)$	& 56.9895 & 76.9734 & 84.2696 & 87.4011 & 89.2655 & 90.3148 & 90.8232 & 91.1703 & 91.3479 & 91.4366\\
$\mathscr{L}(50,0,60\%)$	& 63.8257 & 77.4980 & 82.1711 & 84.3099 & 85.2785 & 85.8273 & 86.1259 & 86.3519 & 86.4084 & 86.4326\\
\rowcolor{mygray}
$\mathscr{L}(50,0.25,60\%)$	& 58.7571 & 77.5948 & 84.3987 & 87.2801 & 88.7732 & 89.8467 & 90.4923 & 90.7829 & 91.0089 & 91.0734 \\
$\mathscr{L}(50,0.50,60\%)$	& 52.8733 & 75.1816 & 84.4068 & 89.5319 & 92.5504 & 94.4310 & 95.6659 & 96.5214 & 96.9976 & 97.2074\\
$\mathscr{L}(50,0.75,60\%)$	& 45.4883 & 66.6990 & 77.7966 & 85.2220 & 90.4681 & 93.6804 & 95.7143 & 97.0218 & 98.1679 & 98.8216 \\
\hline
$\mathscr{T}(10,0,40\%)$	&35.6336 & 50.7506 & 61.3156 & 69.5238 & 76.5698 & 82.9701 & 88.1921 & 91.4851 & 94.4391 & 96.5214 \\
$\mathscr{T}(20,0,40\%)$	&34.7538 & 49.5480 & 59.8305 & 67.5787 & 74.6812 & 80.5650 & 85.4964 & 89.7498 & 92.8410 & 94.9314  \\
$\mathscr{T}(30,0,40\%)$	&35.2623 & 49.7094 & 59.9274 & 68.0307 & 75.3592 & 81.3963 & 85.8596 & 89.6852 & 92.7603 & 94.7619 \\
$\mathscr{T}(40,0,40\%)$	&32.1792 & 44.0597 & 54.6166 & 63.0670 & 70.2018 & 76.5133 & 81.8563 & 86.3842 & 90.0565 & 92.6069 \\
$\mathscr{T}(50,0,40\%)$	&30.5892 & 43.0508 & 53.3737 & 61.8967 & 68.7248 & 74.4713 & 79.5480 & 83.0831 & 86.3115 & 88.4181 \\
\rowcolor{mygray}
$\mathscr{T}(50,0.25,40\%)$	&31.3721 & 45.0202 & 54.7619 & 63.3737 & 71.1461 & 77.5787 & 82.3648 & 86.3519 & 89.6287 & 91.8563 \\
$\mathscr{T}(50,0.50,40\%)$	&32.8814 & 45.7224 & 56.3519 & 64.9798 & 72.0097 & 78.2324 & 83.2768 & 87.7643 & 91.1299 & 93.9629 \\
$\mathscr{T}(50,0.75,40\%)$	&34.0113 & 47.1832 & 57.7805 & 66.9895 & 74.4310 & 80.7425 & 86.0775 & 89.9031 & 93.1638 & 95.2139 \\
\bottomrule
\end{tabular}
\end{center}
\end{table}


\clearpage
\begin{table}[!htbp]
\caption{Maximum accuracy rates of dimension-by-dimension recognition with SVM and top N-Best on DII.}
\label{table:4}
\renewcommand{\tabcolsep}{0.2pc} 
\renewcommand{\arraystretch}{1}
\begin{center}
\begin{tabular}{cllllllllll}
\toprule
& & & & & & & & & &\\
\hline
\multirow{2}{*}{Data set} & \multicolumn{10}{c}{N--candidate} \\
\cmidrule{2-11}
&1 &2 &3 &4 &5 &6 &7 &8 &9 &10\\
\midrule
$\mathscr{L}(10,0,80\%)$	&29.3438 & 45.0915 & 55.0882 & 63.9969 & 70.9697 & 76.6488 & 81.5802&  85.8802&  89.7327&  93.0207 \\
$\mathscr{L}(20,0,80\%)$	&29.3552 & 47.3296 & 57.4534 & 67.0448 & 73.5479 & 79.2198 & 84.0477&  88.0165 & 91.2125&  94.4000 \\
$\mathscr{L}(30,0,80\%)$	&30.5824 & 49.0668 & 61.6422 & 70.3588 & 76.4742 & 81.6552 & 86.1316 & 90.2112 & 93.0098 & 95.5282 \\
$\mathscr{L}(40,0,80\%)$	&31.3881 & 51.0709 & 61.8799 & 70.5493 & 78.1694 & 83.2345 & 87.7401 & 91.4764 & 94.3804 & 96.6437  \\
$\mathscr{L}(50,0,80\%)$	&35.3590 & 52.9908 & 65.4033 & 74.5144 & 81.4203 & 86.4866 & 90.1496 & 93.3702 & 95.9525 & 97.7147 \\
\hline
$\mathscr{T}(10,0,20\%)$	&30.3086 & 46.2396 & 54.6489 & 63.0856 & 69.1697 & 74.5647 & 79.6372 & 83.9981 & 88.0772 & 91.2766\\
$\mathscr{T}(20,0,20\%)$	&29.2051 & 46.0697 & 55.1003 & 62.7021 & 68.8629 & 74.2264 & 79.0846 & 83.8409 & 87.8995  &90.9838\\
$\mathscr{T}(30,0,20\%)$	&29.5038 & 46.0827 & 55.0661 & 62.0462 & 68.2403 & 73.6785 & 78.8558 & 83.4111 & 87.6787 & 91.3259\\
$\mathscr{T}(40,0,20\%)$	&28.5826 & 45.2497 & 53.3937 & 60.1408 & 66.9529 & 73.0946 & 78.3932 & 82.8464 & 87.2144  &91.0958\\
$\mathscr{T}(50,0,20\%)$	&26.8566 & 44.1238 & 53.0242 & 60.4083 & 67.1062 & 72.7848 & 78.6045 & 83.8256 & 88.2956  &91.9439\\
\hline
\hline
$\mathscr{L}(10,0,60\%)$	&29.6971 & 45.4136 & 56.6370 & 65.3096 & 72.1330 & 78.4000 & 83.3930 & 87.6122 & 91.1014  &94.0061\\
$\mathscr{L}(20,0,60\%)$	&29.8586 & 48.0800 & 61.0491 & 70.1224 & 76.3722 & 81.3218 & 85.6909 & 89.5342 & 92.9098  &95.2152\\
$\mathscr{L}(30,0,60\%)$	&31.7213 & 50.4551 & 62.1493 & 72.1824 & 79.1184 & 84.5586 & 88.9960 & 92.3518 & 94.7919  &96.7793\\
$\mathscr{L}(40,0,60\%)$	&34.0660 & 54.5520 & 66.7263 & 75.5318 & 82.2952 & 87.2818 & 91.1775 & 94.1952 & 96.3349  &97.9076\\
$\mathscr{L}(50,0,60\%)$	&39.8064 & 57.2994 & 70.1210 & 79.5504 & 86.0678 & 90.7522 & 93.9192 & 96.1972 & 97.9287  &98.9955\\
\hline
$\mathscr{T}(10,0,40\%)$	&29.4029 & 45.2738 & 54.6847 & 62.3008 & 68.9545 & 74.4504 & 79.6967 & 84.0683 & 88.0486  &91.5156\\
$\mathscr{T}(20,0,40\%)$	&27.8420 & 45.2155 & 54.0493 & 61.6988 & 67.9848 & 73.4767 & 78.4582 & 83.3492 & 87.3313  &90.7006\\
$\mathscr{T}(30,0,40\%)$	&27.4326 & 44.8435 & 53.7272 & 60.4323 & 67.7084 & 73.4091 & 78.5520 & 83.4251 & 87.7196  &91.3985 \\
$\mathscr{T}(40,0,40\%)$	&26.3455 & 43.5107 & 52.5305 & 61.0567 & 67.4679 & 73.0270 & 78.2297 & 82.8868 & 87.4177  &91.2270\\
$\mathscr{T}(50,0,40\%)$	&25.8717 & 42.9881 & 52.4357 & 60.2326 & 67.0739 & 73.7870 & 79.8593 & 84.8659 & 89.1454  &92.6294 \\
\bottomrule
\end{tabular}
\end{center}
\end{table}


\begin{table}[!htbp]
\caption{Maximum accuracy rates of sequence-by-sequence recognition with SVM and top N-Best on DII.}
\label{table:5}
\renewcommand{\tabcolsep}{0.2pc} 
\renewcommand{\arraystretch}{1}
\begin{center}
\begin{tabular}{cllllllllll}
\toprule
& & & & & & & & & &\\
\hline
\multirow{2}{*}{Data set} & \multicolumn{10}{c}{N--candidate} \\
\cmidrule{2-11}
&1 &2 &3 &4 &5 &6 &7 &8 &9 &10\\
\midrule

$\mathscr{L}(10,0,80\%)$	&38.5928 & 59.4387 & 71.6031 & 79.3257 & 85.3321 & 90.3392 & 95.7601 & 97.8397 & 98.6473  &99.2833\\
$\mathscr{L}(20,0,80\%)$	&53.3313 & 71.7141 & 81.8393 & 88.3909 & 92.3178 & 95.0333 & 96.9110 & 97.8397 & 98.6574  &99.2631 \\
$\mathscr{L}(30,0,80\%)$	&60.3170 & 76.0044 & 87.4823 & 92.7519 & 95.3463 & 97.3047 & 98.4757 & 98.9804 & 99.4246  &99.7072\\
$\mathscr{L}(40,0,80\%)$	&61.0337 & 79.8506 & 88.5524 & 93.1052 & 96.0933 & 97.6782 & 98.6170 & 99.1016 & 99.4044  &99.6164\\
$\mathscr{L}(50,0,80\%)$	&65.0616 & 85.1605 & 92.1260 & 95.0030 & 96.7494 & 97.8498 & 98.6069 & 99.1016 & 99.2227  &99.3438\\
\hline
$\mathscr{T}(10,0,20\%)$	&37.1981 & 53.6232 & 63.6876 & 71.6989 & 79.2673 & 84.0580 & 88.6071 & 92.9549 & 95.8132  &96.9807 \\
$\mathscr{T}(20,0,20\%)$	&36.7552 & 52.1337 & 63.8889 & 74.1143 & 80.7166 & 87.1578 & 90.4589 & 93.7601 & 95.6119  &96.9002\\
$\mathscr{T}(30,0,20\%)$	&37.6409 & 53.3414 & 63.4461 & 72.2222 & 79.7504 & 85.5475 & 90.1369 & 92.9147 & 95.2496  &96.6586 \\
$\mathscr{T}(40,0,20\%)$	&34.7826 & 48.7923 & 58.2931 & 67.0692 & 75.4026 & 81.6828 & 86.5539 & 90.5797 & 93.5588  &95.6522\\
$\mathscr{T}(50,0,20\%)$	&33.9372 & 48.5910 & 59.4605 & 68.0757 & 76.0870 & 81.2802 & 86.0306 & 90.3784 & 93.2770  &95.6522\\
\hline
\hline
$\mathscr{L}(10,0,60\%)$	&40.6839 & 65.4685 & 78.3522 & 86.3893 & 91.1820 & 94.0899 & 96.6074 & 98.0075 & 98.8422  &99.3538 \\
$\mathscr{L}(20,0,60\%)$	&60.9989 & 77.4906 & 86.1201 & 91.7744 & 94.8438 & 96.9844 & 98.0614 & 98.7076 & 99.1788  &99.6904 \\
$\mathscr{L}(30,0,60\%)$	&66.7340 & 86.1470 & 92.9052 & 95.9208 & 97.4690 & 98.5460 & 99.2057 & 99.5961 & 99.7981  &99.9192  \\
$\mathscr{L}(40,0,60\%)$	&71.9171 & 88.3549 & 94.3726 & 97.0382 & 98.1556 & 98.8018 & 99.2326 & 99.5019 & 99.6634  &99.7711 \\
$\mathscr{L}(50,0,60\%)$	&75.8481 & 91.5590 & 95.6247 & 97.3748 & 98.4787 & 99.0980 & 99.3134 & 99.4884 & 99.5019  &99.5288 \\
\hline
$\mathscr{T}(10,0,40\%)$	&37.4244 & 55.0786 & 65.7195 & 73.9218 & 81.1769 & 86.5175 & 90.7900 & 93.7324 & 95.9492  &97.3398  \\
$\mathscr{T}(20,0,40\%)$	&37.2229 & 51.5316 & 62.6562 & 72.2289 & 80.6328 & 85.7316 & 89.8428 & 92.5836 & 95.0020  &97.0173\\
$\mathscr{T}(30,0,40\%)$	&37.1826 & 53.4462 & 63.9863 & 72.8537 & 79.4236 & 85.3688 & 89.8025 & 93.0270 & 95.0625  &96.7352\\
$\mathscr{T}(40,0,40\%)$	&34.6030 & 49.4156 & 59.6131 & 68.1378 & 75.5341 & 81.7009 & 86.8601 & 90.5885 & 93.3495  &95.9089 \\
$\mathscr{T}(50,0,40\%)$	&33.1520 & 47.2189 & 58.7666 & 67.6340 & 74.6070 & 80.9754 & 86.0339 & 90.1451 & 93.2890  &95.4454\\
\bottomrule
\end{tabular}
\end{center}
\end{table}


\begin{thebibliography}{56}


\bibitem{1}Ku CS,Roukos DH.(2013) From next-generation sequencing to nanopore sequencing technology: paving the way to personalized genomic medicine. \textit{Expert Rev. Med. Devices}, \textbf{10(1)},1--6.

\bibitem{2} Lipman DJ,Pearson WR.(1985) Rapid and sensitive protein similarity searches. \textit{Science}, \textbf{227 (4693)}, 1435--1441.

\bibitem{3}Pearson WR,Lipman DJ.(1988) Improved tools for biological sequence comparison. \textit{Proc. Natl. Acad. Sci. USA}, \textbf{85(8)}, 2444--2448.

\bibitem{4} Altschul SF,Gish W,Miller W,Myers EW,Lipman DJ.(1990) Basic local alignment search tool. \textit{Journal of Molecular Biology}, \textbf{215}, 403--410.

\bibitem{5}Krogh A,Mian IS,Haussler D.(1994) A hidden Markov model that finds genes in E. coli DNA. \textit{Nucleic Acids Research}, \textbf{22}, 4768--4778.

\bibitem{6}Burge C,Karlin S.(1997) Prediction of complete gene structures in human genomic DNA. \textit{Journal of Molecular Biology}, \textbf{268}, 78--94.

\bibitem{7}Salzberg SL,Delcher AL,Kasif S,White O.(1998) Microbial gene identification using interpolated Markov models. \textit{Nucleic Acids Research}, \textbf{26(2)}, 544-548.

\bibitem{8}Lukashin AV,Borodovsky M.(1998) GenMark.hmm: new solutions for gene finding. \textit{Nucleic Acids Research}, \textbf{26(4)}, 1107-1115.

\bibitem{9}Pedersen JS,Hein J.(2003) Gene finding with a hidden Markov model of genome structure and evolution. \textit{Bioinformatics}, \textbf{19(2)}, 219-227.

\bibitem{10}Cawley SL,Pachter L.(2003) HMM sampling and applications to gene conding and alternative splicing. \textit{Bioinformatics}, \textbf{19 Suppl. 2}: ii36--ii41.

\bibitem{11}DePristo MA,Banks E,Poplin RE,Garimella KV,Maguire JR,Hartl C,Philippakis AA,del Angel G,Rivas MA,  Hanna M, et al.(2011) A framework for variation discovery and genotyping using next-generation DNA sequencing data. \textit{Nat Genet.} , \textbf{43(5)}, 491--498.

\bibitem{12}Abubucker S,Segata N,Goll J,Schubert AM,Izard J,Cantarel BL,Rodriguez-Mueller B,Zucker J,
Thiagarajan M,Henrissat B, et al. (2012) Metabolic reconstruction for metagenomic data and its application to the human microbiome. \textit{PLoS Comput Biol}, \textbf{8(6)}: e1002358.
	
\bibitem{13}Segata N,Izard J,Waldron L,Gevers D,Miropolsky L,Garrett WS, Huttenhower C.(2011)Metagenomic biomarker discovery and explanation. \textit{Genome Biology}, \textbf{12(6)}:R60.

\bibitem{14}Segata N,Waldron L,Ballarini A,Narasimhan V,Jousson O,Huttenhower C.(2013) Metagenomic microbial community profiling using unique clade-specific marker genes. \textit{Nature Methods},\textbf{9(8)}, 811--814.

\bibitem{15}Skewes-Cox P,Sharpton TJ,Pollard KS,DeRisi JL.(2014) Profile hidden Markov models for the detection of viruses within metagenomic sequence data. \textit{PLoS ONE}, \textbf{9(8)}: e105067.


\bibitem{16}Brown MP,Grundy WN,Lin D,Cristianini N,Sugnet CW,Furey TS,Ares M,Haussler D.(2000) Knowledge-based analysis of microarray gene expression data by using support vector machines. \textit{Proc. Natl Acad. Sci. USA}, \textbf{97}, 262--267.

\bibitem{17}Ramaswamy S,Tamayo P,Rifkin R,Mukherjee S,Yeang CH,Angelo M,Ladd C,Reich M,Latulippe E,Mesirov JP, et al.(2001) Multiclass cancer diagnosis using tumor gene expression signatures. \textit{Proc. Natl Acad. Sci. USA}, \textbf{98}, 15149--15154.

\bibitem{18}Guyon I,Weston J,Barnhill S,Vapnik V.(2002) Gene selection for cancer classification using support vector machines. \textit{Machine Learning}, \textbf{46}, 389--422.
\bibitem{19} Bao L,Sun ZR.(2002) Identifying genes related to drug anticancer mechanisms using support vector machine. \textit{FEBS Letters}, \textbf{521}, 109--114.
\bibitem{20}Cho SB,Won HH.(2003) Machine learning in DNA microarray analysis for cancer classification. \textit{Proceedings of the First Asia-Pacific bioinformatics conference on Bioinformatics}, 189-198.

\bibitem{21}Su Y,Murali TM,Pavlovic V,Schaffer M,Kasif S.(2003) RankGene: identification of diagnostic genes based on expression data. \textit{Bioinformatics}, \textbf{19(12)}, 1578--1579.

\bibitem{22}Li F,Yang YM.(2005) Using recursive classification to discover predictive features. \textit{ACM Symposium on Applied Computing}. March 13-17, 2005, Santa Fe, New Mexico, USA. 104--1058.

\bibitem{23}Krause L,McHardy AC,Nattkemper TW,Puhler A,Stoye J,Meyer F.(2007) GISMO--gene identification using a support vector machine for ORF classification. \textit{Nucleic Acids Research}, \textbf{35(2)}, 540--549.

\bibitem{24}Zhou X,Tuck DP.(2007) MSVM-RFE: extensions of SVM-RFE for multiclass gene selection on DNA microarray data. \textit{Bioinformatics}, \textbf{23(9)}, 1106--1114.

\bibitem{25}Tsai MH,Chang JD,Chiu SH,Lai CH.(2007) Identification of marker genes discriminating the pathological stages in ovarian carcinoma by using support vector machine and systems biology. in  Randall M, Abbass HA, Wiles J (eds): \textit{ACAL 2007}, LNAI 4828, 381--389.

\bibitem{26}Wu S,Zhang Y.(2008) A comprehensive assessment of sequence-based and template-based methods for protein contact prediction. \textit{Bioinformatics}, \textbf{24}, 924--931.

\bibitem{27}Sinha S,Vasulu TS,De RK.(2009) Performance and evaluation of microRNA gene identification tools. \textit{Journal of Proteomics \& Bioinformatics}, \textbf{2}, 336--343.

\bibitem{28}Yousef M,Ketany M,Manevitz L,Showe LC,Showe MK.(2009) Classification and biomarker identification using gene network modules and support vector machines. \textit{BMC Bioinformatics}, \textbf{10}: 337.

\bibitem{29}Liang Y,Zhang F,Wang J,Joshi T,Wang Y,Xu D. (2011) Prediction of drought-resistant genes in arabidopsis thaliana Using SVM-RFE. \textit{PLoS ONE}, \textbf{6(7)}: e21750.

\bibitem{30}Chen ZY,Li JP,Wei LW,Xu WX,Shi Y.(2011) Multiple-kernel SVM based multiple-task oriented data mining system for gene expression data analysis. \textit{Expert Systems with Applications}, \textbf{38}, 12151-12159.

\bibitem{31}Liu YC,Guo JT,Hu GQ,Zhu HQ.(2013) Gene prediction in metagenomic fragments based on the SVM algorithm. \textit{BMC Bioinformatics}, \textbf{14(Suppl 5)}:S12.

\bibitem{32}Lu TP,Hsu YY,Lai LC,Tsai MH,Chuang EY.(2014) Identification of gene expression biomarkers for predicting radiation exposure. \textit{Scientific Reports}, \textbf{4} : 6293.

\bibitem{33}Maji S,Garg D.(2014)Hybrid approach using SVM and MM2 in splice site junction identification. \textit{Current Bioinformatics}, \textbf{9}, 76--85 .

\bibitem{34}Zhang C,Yeung P,Beviglia L,Cancilla B,Tang T,Yen WC,Gurney A,Lewicki J,Hoey T,Kapoun AM.(2014) Predictive biomarker identification for response to vantictumab (OMP-18R5; anti-Frizzled) by mining gene expression data of human breast cancer xenografts. \textit{Cancer Res}, \textbf{74(19 Suppl)}: Abstract nr 2830.

\bibitem{35}Hoff KJ,Tech M,Lingner T,Daniel R, Morgenstern B, Meinicke P. (2008) Gene prediction in metagenomic fragments: a large scale machine learning approach. \textit{BMC Bioinformatics}, \textbf{9}:217.

\bibitem{36}Cortes C,Vapnik V.(1995) Support-vector networks. \textit{Machine Learning}, \textbf{20(3)}, 273--279.

\bibitem{37}Hsu CW,Lin CJ.(2002) A Comparison of Methods for Multiclass Support Vector Machines. \textit{IEEE Transactions on Neural Networks},\textbf{13}, 415--425.
\bibitem{38}Duan KB, Keerthi SS. (2005) Which is the best multiclass SVM method? An empirical study. \textit{Multiple Classifier Systems}, LNCS 3541. 278--285.

\bibitem{39}Chang CC,Lin CJ.(2011)LIBSVM : a library for support vector machines. \textit{ACM Transactions on Intelligent Systems and Technology},\textbf{2}, 27:1--27:27.



\bibitem{40}McHardy AC,Mart\'in HG,Tsirigos A,Hugenholtz P,Rigoutsos I.(2007) Accurate phylogenetic classification of variable-length DNA fragments. \textit{Nature methods}, \textbf{4(1)}, 63--72.

\bibitem{41}Chan CKK,Hsu AL,Halgamuge SK,Tang SL.(2008) Binning sequences using very sparse labels within a metagenome. \textit{BMC Bioinformatics}, \textbf{9}:215.

\bibitem{42}Chikhi R,Medvedev P.(2014)Informed and automated k-mer size selection for genome assembly.
\textit{Bioinformatics}, \textbf{30(1)}, 31--37.

\bibitem{43}Wu YW,Tang YH,Tringe SG,Simmons BA,Singer SW.(2014)MaxBin: an automated binning method to recover individual genomes from metagenomes using an expectation-maximization algorithm. \textit{Microbiome}, \textbf{2}:26.


\bibitem{44}Powers DMW.(2011) Evaluation: from precision, recall and F-Factor to ROC, informedness, markedness \& correlation. \textit{Journal of Machine Learning Technologies}, \textbf{2(1)}, 37--63.
\bibitem{45} Fawcelt T.(2006) An introduction to ROC analysis. \textit{Pattern Recognition Letters}, \textbf{27(8)}, 861--874.

\bibitem{46}Liu JW, Qian MP.(2011) Protein function prediction using kernal logistic regresssion with ROC curves. \textit{Computing and Intelligent Systems}, 491-502. Springer Berlin Heidelberg.

\bibitem{47}Jelizarow M,Guillemot V,Tenenhaus A,Strimmer K,Boulesteix AL. (2010) Over-optimism in bioinformatics: an illustration.\textit{Bioinformatics}, \textbf{26(16)}, 1990--1998.



\bibitem{48}Broberg P.(2003) Statistical methods for ranking differentially expressed genes. \textit{Genome Biology}, \textbf{4}:R41.
\bibitem{49}Boulesteix AL,Slawski M.(2009)Stability and aggregation of ranked gene lists. \textit{Briefings in Bioinformatics}, \textbf{10(5)}, 556-568.

\bibitem{50}Schwartz R,Chow YL.(1990) The N-best algorithms: an efficient and exact procedure for finding the N most likely sentence hypotheses. \textit{ICASSP-90}, Apr.3-6,1990, Albuquerque, NM. \textbf{1},81-84.

\bibitem{51}Pusateri E,Thong JMV.(2001) N-best list generation using word and phoneme recognition fusion. in Dalsgaard P, Lindberg B, Benner H, Tan ZH (eds), \textit{INTERSPEECH}, ISCA. 1817--1820 .

\bibitem{52}Williams JD,Balakrishnan S.(2009) Estimating probability of correctness for ASR N-best lists. in Healey PGT,Pieraccini R,Byron DK,Young S,Purver M (eds) \textit{SIGDIAL Conference, The Association for Computer Linguistics}, 132--135 .


\bibitem{53} Huang K,Brady A,Mahurkar A,White O,Gevers D,Huttenhower C,Segata N.(2014) MetaRef: a pan-genomic database for comparative and community microbial genomics. \textit{Nucleic Acids Research}, \textbf{42}, Database issue D617--D624.

\bibitem{54} Liu JW.(2014) Statistical analysis of microbial genome sequence on MetaRef [E-letter]. \textit{Nucleic Acids Research}(Dec.4,2014).

\bibitem{55}Futuyma DJ.(2013) \textit{Evolution} (3rd ed.). Sinauer Associates, Inc, Sunderland, Massachusetts.

\bibitem{56} Lande R,Arnold SJ.(1983)The measurement of selection on correlated characters. \textit{Evolution}, \textbf{37}, 1210--1226.



\end{thebibliography}
\end{document}